\documentclass[submission,copyright,creativecommons]{eptcs}

\usepackage{amssymb}

\usepackage{hyperref}

\usepackage{amsmath}
\usepackage[latin1]{inputenc}
\usepackage{amssymb}
\usepackage{longtable}
\usepackage{color}
\usepackage{graphicx}
\usepackage{fontenc}
\usepackage{url}
\usepackage{breakurl}             
\usepackage{psfrag}
\usepackage{listings}
\usepackage{rotating}
\usepackage{latexsym,epsfig}
\usepackage{bm}

\usepackage{diagrams}

\usepackage{ifthen}

\newtheorem{theorem}{Theorem}[section]
\newtheorem{lemma}[theorem]{Lemma}
\newtheorem{proposition}[theorem]{Proposition}
\newtheorem{corollary}[theorem]{Corollary}

\newtheorem{definition}[theorem]{Definition}
\newtheorem{example}[theorem]{Example}
\newtheorem{remark}[theorem]{Remark}

\newenvironment{proof}[1][\!\!\,]{\vspace{1ex}\noindent\textbf{Proof #1: }}{\hfill$\Box$\vspace{2ex}}

\newcommand{\refeq}[1]{(\ref{#1})}

\newcommand{\cp}[1]{}
\newcommand{\notforsubmission}[1]{}
\newcommand{\private}[1]{\notforsubmission}

\newcommand{\notforotherstosee}[1]{}
\newcommand{\unclearAppendix}[1]{}

\newcounter{case}

\newcommand\evlist[1]{\ensuremath{\overrightarrow{#1}}}

\newcommand\labelH{\ensuremath{l}}
\newcommand\executing{\ensuremath{\frac{1}{2}}}

\newcommand\startUniv[2]{\ensuremath{[\hspace{-2.5px}\{#1\}\hspace{-2.5px}]#2}}

\newcommand\start[2]{\ensuremath{\{#1\}#2}}
\newcommand\terminate[2]{\ensuremath{\langle #1\rangle #2}}

\newcommand\startback[2]{\ensuremath{\overline{\{#1\}}#2}}
\newcommand\terminateback[2]{\ensuremath{\overline{\langle #1\rangle}#2}}

\newcommand\hHDML{\ensuremath{\mathit{hHDML}}}
\newcommand\HDA{\ensuremath{\mathit{HDA}}}
\newcommand\HDAs{\ensuremath{\mathit{HDAs}}}

\newcommand\modelH{\ensuremath{\mathcal{H}}}

\newcommand\defequal{\ensuremath{\stackrel{\vartriangle}{=}}}
\newcommand\conflict{\ensuremath{\,\#\,}}

\newcommand\atomicformulas{\ensuremath{\Phi_{B}}}

\newcommand\chu[2]{\ensuremath{(#1,#2)}}
\newcommand\C{\ensuremath{\mathsf{C}}}
\newcommand\E{\ensuremath{\mathsf{E}}}
\newcommand\allC{\ensuremath{\mathbb{C}}}
\newcommand\ST{\ensuremath{\mathsf{ST}}}
\newcommand\allST{\ensuremath{\mathbb{ST}}}
\newcommand\allHDA{\ensuremath{\mathbb{HDA}}}
\newcommand\STC{\ensuremath{\mathsf{STC}}}

\newcommand\allEv{\ensuremath{\mathbb{E}}}
\newcommand\cancellation{\ensuremath{\times}}
\newcommand\stepTransConfGlabbeek{\ensuremath{\rightarrow_{C}}}
\newcommand\stepTransEvGlabbeek{\ensuremath{\rightarrow_{E}}}

\newcommand\enableRelEv{\ensuremath{\vdash}}

\newcommand\seqSTC{\ensuremath{\cdot}}

\newcommand\frestrict[1]{\ensuremath{\upharpoonright_{#1}}}
\newcommand\reffun{\ensuremath{\mathit{ref}}}
\newcommand\refinement[1]{\ensuremath{\widetilde{#1}}}
\newcommand\isomorphic{\ensuremath{\cong}}
\newcommand\isomorphicHDA{\ensuremath{\cong}}

\newcommand\parallelSet[1]{\ensuremath{||(#1)}}
\newcommand\pomset[1]{\ensuremath{\mathit{pomset}(#1)}}
\newcommand\stintoh{\ensuremath{\mathsf{H}}}
\newcommand\stintosculpture{\ensuremath{\stintoh^{\mathsf{s}}}}
\newcommand\hintost{\ensuremath{\mathsf{ST}}}
\newcommand\unfolding{\ensuremath{\mathsf{U}}}
\newcommand\equiatingEvents[1]{\ensuremath{\mathsf{Eq}^{#1}}}
\newcommand\sculpintost{\ensuremath{\mathsf{ST}_{\!\mathsf{s}}}}
\newcommand\hintostScultures{\ensuremath{\mathsf{ST}_{\!\mathsf{b}}}}
\newcommand\cintost{\ensuremath{\mathsf{ST}}}
\newcommand\cintostSecond{\ensuremath{\mathsf{ST}_{\hspace{-0.4ex}2}}}
\newcommand\cintostThird{\ensuremath{\mathsf{ST}_{\hspace{-0.4ex}3}}}
\newcommand\stintoc{\ensuremath{\mathsf{C}}}
\newcommand\stintoe{\ensuremath{\mathsf{E}}}
\newcommand\eintost{\ensuremath{\cintost}}
\newcommand\STofC[1]{\ensuremath{\cintostSecond(#1)}}
\newcommand\stEnc{\ensuremath{\ST}}

\newcommand{\chuPrat}{\ensuremath{\mathsf{Chu}}}
\newcommand\counit{\ensuremath{\epsilon}}

\newcommand\restrictedToSet[1]{\ensuremath{\!\!\upharpoonright_{#1}}}
\newcommand\quotientofwrt[2]{\ensuremath{#1\!/\!_{#2}}}
\newcommand\applyChainList[2]{\ensuremath{#1[#2]}}
\newcommand\sculpture[2]{\ensuremath{#1^{#2}}}
\newcommand\alphachain[1]{\ensuremath{\alpha_{#1}}}
\newcommand\embedMorphism{\ensuremath{\mathit{em}}}
\newcommand\finishPath[1]{\ensuremath{\mathit{en}(#1)}}
\newcommand\startPath[1]{\ensuremath{\mathit{st}(#1)}}
\newcommand\eventEquivHDAs{\ensuremath{\stackrel{\mathsf{ev}}{\sim}}}
\newcommand\eventEquivHDAsculpture{\ensuremath{\overset{\mathsf{ev}}{\underset{\mathsf{b}}{\sim}}}}
\newcommand\eventEquivFromBulk[1]{\ensuremath{\underset{\mathsf{#1}}{\sim}}}
\newcommand\chainEquivHDAsculpture{\ensuremath{\overset{\mathsf{c}}{\sim}}}
\newcommand\cellEquivBulk{\ensuremath{\underset{\mathsf{b}}{\sim}}}
\newcommand{\equivClass}[2][]{\ensuremath{[#2]\ifthenelse{\equal{#1}{}}{}{_{#1}}}}
\newcommand\homotopyClass[1]{\ensuremath{\overleftrightarrow{[#1]}}}
\newcommand\adjacentHDA{\ensuremath{\!\stackrel{adj}{\longleftrightarrow}\!}}
\newcommand\ladjacentHDA[1]{\ensuremath{\!\stackrel{#1}{\longleftrightarrow}\!}}
\newcommand\homotopicHDA{\ensuremath{\!\stackrel{hom}{\longleftrightarrow}\!}}
\newcommand\ccequiv{\ensuremath{\stackrel{cc}{\sim}}}
\newcommand\hhequiv{\ensuremath{\stackrel{hh}{\sim}}}
\newcommand\hequiv{\ensuremath{\stackrel{h}{\sim}}}
\newcommand\modalequiv{\ensuremath{\stackrel{\hHDML}{\sim}}}

\newcommand\concurr{\ensuremath{||}}
\newcommand\causes{\ensuremath{<}}
\newcommand\causeseq{\ensuremath{\leq}}
\newcommand\ststruct{\ensuremath{\ST}}
\newcommand\sttrace[1]{\ensuremath{\mathit{st}(#1)}}
\newcommand\sttraceGlabbeek[1]{\ensuremath{\mathit{st}^{G}(#1)}}
\newcommand\categoryST{\ensuremath{\allST}}

\newcommand\categoryHDA{\ensuremath{\allHDA}}

\newcommand{\transition}[1]{\ensuremath{\xrightarrow{#1}}}
\newcommand{\transitions}[1]{\ensuremath{\xrightarrow[s]{#1}}}
\newcommand{\transitiont}[1]{\ensuremath{\xrightarrow[t]{#1}}}
\newcommand{\transitionUpDown}[2]{\ensuremath{\xrightarrow[#2]{#1}}}
\newcommand{\homotopic}[1]{\ensuremath{\stackrel{#1}{\longleftrightarrow}}}

\newcommand{\imply}{\ensuremath{\,\rightarrow\,}}

\newcommand{\bottom}{\perp}

\begin{document}

\title{Extensions of Configuration Structures}

\author{Cristian Prisacariu
\institute{Dept. of Informatics, University of Oslo, \ -- \ P.O.\ Box 1080 Blindern, N-0316 Oslo, Norway.}
\email{cristi@ifi.uio.no}
}
\def\titlerunning{Extensions of Configuration Structures}
\def\authorrunning{C.~Prisacariu
}

\maketitle

\begin{abstract}
The present paper defines ST-structures (and an extension of these, called STC-structures). The main purpose is to provide concrete relationships between highly expressive concurrency models coming from two different schools of thought: the higher dimensional automata, a \textit{state-based} approach of Pratt and van Glabbeek; and the configuration structures and (in)pure event structures, an \textit{event-based} approach of van Glabbeek and Plotkin. In this respect we make comparative studies of the expressive power of ST-structures relative to the above models. Moreover, standard notions from other concurrency models are defined for ST(C)-structures, like steps and paths, bisimilarities, and action refinement, and related results are given.
These investigations of ST(C)-structures are intended to provide a better understanding of the \textit{state-event duality} described by Pratt, and also of the (a)cyclic structures of higher dimensional automata.
\end{abstract} 

\tableofcontents

\section{Introduction}\label{sec:intro}
%


The geometric model of concurrency, studied by Pratt and van Glabbeek \cite{pratt91hda,Pratt00HDArev,Glabbeek06HDA}, is of high expressive power, thus providing a general framework for studying the differences and common features of various other models of concurrency (as done in \cite{Glabbeek06HDA} and \cite{Goubault12Category_Cubical}). This model was named Higher Dimensional Automata (\HDA) by Pratt \cite{pratt91hda}. An attractive aspect of \HDA\ is the automata-like presentation, which emphasizes the \textit{state} aspect of the modeled system (and transitions between states). This aspect is opposed to the event-based models of concurrency, like (prime, flow, (non-)stable) event structures \cite{NielsenPW79eventstructures,BoudolC88flowEventStruct,Winskel86} or configuration structures and (in)pure event structures \cite{GlabbeekP95config,GlabbeekP09configStruct}.

We see the notion of \textit{configuration} (in its various guises \cite{NielsenPW79eventstructures,GlabbeekP09configStruct,HoareMSW11CKA_foundationsJLAP}) as fundamental to event-based models.
The configuration structures, introduced in \cite{GlabbeekP95config}, are a rather general model of concurrency based on sets of events (forming the configurations of the modeled system). A thorough study of the generality and expressiveness of configuration structures is carried out in \cite{GlabbeekP09configStruct} where relations with general forms of event structures are made (where the \textit{pure event structures} are the more well behaved, instances of which are found in the literature).
The configuration structures lend themselves easily to action refinement, as studied in \cite{GlabbeekG01refinement}, which makes them an ideal candidate for incremental development of concurrent systems where the system architect starts with an abstract model which is subsequently refined to more concrete instances.

We are interested in studying such models based on sets of events, but in relation to the state-based model of higher dimensional automata. This study of event-state duality is argued for by Pratt \cite{Pratt02duality}, and the model of Chu spaces has been developed in response \cite{gupta94phd_chu,pratt95chu}. Here we take the challenge of Pratt, with insights from Chu spaces, and develop models based on sets of events, in the spirit of van Glabbeek and Plotkin \cite{GlabbeekP09configStruct}. We call this model \textit{ST-structures}. We investigate the expressiveness and relationships of this new model with the ones we mentioned above  (i.e., with (in)pure event structures and configuration structures of \cite{GlabbeekP09configStruct} and with the triadic event structures, Chu spaces, and \HDA\ of Pratt). 
We also investigate how definitions that one finds for configuration structures are extended to this new setting. 
In particular, we define for ST-structures a hereditary history preserving bisimulation, which in \cite{GlabbeekG01refinement} is the most expressive equivalence presented for configuration structures.
We also investigate the notion of action refinement (and properties of this) for ST-structures.  

We point out shortcomings in the expressiveness of the ST-structures using examples from the literature. We then present an extension with the notion of \textit{cancellation}, advocated by Pratt \cite{Pratt03trans_cancel}. In this extension, called \textit{STC-structures}, we are able to investigate closer the \HDA\ with cycles. This extended model opens the way to tackling the problem posed by Pratt in \cite{Pratt00HDArev} about the expressive power of \HDAs\ with cycles wrt.\ event-based models.

The notion of an ST-configuration  has been used in \cite{GlabbeekV97splitting} to define ST-bisimulation and in \cite{Glabbeek06HDA} in the context of \HDA. But the model of \textit{ST-structures}, as we define here for capturing concurrency, does not appear elsewhere.\footnote{I am thankful for having been made aware of the invited talk of van Glabbeek at CONCUR'99 \cite{Glabbeek99invitedCONCUR} where it is mentioned (at the end of Sec.1) as future work the investigation, on the same lines as the work of \cite{GlabbeekP95config}, of ``translations between arbitrary Petri nets and ST-structures, showing that also these models are equally expressive''; nevertheless, their recent work \cite{GlabbeekP09configStruct} does not present such an investigation yet.} We think that a main characteristic of higher dimensional automata is captured by ST-structures, opposed to the standard configuration structures; this is the power to look at the currently executing concurrent events (not only observe their termination). 
In other words, we can now talk about what happens \emph{during} the concurrent execution of one or more events. This is opposed to standard models that talk only about what happens \emph{after} the execution (which may have duration and complex structure, apparent only after subsequent refinements of an initial abstract model).



\section{ST-structures}\label{sec_st_configs}

We define ST-structures, showing in Section~\ref{subsec_expressST} that they are a natural extension of configuration structures \cite{GlabbeekP09configStruct}, and define related notions that stem from the latter. 
The classical notions of concurrency, causality, and conflict are not interdefinable as in the case of event structures or stable configuration structures; but are more loose, as is the case with \HDAs.
In Section~\ref{subsec_expressST} we relate ST-structures also to \HDAs\ by identifying a corresponding class of ST-structures, i.e., with the particular property of \textit{adjacent-closure}. We also define the class of stable ST-structures and relate this with their counterpart in stable configuration structures. 
We define the (hereditary) history preserving bisimulation in the context of ST-structures, which when \textit{stability} is imposed on adjacent-closed ST-structures it corresponds to the same bisimulation for stable configuration structures. 
In Section~\ref{subsec_actref} we define action refinement for ST-structures and investigate properties of it, like being preserved under the above bisimulation, or that it preserves the properties of the refined ST-structures.

\begin{definition}[ST-configuration]\label{def_STconfig}
An \emph{ST-configuration} is a pair of finite sets $(S,T)$ 
 of events (i.e., $S,T\subseteq E$) 
respecting the property:
\begin{center}
(start before terminate)\ \ $T\subseteq S$.
\end{center}
\end{definition}

Intuitively $S$ contains the events that have \textit{started} and $T$ the events that have \textit{terminated}.
Define the \emph{dimension} of an ST-configuration to be $|(S,T)|=|S|+|T|$.

\begin{definition}[ST-structures]\label{def_st_structs}
An \emph{ST-configuration structure} (also called \emph{ST-structure}) is a tuple $\ST=(E,ST,l)$ with $ST$ a set of ST-configurations satisfying the \emph{constraint}:
\begin{equation}
\hspace{6ex}\mbox{ if } (S,T)\in ST \mbox{ then }(S,S)\in ST, \label{eq_ST_constraint}
\end{equation}
and  $l:E\rightarrow \Sigma$ a labelling function with $\Sigma$ the set of labels.
We often omit the set of events $E$ from the notation when there is no danger of confusion.
\end{definition}

The constraint \refeq{eq_ST_constraint} above is a closure, ensuring that we do not represent events that are started but never terminated. The set of all ST-structures is denoted \allST.

\begin{definition}[stable ST-structures]\label{def_stableST}
An ST-structure $(ST,l)$ is called:
\begin{enumerate}
 \item\label{def_stableST_rooted} \emph{rooted} iff $(\emptyset,\emptyset)\in ST$;
\item\label{def_stableST_connected} \emph{connected} iff\ \, for any non-empty $(S,T)\!\in\!ST$, either $\exists e\!\in\!S:(S\setminus\!e,T)\!\in\!ST$ or $\exists e\!\in\!T:(S,T\setminus\!e)\!\in\!ST$;
\item \emph{closed under bounded unions} iff for any $(S,T),(S',T'),(S'',T'')\in ST$ if $(S,T)\cup(S',T')\subseteq(S'',T'')$ then $(S,T)\cup(S',T')\in ST$;
\item \emph{closed under bounded intersections} iff for $(S,T),(S',T'),(S'',T'')\in ST$ if $(S,T)\cup(S',T')\subseteq(S'',T'')$ then $(S,T)\cap(S',T')\in ST$.
\end{enumerate}
An ST-structure is called \emph{stable} iff it is rooted, connected, and closed under bounded unions and intersections.
\end{definition}

ST-structures have a natural \textit{computational interpretation} (on the same lines of configuration structures) as \textit{steps} between ST-configurations, and \textit{paths}. Results below, like Theorem~\ref{th_configtoSTsteps}, show that this computational interpretation is more fine-grained than for  other models we compare with. 
Intuitively, opposed to standard event-based models, the computational interpretation of ST-structures naturally captures the ``during'' aspect of the events, i.e., what happens while an event is executing (before it has finished). Action refinement and bisimulation are well behaved wrt.\ this interpretation. The model of \HDAs\ do the same job but in the state-based setting.
Besides, ST-structures exhibit a natural \textit{observable information} (on the same lines as for \HDAs) as \textit{ST-traces}, which, cf.~\cite[Sec.7.3]{Glabbeek06HDA}, constitute the best formalization of observable content.

\begin{definition}[ST steps]\label{def_STsteps}
A step between two ST-configurations is defined as either:
\begin{description}
\item[s-step] $(S,T)\transitions{a}(S',T')$ when $T=T'$, $S\subset S'$, $S'\setminus S=\{e\}$ and $l(e)=a$; or
\item[t-step] $(S,T)\transitiont{a}(S',T')$ when $S=S'$, $T\subset T'$, $T'\setminus T=\{e\}$ and $l(e)=a$.
\end{description}
When the type is unimportant we denote a step by \, $\transition{a}$ \, for \, $\transitions{a}\cup\transitiont{a}$.
\end{definition}

\begin{definition}[paths and traces]\label{def_pathstrace}
A \emph{path} of an ST-structure, denoted $\pi$, is a sequence of steps, where the end of one is the beginning of the next, i.e.,
\[
\pi\defequal(S,T)\transition{a}(S',T')\transition{b}(S'',T'')\dots
\]
A path is \emph{rooted} if it starts in $(\emptyset,\emptyset)$.\footnote{We generally work with rooted paths.}
The \emph{ST-trace of a rooted path $\pi$}, denoted $\sttrace{\pi}$, is the sequence of labels of the steps of $\pi$ where each label is annotated as $a^{0}$ if it labels an s-step or as $a^{n}$ if it labels a t-step, where $n\in\mathbb{N}^{+}$ is determined by counting the number of steps until the s-step that has added the event $e$ to the $S$ set, with $e$ being the event that has been added to $T$ in the current t-step.
\end{definition}

For rooted and connected ST-structures the notion of \emph{ST-trace} conforms with the one defined in \cite[def.2.5]{GlabbeekV97splitting} or \cite[sec.7.3]{Glabbeek06HDA}.

\begin{proposition}[connectedness through paths]\label{prop_connectPaths}\ 
\begin{enumerate}
\item\label{prop_connectPaths_1} For a rooted ST-structure $\ST$ the following are equivalent:
\begin{enumerate}
\item\label{prop_connectPaths_1a} \ST\ is connected;
\item\label{prop_connectPaths_1b} For any $(S,T)\in\ST$ there exists a rooted path ending in $(S,T)$.
\end{enumerate}
\item\label{prop_connectPaths_2} For a rooted $\ST$ that is closed under bounded unions the following are equivalent:
\begin{enumerate}
\item\label{prop_connectPaths_2a} \ST\ is connected;
\item\label{prop_connectPaths_2b} For any two ST-configurations s.t.\ $(S,T)\!\subseteq\!(S',T')$, there exists a path starting in $(S,T)$ and ending in $(S',T')$.
\end{enumerate}
\end{enumerate}
\end{proposition}

\begin{proof}
To prove the implication (\ref{prop_connectPaths_1a})$\Rightarrow$(\ref{prop_connectPaths_1b}) use induction on the dimension of $(S,T)$ applying subsequently to smaller ST-configurations the connectedness property of Definition~\ref{def_stableST}.\ref{def_stableST_connected}.

To prove the implication (\ref{prop_connectPaths_1b})$\Rightarrow$(\ref{prop_connectPaths_1a}) is easier by using the definition of a path which implies the connectedness Definition~\ref{def_stableST}.\ref{def_stableST_connected}.

The proof of (\ref{prop_connectPaths_2a})$\Rightarrow$(\ref{prop_connectPaths_2b}) makes unions of the ST-configurations on the two rooted paths corresponding to $(S',T')$ respectively $(S,T)$. Since the paths evolve through simple steps (i.e., which remove one event at a time) and since $(S',T')$ includes $(S,T)$ we slowly reach configurations that include events not part of $(S,T)$. Union with these intermediate configurations will make up the configurations on the path we are looking for.

To prove the implication (\ref{prop_connectPaths_2b})$\Rightarrow$(\ref{prop_connectPaths_2a}) observe that (\ref{prop_connectPaths_2b}) implies (\ref{prop_connectPaths_1b}).
\end{proof}

\begin{proposition}\label{prop_pathsEqualLength}
For any ST-configuration $(S,T)$, all the rooted paths ending in  $(S,T)$ have the same length.
\end{proposition}

\begin{proof}
Each single step adds one single new event to either the $S$ or the $T$ sets. Therefore, since the number of events in the goal ST-configuration $(S,T)$ is fixed, no matter the order of adding these events, there will be the same number of steps, or event addition operations, that can be performed from the root.
\end{proof}

\begin{definition}[concurrency and causality]\label{def_ConcCausal}\ 

For a particular ST-configuration $(S,T)\in\ST$ define the relations of \emph{concurrency} and \emph{causality} on the events in $S$ as:
\begin{description}
\item[concurrency] for $e,e'\!\in\!S$ then $e\concurr e'$ iff \, exists $(S',T')\!\subseteq\!(S,T)$ s.t.\ $(S',T')\!\in\!\ST$ and $\{e,e'\}\!\subseteq\!S'\setminus T'$;
\item[causality] for $e,e'\!\in\!S$ then $e\!\causes\!e'$ iff \, $e\!\neq\!e'$ and for any $(S',T')\!\subseteq\!(S,T)$ s.t.\ $(S',T')\!\in\!\ST$, is the case that $e'\!\in\!S'\Rightarrow e\!\in\!T'$.
 \end{description}
\end{definition}

ST-structures represent \textit{concurrency} in a way that is different than other event-based models in the sense that each ST-configuration gives information about the currently concurrent events, and this information is persistent throughout the whole execution.
Two events are considered concurrent wrt.\ a particular ST-configuration if and only if at some point in the past (i.e., in some sub-configuration) both events appeared as executing (i.e., in $S'$) and none was terminated yet (i.e., not in $T'$); they were both executing concurrently.
In event structures or configuration structures in order to decide whether two events are concurrent one needs to look at many configurations or many events to decide this. For example, in event structures the concurrency is defined as not being dependent nor conflicting; which requires to inspect all configurations to decide.
An ST-configuration does not give complete information about the concurrency relation in the whole system. In consequence one could view the information about concurrency that an ST-configuration provides as being sound but not complete.

The above two notions of concurrency and causality are defined for one particular ST-configura\-tion; in consequence one could emphasize this by indexing the relation symbol by the particular ST-configuration (similar to what is done in \cite[Sec.5.3]{GlabbeekG01refinement}), but the aesthetics would not be so nice in our case.
The above two notions are lifted naturally to the whole ST-structure.

An event \textit{$e$ is a cause of $e'$} 
iff in all the past $e'$ is never started without $e$ having terminated. In other words, whenever in the past the event $e'$ is to be started (i.e., $e'\in S'$), the event $e$ on which it depends must have terminated already (i.e., $e\in T'$).

\begin{figure}[tp]
\psfrag{ee}{{\scriptsize  $(\emptyset,\emptyset)$}}
\psfrag{ae}{{\scriptsize $(a,\emptyset)$}}
\psfrag{be}{{\scriptsize $(b,\emptyset)$}}
\psfrag{0e}{{\scriptsize $(0,\emptyset)$}}
\psfrag{1e}{{\scriptsize $(1,\emptyset)$}}
\psfrag{01e}{{\scriptsize $(01,\emptyset)$}}
\psfrag{00}{{\scriptsize $(0,0)$}}
\psfrag{010}{{\scriptsize $(01,0)$}}
\psfrag{0101}{{\scriptsize $(01,01)$}}
\psfrag{011}{{\scriptsize $(01,1)$}}
\psfrag{11}{{\scriptsize $(1,1)$}}
\psfrag{b11}{{\scriptsize $(1b,1)$}}
\psfrag{b1b1}{{\scriptsize $(1b,1b)$}}
\psfrag{b01b1}{{\scriptsize $(01b,1b)$}}
\psfrag{b0101}{{\scriptsize $(01b,01)$}}
\psfrag{b010}{{\scriptsize $\mathbf{(01b,0)}$}}
\psfrag{b01b0}{{\scriptsize $(01b,0b)$}}
\psfrag{b01b01}{{\scriptsize $(01b,01b)$}}
\psfrag{b011}{{\scriptsize $\mathbf{(01b,1)}$}}
\psfrag{b00}{{\scriptsize $(0b,0)$}}
\psfrag{b0b0}{{\scriptsize $(0b,0b)$}}
\psfrag{b}{{\scriptsize $b$}}
\psfrag{a}{{\scriptsize $a$}}
\psfrag{c}{{\scriptsize $c$}}
\psfrag{bc0}{{\scriptsize $\mathbf{(bc,\emptyset)}$}}
\psfrag{ac0}{{\scriptsize $\mathbf{(ac,\emptyset)}$}}
\psfrag{s}{{\scriptsize $s$}}
\psfrag{t}{{\scriptsize $t$}}
\psfrag{E}{{$I_{RC}$}}
\psfrag{F}{{$I_{W}$}}
\begin{center}
    \includegraphics[height=4.5cm]{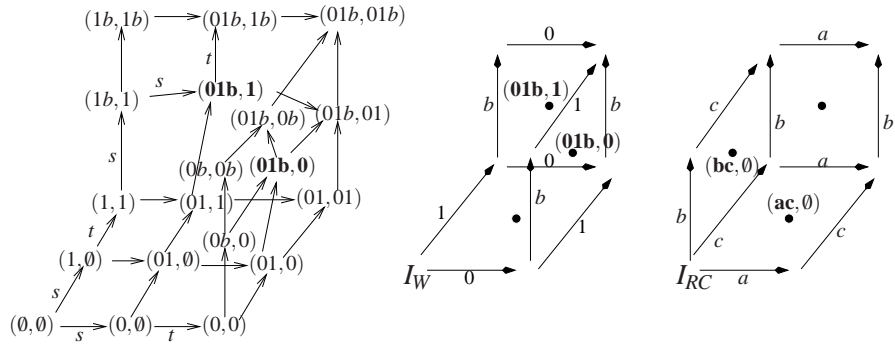}
  \end{center}
\caption{ST-structures (and acyclic \HDAs) representing, on the left, the ``parallel switch'' of Winskel \cite[Ex.1.1.7]{Winskel86} (not closed under intersections) and, on the right, the ``resolved conflict'' of \cite[Ex.2]{GlabbeekP09configStruct} (not closed under unions).}
\label{fig_ex_winskel}
\end{figure}

This is a specific notion of causality, that comes from the tradition of viewing causality as a partial order (in fact this definition makes a partial order only when the structure is rooted and connected). The definition of event structures from \cite{GlabbeekP09configStruct} define a \textit{dependency relation} that can characterize \textit{conjunctive causality} in the sense that one event depends on several events (i.e., not a binary relation any more). Besides this (rather common) causality as dependency, there is a notion of \textit{disjunctive causality} which is nicely exemplified by the ``parallel switch of Winskel'' (see Example~\ref{ex_Winskel_switch_resolved_conflict} and Figure~\ref{fig_ex_winskel} on page~\pageref{fig_ex_winskel}) where an event $b$ is caused by either of the two events $0$ or $1$ having happened.

On arbitrary ST-structures the concurrency and causality are not interdefinable (in a standard way e.g.\ \cite[Def.5.6]{GlabbeekG01refinement} where concurrency is the negation of causality). Nevertheless, concurrency and causality are disjoint on every ST-configuration of an arbitrary ST-structure.
For the more well behaved stable ST-structures the concurrency and causality are interdefinable. Even more, results similar to the ones in \cite[Sec.5.3]{GlabbeekG01refinement} can be stated and proven about stable ST-structures and their causality partial order.


\begin{proposition}\label{prop_concDisjCausal}
On arbitrary ST-structures 
\begin{enumerate}
\item concurrency and causality are \emph{disjoint};
\item concurrency and causality are \emph{not interdefinable} (in a standard way e.g.\ \cite[Def.5.6]{GlabbeekG01refinement} where concurrency is the negation of causality).
\end{enumerate}
\end{proposition}

\begin{proof}
The counterexample for the second part of the proposition consists of the empty square from Figure~\ref{fig_ex_hda}(middle-right) 
\[
\ST=(\{a,b\},\{(\emptyset,\emptyset),(a,\emptyset),(b,\emptyset),(a,a),(b,b),(ab,a),(ab,b),(ab,ab)\})
\]
with the upper right ST-configuration $(\{a,b\},\{a,b\})$. For this configuration the two events $a$ and $b$ are not causal in any order, because of the existence of the two ST-configurations $(a,\emptyset)$ and $(b,\emptyset)$. The two events $a$ and $b$ are neither concurrent, because the ST-configuration $(ab,\emptyset)$ is missing. Moreover, the two events are not conflicting in the sense of the Definition~\ref{def_conflict}.

This counterexample clearly shows how \HDAs\ and ST-structures model a notion that is eluding the standard notions of causality, concurrency, or conflict. I would call this notion \textit{interleaving} (i.e., events $a$ and $b$ are interleaving) and in such models, using this example, interleaving is thus different than concurrency, i.e., making it a standalone notion. The empty square of Pratt proves once again very good at exemplifying true concurrency, and the ST-structures and the above notions, only use it in a new light.

To show disjointness one can notice that if two events are concurrent then they cannot be causally depended in any order. The witnessing ST-configuration is exactly the configuration that witnesses the concurrency, i.e., the $(S,T)$ with $\{e,e'\}\subseteq S\setminus T$. This configuration breaks the $e\causes e'$ because $e'\in S$ and $e\not\in T$; and analogous for $e'\not\causes e$.
\end{proof}

\begin{proposition}\label{prop_concCausalStable}
Let \ST\ be a \emph{stable} ST-structure. For some $(S,T)\in\ST$ and two events $e,e'\in S$ we have: 

\centerline{$e\concurr e'$ if not $(e\causes e'\mbox{ or }e'\causes e)$.}
\end{proposition}

\begin{proof}
Knowing that $e\not\causes e'$ and $e'\not\causes e$ we show the existence of some ST-configuration $(S',T')\subseteq(S,T)$ for which $\{e,e'\}\subseteq S'\setminus T'$, hence that $e\concurr e'$.

The two assumptions are equivalent to 
\begin{itemize}
\item $\exists (S_{1},T_{1})\subseteq(S,T):e'\in S_{1} \vee e\not\in T_{1}$; and
\item $\exists (S_{2},T_{2})\subseteq(S,T):e\in S_{2} \vee e'\not\in T_{2}$.
\end{itemize}
From the fact that \ST\ is connected and closed under bounded unions, using Proposition~\ref{prop_connectPaths} we know that $(S_{1},T_{1})\transition{}^{*}(S,T)$ and $(S_{2},T_{2})\transition{}^{*}(S,T)$. This means that from $(S_{1},T_{1})$ we can reach a configuration $(S'_{1},T'_{1})\subseteq (S,T)$ where $S'_{1}$ contains both $e,e'$ but still $e\not\in T'_{1}$. The same for some $(S'_{2},T'_{2})$ s.t.\ $e,e'\in S'_{2}$ and $e'\not\in T'_{2}$. This implies that $e,e'\in S'_{1}\cap S'_{2}$, and that $e\not\in T'_{1}\cap T'_{2}$ and $e'\not\in T'_{1}\cap T'_{2}$. Because \ST\ is closed under bounded intersections it means that we have found $(S'_{1}\cap S'_{2},T'_{1}\cap T'_{2})$ which is an ST-configuration of \ST\ that is included in the original $(S,T)$ and which satisfies $\{e,e'\}\subseteq (S'_{1}\cap S'_{2})\setminus(T'_{1}\cap T'_{2})$.
\end{proof}

The notion of conflicting events is not definable for a specific ST-configuration because it is a general notion definable only on the whole ST-structure. Essentially, conflicting events can never appear in the same configuration.

\begin{definition}[conflict]\label{def_conflict}
For an ST-structure $\ST$ the notion of \emph{global conflict} is defined as a predicate over sets of events $E'\!\subseteq\! E$: 

\centerline{$\conflict\!E'$\,\ iff\ \,$\nexists(S,T)\!\in\!\mathsf{ST}$ with $E'\subseteq S$.}
\end{definition}

The standard notion of binary conflict is an instance of the above, where $E=\{e,e'\}$.
Moreover, a particular ST-configuration cannot contain conflicting events.

\begin{proposition}[partial order causality]\label{prop_partialOrderCausality}
The causality relation of Definition~\ref{def_ConcCausal} when extended with equality is a partial order iff the ST-structure \ST\ from which the ST-configuration $(S,T)$ on which \causes\ is defined, is rooted and connected.
\end{proposition}

\begin{proof}
Extend the causality relation with equality by defining $e\causeseq e'\defequal e\causes e' \vee e=e'$.

Clearly \causeseq\ is reflexive.

To prove that \causeseq\ is transitive take three events $e_{1}\causeseq e_{2}\causeseq e_{3}$ and show $e_{1}\causeseq e_{3}$. The proof is immediate when any two of the three events are equal. Thus work with the assumption $e_{1}\causes e_{2}\causes e_{3}$. By applying two times Definition~\ref{def_ConcCausal} we have that $\forall(S',T')\subseteq(S,T)$ a ST-configuration of \ST\ then $e_{3}\in S'\Rightarrow e_{2}\in T'\subseteq S'\Rightarrow e_{1}\in T'$; thus having the desired result.

To prove antisymmetry assume $e_{1}\causes e_{2}$ and $e_{2}\causes e_{1}$. Applying two times the Definition~\ref{def_ConcCausal} we get that $\forall(S',T')\subseteq(S,T)$ a ST-configuration of \ST\ $e_{2}\in S'\Rightarrow e_{2}\in T'$. But this contradicts that fact that \ST\ is rooted and connected, which implies that there is a rooted path to $(S',T')$. Hence this path coming from the root through single steps must necessarily pass through an ST-configuration that has $e_{2}$ started by not terminated.
\end{proof}

We can define a notion of equivalence that extends that of pomset-trace equivalence of \cite{GlabbeekV97splitting,GlabbeekG01refinement} to the setting of general ST-structures. This notion of equivalence, when interpreted over stable ST-structures becomes exactly the pomset-trace equivalence of \cite{GlabbeekG01refinement}.

\begin{definition}[cc-equivalence]\label{def_cc_equiv}
For an ST-configuration $(S,T)$ of some \ST\ define:
\begin{itemize}
\item the \emph{pomset} as $\pomset{(S,T)}\defequal[(S,\causes\,,\,l\!\!\upharpoonright_{S})]_{\isomorphic}$ the isomorphism class of the set $S$ where the causal relation $\causes$ of $(S,T)$ is preserved and the labeling function of the ST-structure \ST\ is restricted to $S$.

\item the \emph{parallel set} as $\parallelSet{(S,T)}\defequal\{\{e,e'\}\mid e||e'\mbox{ in }(S,T)\}$.
\end{itemize}
Two ST-configurations are cc-equivalent, written $(S,T)\ccequiv(S',T')$, iff

\centerline{$\pomset{(S,T)}=\pomset{(S',T')}$ and $\parallelSet{(S,T)}=\parallelSet{(S',T')}$.}

\noindent We say that one ST-structure \ST\ cc-simulates another $\ST'$ iff\ \ $\forall (S',T')\in\ST',\exists (S,T)\in\ST:(S,T)\ccequiv(S',T')$. Two structures are cc-equivalent, written $\ST\ccequiv\ST'$ iff\ \ they cc-simulate each other.
\end{definition}

\begin{definition}[adjacent-closure]\label{def_adj_ST}
We call an ST-structure \ststruct\ \emph{adja\-cent-closed} if the following are respected:
\begin{enumerate}
  \item if $(S,T),(S\cup e,T),(S\cup\{e,e'\},T)\!\in\!\ststruct$, with $(e\!\neq\!e')\!\not\in\!S$, then $(S\cup e',T)\!\in\!\ststruct$;
  \item if $(S,T),(S\cup e,T),(S\cup e,T\cup e')\!\in\!\ststruct$, with $e\!\not\in\! S\wedge e'\!\not\in\! T\wedge e\!\neq\!e'$, then $(S,T\cup e')\!\in\!\ststruct$;
  \item if $(S,T),(S\cup e,T),(S,T\cup e')\!\in\!\ststruct:e\!\not\in\! S\wedge e'\!\not\in\! T\wedge e\!\neq\!e'$, then $(S\cup e,T\cup e')\!\in\!\ststruct$;
  \item if $(S,T),(S,T\cup e),(S,T\cup\{e,e'\})\!\in\!\ststruct$, with $(e\!\neq\!e')\!\not\in\!T$, then $(S,T\cup e')\!\in\!\ststruct$.
\end{enumerate}
\end{definition}

Anticipating the definition of higher dimensional automata (see \cite{pratt91hda,Pratt00HDArev,Glabbeek06HDA} and the Definition~\ref{def_hda} on page~\pageref{def_hda}) one can see a correlation of the above definition of adjacent-closure on ST-structures and the cubical laws of higher dimensional automata. This correlation is even more visible in the definition of \textit{adjacency} of \cite[Def.19]{Glabbeek06HDA} which is used to define homotopy over higher dimensional automata (see Definition~\ref{def_history_HDA} on page~\pageref{def_history_HDA}). Since homotopy classes essentially define histories, then the above adjacent-closure on ST-structures intuitively makes sure that the histories of ST-configurations are not missing anything.

\begin{definition}[closure under single events]\label{def_closeSingleEv}
An ST-structure \ststruct\ is called \emph{closed under single events} 
iff $\forall (S,T)\in\ststruct,\forall e \mbox{ s.t. }e\in S\setminus T \mbox{ then }$
\begin{enumerate}
\item $(S,T\cup\{e\})\in\ststruct$ and
\item $(S\setminus\{e\},T)\in\ststruct$.
\end{enumerate}
\end{definition}

\begin{proposition}[equivalent with adjacent-closure]\label{prop_adj_equiv}
A rooted and connected ST-con\-figuration structure 

\centerline{is closed under single events\ \ iff\ \ is adjacent-closed.}
\end{proposition}

\begin{proof}
The left-to-right implication is simple. For the first condition in Def.~\ref{def_adj_ST} use the second restriction of this proposition. For the second condition we may use any of the two restrictions, as we know that $e'\in S\setminus T$.  For the third and forth condition use the first restriction, knowing that $e'\in S\setminus T$.

The right-to-left implication is more involved.

We first use induction on the reachability path to show that: 
\begin{itemize}
\item[] for every ST-configuration $(S,T)$ with $|S\setminus T|\neq 0$ then all the immediately lower ST-configurations that can reach $(S,T)$ through an s-step exist in \ststruct, i.e., 
\[
\forall e\in S\setminus T:(S\setminus \{e\},T)\in\ststruct.
\]
\end{itemize}
This would prove the second requirement for closure under single events.

Since the ST-structure that we work with is rooted and connected, then every ST-configuration is reachable from the root $(\emptyset,\emptyset)$ through a series of single steps, i.e., through a rooted path, cf.\ Proposition~\ref{prop_connectPaths}.

Because of Proposition~\ref{prop_pathsEqualLength} we can use induction on the reachability path, because there exists at least one such path, and any other path has the same distance.

\textit{Base step:} is for reachability paths of distance $1$. This means when $(S,T)=(\{e\},\emptyset)$; trivial.

For the \textit{Induction case} use the proof principle \textit{reductio ad absurdum} and assume for some $e\in S\setminus T$ the ST-configuration $(S\setminus\{e\},T)\not\in\ststruct$. From connectedness we know that  $(S,T)$ is reachable through either an s- or a t-step from an ST-configuration that has lower reachability distance.

Assume that $(S,T)$ is reachable through an s-step, thus $\exists e'\in S\setminus T$ s.t\ $(S\setminus\{e'\},T)\in\ststruct$ and $e\neq e'$. Since $e\in(S\setminus e')\setminus T$ we may apply the induction hypothesis to $(S\setminus\{e'\},T)$ to get that $(S\setminus\{e,e'\},T)$. We now can apply the first adjacent-closure requirement of Definition~\ref{def_adj_ST} to get that $(S\setminus\{e\},T)\in\ststruct$, which is a contradiction.

Assume now that no s-steps are possible, and thus only a t-step is possible from some $(S,T\setminus\{f\})$ with $f\not\in S\setminus T$, hence $f\neq e$. By applying the induction hypothesis to $(S,T\setminus\{f\})$ we get that $(S\setminus\{e\},T\setminus\{f\})\in\ststruct$, since $e\in S\setminus(T\setminus\{f\})$. We can now apply the second condition for adjacent-closure to get that $(S\setminus\{e\},T)\in\ststruct$, which is a contradiction.

It remains to show that the first requirement of closure under single events is satisfied.
Thus, for some arbitrary $(S,T)$ we use induction on the dimension of $|S\setminus T|$ to show that 
\[
\forall e\in S\setminus T:(S,T\cup e)\in\ST.
\]
We could also use induction on the reachability path (as before).

\textit{Base step} is for $|S\setminus T|=1$, i.e., when $S=T\cup e$ for some $e\not\in T$. By the definition of ST-structures we have that for our $(T\cup e,T)$ there also exists $(T\cup e,T\cup e)\in\ST$.

For the inductive case, i.e., when $\exists e,e'\in S\setminus T$ distinct, we know from the previous step of the proof that for all $f\in S\setminus T$ we have $(S\setminus f,T)\in\ST$. Pick one of these which is different than $e$, as at least one exists $e'\neq e$. Since $|(S\setminus e')\setminus T|$ is smaller than the initial $|S\setminus T|$ we can apply the induction hypothesis to obtain that for $e\in ((S\setminus e')\setminus T)$ we have $(S\setminus e',T\cup e)\in\ST$. We may now apply the third requirement in the definition of adjacent-closure to obtain that also $(S,T\cup e)\in\ST$.
\end{proof}

One may assume to work with rooted and connected structures, not only because these are natural, but also because we can obtain them using the notion of \textit{reachability}.

\begin{definition}[reachable part]\label{def_reachability}
An ST-configuration $(S,T)$ is said to be \emph{reachable} iff there exists a rooted path ending in $(S,T)$. The \emph{reachable part} of some arbitrary ST-structure is formed of all and only the reachable configurations. 
\end{definition}

The reachable part of a structure is connected, cf.~Prop.~\ref{prop_connectPaths}.\ref{prop_connectPaths_1}. Therefore, assuming connectedness is the same as assuming to work with the reachable part of a structure. 

\begin{definition}[morphisms of ST-structures]\label{def_morphism_ST}
A \emph{morphism} $f:\ST\rightarrow\ST'$ between two ST-structures $\ST=(E,ST,l)$ and $\ST'=(E',ST',l')$ with the sets of events as in Definition~\ref{def_st_structs}, is defined as a partial function on the events $f:E\rightarrow E'$ which: 
\begin{itemize}
\item 
preserve ST-configurations, if $(S,T)\!\in\!\ST$ then $f(S,T)\!=\!(f(S),f(T))\!\in\!\ST'$,
\item 
preserve the labeling when defined, i.e., $l'(f(e))=l(e)$ if $f$ is defined for $e$, and 
\item
are locally injective, i.e., for any $(S,T)\in\ST$ the restriction $f\restrictedToSet{S}$ is injective.
\end{itemize}
\end{definition}

Note that if $T\subseteq S$ then $f(T)\subseteq f(S)$.

\begin{proposition}
The morphisms of Definition~\ref{def_morphism_ST} preserve steps.
\end{proposition}

\begin{proof}
We prove that for a step $(S,T)\transitionUpDown{e}{s}(S\cup e,T)\in\ST$ then $f(S,T)$ can make an s-step with the event $f(e)$ into the corresponding ST-configuration $f(S\cup e,T)$. Since $(S\cup e,T)\in\ST$ then $f(S\cup e,T)$ is also a configuration in $\ST'$ and thus $f$ is defined for $e$. Since $e\not\in S$ (by definition of an s-step) it means that $e$ is different than any other event $g$ from $S$, and by the injective property of $f$ it means that $f(e)\not\in f(S)$. Moreover, the label is preserved. Therefore we have the s-step in $\ST'$ with the same label and the corresponding event, $(f(S),f(T))\transitionUpDown{f(e)}{s}(f(S)\cup f(e),f(T))$.
\end{proof}

We can define a \textit{category \categoryST} to have objects ST-structures and the morphisms from Definition~\ref{def_morphism_ST} because composition of morphism is well defined and for any ST-structure there exists a unique identity morphism which is the total function taking an event to itself.

\begin{definition}[isomorphic ST-structures]\label{def_isomorphism}
A function $f$ is an \emph{isomorphism} of two ST-configurations $(S,T)f(S',T')$\,\ iff\,\ $f$ is an isomorphism of $S$ and $S'$ that agrees on the sets $T$ and $T'$ (i.e., $f\!\!\!\upharpoonright_{T}=T'$).
Two ST-structures $\ST$ and $\ST'$ are isomorphic, denoted $\ST\isomorphic\ST'$, iff there exists a bijection $f$ on their events that is also a morphism between the two ST-structures; in particular, $f$ takes an ST-configuration into an isomorphic ST-configuration, and agrees on the labeling.
\end{definition}

\begin{definition}[hh-bisimulation for ST-structures]\label{def_hh_ST}\ 

For two ST-structures $\mathsf{ST}$ and $\mathsf{ST}'$, a relation $R\!\subseteq\!ST\!\!\times\!\!ST'\!\!\times\!\!\mathcal{P}(ST\!\!\times\!\!ST')$ is called a history preserving bisimulation between $\mathsf{ST}$ and $\mathsf{ST}'$ iff $(\emptyset,\emptyset,\emptyset)\!\in\!R$ and whenever $((S,T),(S',T'),f)\!\in\!R$

\begin{enumerate}
\item $f$ is an isomorphism between $(S,T)$ and $(S',T')$; and
\item if $(S,T)\transition{a}(S_a,T_a)$ then exists $(S'_a,T'_a)\in\mathsf{ST}'$ and $f'$ extending $f$ (i.e., $f'\!\!\upharpoonright_{(S,T)}=f$) with $(S',T')\transition{a}(S'_a,T'_a)$ and  $((S_a,T_a),(S'_a,T'_a),f')\in R$; and
\item if $(S',T')\transition{a}(S'_a,T'_a)$ in $\mathsf{ST}'$ then exists $(S_a,T_a)\in\mathsf{ST}$ and $f'$ extending $f$ with $(S,T)\transition{a}(S_a,T_a)$ and $((S_a,T_a),(S'_a,T'_a),f')\in R$.
\end{enumerate}
$R$ is moreover called \emph{hereditary} if the following back condition holds:
\begin{enumerate}
\setcounter{enumi}{3}
\item if $(S_a,T_a)\transition{a}(S,T)$ in $\mathsf{ST}$ then exists $(S'_a,T'_a)\in\mathsf{ST}'$ and $f'$ with $f\!\!\upharpoonright_{(S_a,T_a)}=f'$ and $(S'_a,T'_a)\transition{a}(S',T')$ and  $((S_a,T_a),(S'_a,T'_a),f')\in R$.
\item if $(S'_a,T'_a)\transition{a}(S',T')$ in $\mathsf{ST}'$ then exists $(S_a,T_a)\!\in\!\mathsf{ST}$ and $f'$ with $f\!\!\upharpoonright_{(S'_a,T'_a)}=f'$) and $(S_a,T_a)\transition{a}(S,T)$ and  $((S_a,T_a),(S'_a,T'_a),f')\in R$.
\end{enumerate}
A history preserving bisimulation between two ST-structures is denoted $\ST\hequiv\ST'$, and a hereditary one is denoted $\ST\hhequiv\ST'$. We usually abbreviate to hh-bisimulation.
\end{definition}

Because of symmetry of the requirements for history preserving bisimulation (i.e., the points 2 and 3 above), the two conditions for hereditary are redundant together, and we could well use only one of them. In our proofs we will consider only condition 4.

\begin{example}\label{ex_Winskel_switch_resolved_conflict}
The parallel switch of Winskel \cite[Ex.1.1.7]{Winskel86} consists of an event $b$ (lighting a light bulb) that depends on either of the two parallel switches being closed (i.e., the two events $0$ and $1$). This example emphasizes \textit{disjoint causality}, where event $b$ depends on either $0$ or $1$, and hence the fact that there is no \textit{unique causal history}, as opposed to \textit{stable structures}.
The ST-structure for this example, in Figure~\ref{fig_ex_winskel}(left and middle), is adjacent-closed and closed under unions, but not closed under intersections, i.e., the ST-configurations $(01b,\emptyset)\!=\!(01b,1)\cap(01b,0)\!\subseteq\!(01b,01b)$ but $(01b,\emptyset)\not\in \ST_{W}$. 

The resolved conflict of \cite[Ex.2]{GlabbeekP09configStruct}, pictured in Figure~\ref{fig_ex_winskel}(right), represents the fact that the initial conflict of the two actions $a$ and $b$ is resolved as soon as the action $c$ has finished (i.e., $a$ and $b$ may run concurrently as soon as $c$ has finished). The corresponding ST-structure of Figure~\ref{fig_ex_winskel}(right) is adjacent-closed and closed under intersections but not closed under unions:  $(bc,\emptyset)\cup(ac,\emptyset)=(abc,\emptyset)\not\in \ST_{RC}$.
Both examples can be pictured as three sides of a \HDA\ cube (middle and right) whereas on (left) is an ST-structure. In several cases we use the more clean \HDA\ presentation for ST-structures because of the results below (see the Definition~\ref{def_hda} of a \HDA).
The standard example of a square with the empty inside, as pictured in Fig.~\ref{fig_ex_hda}(middle-right) on page~\pageref{fig_ex_hda}, is adjacent-closed but not closed under unions nor under intersections.
\end{example}

\begin{example}\label{ex_not_adjacent_stable}
For an example of ST-structure that is not adjacent-closed but is stable take the example of the filled square of $a||b$ but where the triangle above the diagonal is removed as in Fig.~\ref{fig_ex_hda}(right) on page~\pageref{fig_ex_hda}. Intuitively, this models a system where both $a$ and $b$ may run concurrently but $a$ is always faster than $b$ (hence starts and also terminates first); in other words $b$ cannot start before $a$ has started and cannot finish before $a$ has finished. 
%
The event $a$ may be a resource allocation mechanism that $b$ may need for running, and thus $a$ must be running when $b$ can start. But $b$ may run concurrently with $a$, e.g., while $a$ finishes all the resource allocation work (like logging or lock setting). Nevertheless, $b$ must wait for all this resource allocation work to properly finish (having all logging in place, etc.) before itself can finish (and maybe do some more logging and lock releasing).
\end{example}

These examples lead to the results in the next section where we isolate the class of ST-structures that corresponds to a popular class of \HDAs, that of acyclic and non-degenerate \HDAs, which are expressive enough to faithfully represent the various examples used in papers like \cite{GlabbeekV97splitting,GlabbeekG01refinement,BaldanC10concur,phillips11express} which studied bisimulations for true concurrency. Nevertheless, towards the end of this paper we give examples that challenge the expressiveness of this class of \HDAs\ (and ST-structures also), and thus justify an extension which we will call \textit{STC-structures} in Section~\ref{sec_STCstruct}.

\section{Expressiveness of ST-structures and correspondences}\label{subsec_expressST}

\subsection{Correspondence with configuration structures}\label{subsec_configStruct}

We investigate the relationship of ST-structures with the configuration structures of \cite{GlabbeekP95config,GlabbeekP09configStruct} and show that ST-structures are a natural extension of the later. This extension also holds when their respective computational aspects are considered, i.e., the concurrent step interpretations are related.

\begin{definition}[{cf.~\cite[Def.5.1]{GlabbeekG01refinement}\cite[Def.1.1]{GlabbeekP09configStruct}}]\label{def_configurationStruct}\ 

A \textit{configuration structure} $\C=(E,C)$, is formed of a set $E$ of \textit{events} and a set of configurations which are subsets of events $C\subseteq 2^{E}$. A \emph{labeled} configuration structure also has a labeling function of its events, $l:E\rightarrow\Sigma$.
\end{definition}

\begin{definition}[\allC\ to\ \allST]\label{def_confInST}
Define a mapping $\cintost:\allC\rightarrow\allST$ that associates to every configuration structure \C\ an ST-structure $\cintost(\C)$ as follows.
Associate to each configuration $X\in\C$ an ST-configuration $\cintost(X)=(X,X)\in\cintost(\C)$. No other ST-configurations are part of $\cintost(\C)$. 
The labeling function is just copied.
\end{definition}

\begin{definition}[morphisms for \allC]\label{def_morphisms_C}
A \emph{morphism} between two labeled configuration structures $\C=(E,C,l)$ and $\C'=(E',C',l')$ is a partial map $f:E\rightharpoonup E'$ between their events that: 
\vspace{-1ex}\begin{itemize}
\item 
preserves the configurations; i.e., if $X\in C$ then $f(X)\in C'$,
\item 
preserves the labeling when defined, i.e., $l'(f(e))=l(e)$ if $f$ is defined for $e$, and 
\item
is locally injective, i.e., for any $X\in\C$ the restriction $f\restrictedToSet{X}$ is injective.
\end{itemize} 
Two configurations structures are called \emph{isomorphic}, denoted $\C\isomorphic\C'$, iff there exists a morphism $f$ that is bijective on the events.
\end{definition}

The set of labeled configurations together with the morphisms form a category, which we will denote the same \allC.

\begin{proposition}\label{prop_functorExtension_cintost}
The mapping \cintost\ can be extended to a functor between the categories \allC\ and \allST\ by defining its application on the morphisms as $\cintost(f)=f$.
\end{proposition}

\begin{proof}
The application of \cintost\ to morphisms is correct because for some configuration $\C$ the events are preserved through the mapping $\cintost(\C)$. Therefore for some morphism $f:E_{1}\rightarrow E_{2}$ the morphism $\cintost(f)$ is well defined from the events of $\cintost(\C_{1})$ to the events of $\cintost(\C_{2})$.

The proof that $\cintost(f)$ preserves ST-configurations and the labeling, and is local injective, follows from the same properties of the morphism $f$ on the configuration structures.
\end{proof}

\begin{proposition}[\cintost\ is embedding]\label{prop_embeding_CtoST}
\ 

\begin{enumerate}
\item The map \cintost\ from Definition~\ref{def_confInST} preserves isomorphic configuration structures and does not identify non-isomorphic configuration structures.

\item There are ST-structures that are not the image of any configuration structure.
\end{enumerate}
\end{proposition}

\begin{proof}
The categorical claim that the functor $\cintost$ between the two categories \allC\ and \allST\ is embedding can be seen from the fact that the morphisms between two configuration structures are the same as those between their associated ST-structures. More precisely, for any two configuration structures the function $\cintost_{\C_{1},\C_{2}}:\mathit{Hom}_{\allC}(\C_{1},\C_{2})\rightarrow \mathit{Hom}_{\allST}(\cintost(\C_{1}),\cintost(\C_{2}))$ that associates to each morphism $f:\C_{1}\rightarrow\C_{2}$ the morphism $\cintost(f)$ is bijective. Injectivity is easy to see since for two different morphisms $f,f'$ between $\C_{1}$ and $\C_{2}$ the function $\cintost_{\C_{1},\C_{2}}$ associates the morphisms $\cintost(f)$ and $\cintost(f')$, which following the definition from Proposition~\ref{prop_functorExtension_cintost} are different. For surjectivity one has to check that for any morphisms $F$ between $\cintost(\C_{1})$ and $\cintost(\C_{2})$, which is a partial map between their events, the same partial map between the events of $\C_{1}$ and $\C_{2}$ also respects the conditions of being a morphism in \allC.

To show that \cintost\ preserves isomorphic configuration structures consider the bijective morphism $f:E_{1}\rightarrow E_{2}$ that witnesses the isomorphism of $\C_{1}$ and $\C_{2}$. This same function between the events of $\cintost(\C_{1})$ and $\cintost(\C_{2})$ is also a bijection and a morphism between the two ST-structures. Checking that $f$ preserves ST-configurations, the labeling, and is local injective is easy based on its properties on the configuration structures.

To show that non-isomorphic configuration structures are not identified by \cintost\ for some arbitrary $\C_{1}\not\isomorphic\C_{2}$ assume that there exists a bijective morphisms $f$ witnessing the isomorphism of their translations $\cintost(\C_{1})\isomorphic\cintost(\C_{2})$. It is easy to show that this same function between the events of $\C_{1}$ and $\C_{2}$ makes these isomorphic as it preserves configurations, the labeling and is locally injective.

For the part (2) of the proposition just take any ST-structure that has ST-configurations of concurrency degree non-zero; these are not the image through the \cintost\ in Definition~\ref{def_confInST} of any configuration structure.
\end{proof}

\begin{definition}[\allST\ to\ \allC]\label{def_STtoC}
Define a mapping $\stintoc:\allST\!\rightarrow\!\allC$ that associates to every ST-structure $\mathsf{ST}$ a configuration structure by keeping only those ST-configu\-rations that have $S=T$; i.e., $\stintoc(\mathsf{ST})\!=\!\{T \mid (S,T)\!\in\!\mathsf{ST}\wedge S\!=\!T\}$, which preserves the labeling.
\end{definition}

\begin{proposition}\label{prop_ST_config_connect}
If an ST-structure $\ST$ is rooted, connected, or closed under bounded unions, or intersections, then the corresponding $\stintoc(\ST)$ is respectively rooted, connected, closed under bounded unions, or intersections.
\end{proposition}

\begin{proof}
A configuration structure is rooted if it contains the configuration $\emptyset$. The definitions of bounded union and intersection for configuration structures are the natural simplification of the respective definitions for ST-structures from Definition~\ref{def_stableST}. Proving the rootedness and closure properties is immediate. 
A configuration structure is connected (cf.\ \cite[Def.5.5]{GlabbeekG01refinement}) if for every configuration $X$ exists an event $e\in X$ s.t.\ $X\setminus\{e\}$ is also a configuration in the structure. For connectedness note that any connected ST-structure is also rooted (and the same holds for configuration structures). Therefore, for any configuration $X\in\stintoc(\ST)$ there exists the ST-configuration $(X,X)\in\ST$ from which it was obtained. Since $\ST$ is connected it means that there is a sequence of ST-configurations, each one event smaller than the previous, which reach the root $(\emptyset,\emptyset)$. This means that on this sequence there must eventually be an ST-configuration $(X\setminus e,Y)$ with $Y\subseteq (X\setminus e)$. By the constraint of the ST-structures it means that also the ST-configuration $(X\setminus e,X\setminus e)\in\ST$ and therefore also the configuration $(X\setminus e)\in\stintoc(\ST)$.
\cp{!!Do this proof with the weaker constraint, so the last line needs redoing.}
\end{proof}

But there is not a one to one correspondence between ST-structures and the configuration structures because there can be several ST-structures that have the same configuration structure. The example is of one \HDA\ square that is filled in and one that is not; both have the same set of corners and hence the same configuration structure. But the two ST-structures are not isomorphic and also not hh-bisimilar (in the sense of Definition~\ref{def_hh_ST}).

\begin{proposition}[\stintoc\ is forgetful]\label{prop_forget_STtoC}
\ 

\begin{enumerate}
\item The map \stintoc\ from Definition~\ref{def_STtoC} preserves isomorphic ST-structures. 

\item The map \stintoc\ may identify non-isomorphic ST-structures (in fact non-hh-bisimilar).
\end{enumerate}
\end{proposition}

\begin{proof}
The part (1) is proven easily, similar to what we did for Proposition~\ref{prop_embeding_CtoST}.

For part (2) take the empty square and the filled-in concurrency square examples. These two are translated in the same configuration structure; i.e., their corners only. But as ST-structures these two examples are not isomorphic and neither hh-bisimiar.
\end{proof}

Next we show that the \textit{asynchronous concurrent step interpretation} of configuration structures is captured by ST-structures (cf.~\cite[Def.2.1]{GlabbeekP09configStruct}, an asynchronous step is defined between two configurations $X\stepTransConfGlabbeek Y$ iff $X\!\subseteq\!Y$ and $\forall Z:X\!\subseteq\!Z\subseteq\!Y \Rightarrow Z\!\in\!\C$).

\begin{lemma}\label{lemma_morph_preserve_steps}
\ 

\begin{enumerate}
\item\label{lemma_morph_preserve_steps_1} Morphisms of \allC\ preserve asynchronous concurrent steps. 

\item\label{lemma_morph_preserve_steps_2} Morphisms of \allST\ preserve (s-/t-)steps.
\end{enumerate}
\end{lemma}

\begin{proof}
For part \refeq{lemma_morph_preserve_steps_1} we take an arbitrary $f:\C_{1}\rightarrow\C_{2}$ and an arbitrary step $X\stepTransConfGlabbeek Y \in\C_{1}$ and show that $f(X)\stepTransConfGlabbeek f(Y)$. The definition of an asynchronous step says that $X\subseteq Y$ which by the local injectivity of $f$ it means that $f(X)\subseteq f(Y)$. Moreover, the injectivity on the larger set $Y$ makes $f$ bijective between $Y$ and $f(Y)$ which means that any subset of $f(Y)$ is the image of some subset of $Y$. The asynchronous step says that all subsets $X\subseteq Z\subseteq Y$ are configurations $Z\in\C_{1}$ and since $f$ preserves configurations it means that $f(Z)\in\C_{2}$. These are all possible subsets $f(X)\subseteq Z'\subseteq f(Y)$, therefore we have the expected step $f(X)\stepTransConfGlabbeek f(Y)$.

The proof of part \refeq{lemma_morph_preserve_steps_2} is easy since the steps in ST-structures involve single events. The proof again uses the fact that the morphisms are locally injective.
\end{proof}

\begin{theorem}\label{th_configtoSTsteps}
Define a mapping $\cintostSecond:\allC\rightarrow\allST$ by extending the one in Definition~\ref{def_confInST} s.t.\ for each asynchronous step $X\stepTransConfGlabbeek Y\!\in\!\C$ add also an ST-configuration $\STofC{X\stepTransConfGlabbeek Y}=(Y,X)\in\STofC{\C}$.
This map \cintostSecond\ preserves the asynchronous concurrent steps of the configuration structure, i.e., for each asynchronous step $X\stepTransConfGlabbeek Y\in \C$ there is a chain of single steps in the ST-structure $\STofC{\C}$ that passes through $(Y,X)$ (thus signifying the concurrent execution of all events in $Y\setminus X$).
\end{theorem}

\begin{proof}
Take $\C$ to be some configuration structure and $\STofC{\C}$ the corresponding ST-structure that we construct for it.
The construction extends the simple encoding from before which associated with each configuration $X\in \C$ an ST-configuration $\STofC{X}=(X,X)\in\STofC{\C}$. 
The function \STofC{.}\ is applied to the configurations of \C\ and does not introduce new events. Thus the labeling of the structures is just copied.

We show that for any $X\stepTransConfGlabbeek Y\in \C$ we have $\cintostSecond(X)\transition{}^{*}(Y,X)\transition{}^{*}\cintostSecond(Y)$ in $\STofC{\C}$. We do this using induction on the number of concurrent events in the concurrent step between the configurations. 

The base case is for $|Y\setminus X|=1$ (we ignore the reflexive steps that are assumed for each configuration in \cite{GlabbeekP09configStruct}).
Essentially, in terms of \HDAs, the \cintostSecond\ adds also the transition between the two states of the \HDA. 
In $\STofC{\C}$ we have one s-step from $(X,X)$ to $(Y,X)$ and one t-step from $(Y,X)$ to $Y,Y$, where $Y=X\cup\{e\}$ as $\{e\}=Y\setminus X$.

Take $|Y\setminus X|=n\geq 2$, thus $\exists e\neq e'\in Y\setminus X$. 
We use the property of asynchronous steps in configuration structures from \cite[Def.2.1]{GlabbeekP09configStruct} which says that if $X\stepTransConfGlabbeek Y$ then $\forall Z:X\subseteq Z\subseteq Y \Rightarrow Z\in\C$. This also implies that there are asynchronous steps from $X\stepTransConfGlabbeek Z$ and $Z\stepTransConfGlabbeek Y$, and both have fewer number of concurrent events. 
We can apply the inductive hypothesis in the following two instances:
(1) $|(Y\setminus e)\setminus X|=n-1$; and (2) $|Y\setminus(X\cup e')|=n-1$. From (1) we get the chain of single steps $(X,X)\transition{}^{*}(Y\setminus e,X)\transition{}^{*}(Y\setminus e,Y\setminus e)$. Since $(Y,X)\in\STofC{\C}$ was added by $\cintostSecond(X\stepTransConfGlabbeek Y)$ we have $(Y\setminus e,X)\transition{}(Y,X)$. By the induction hypothesis on (2) we have $(X\cup e',X\cup e')\transition{}^{*}(Y,X\cup e')\transition{}^{*}(Y,Y)$. 
Because $e'\in Y$ and $e'\not\in X$ we have also the transition $(Y,X)\transition{}(Y,X\cup e')$.
Thus we have the conclusion that there exists the chain of single steps $(X,X)\transition{}^{*}(Y\setminus e,X)\transition{}(Y,X)\transition{}(Y,X\cup e')\transition{}^{*}(Y,Y)$ that passes through $(Y,X)$.

Intuitively, thinking in terms of acyclic \HDAs, for each transition $X\stepTransConfGlabbeek Y\in \C$ we build the \HDA\ cube of dimension $|Y\setminus X|$ with all the faces filled in.
\end{proof}

\begin{corollary}\label{cor_adjacent}
An ST-structure $\STofC{\C}$ generated as in Theorem~\ref{th_configtoSTsteps} is adja\-cent-closed (though not necessarily closed under bounded unions nor intersections).
\end{corollary}

\begin{proof}
Since we work with rooted configurations structures, the \cintostSecond\ function clearly preserves rootedness.

From a connected configuration structure for any sequence of transitions $X\transition{}^{*}Y$ we find a sequence of single steps in the associated ST-structure. This is easy to see from Theorem~\ref{th_configtoSTsteps}. Each individual transition has a corresponding sequence of single steps in the ST-structure.

From the proof of Theorem~\ref{th_configtoSTsteps} we see that an ST-configuration $(Y,X)$ with $X\neq Y$, is introduced only when there is a concurrent transition between the configurations $X$ and $Y$. With this observation it is easy to prove the four adjacency restrictions. Take as example the first restriction (leaving the others as exercise) and infer from $(S,T)$ that there is the transition $T\transition{}S$ and thus from the \cite[Def.2.1]{GlabbeekP09configStruct} it means that $\forall X: T\subseteq X\subseteq S$ then $X\in\C$ is also a configuration. Also we have $T\transition{}S\cup e$ and $T\transition{}S\cup \{e,e'\}$. To prove that $(S\cup e',T)\in\ST$ it is enough to show that there is a transition $T\transition{}S\cup e'$. This is easy from the definition \cite[Def.2.1]{GlabbeekP09configStruct} and the fact that $S\cup e' \subseteq S\cup \{e,e'\}$.

It is easy to check that the parallel switch of Winskel \cite{Winskel86} (not closed under bounded intersections) and the resolved conflict example of \cite[Ex.2]{GlabbeekP09configStruct} (not closed under bounded unions) are expressible as configuration structures.
\end{proof}

\begin{corollary}\label{cor_counit_C_ST}
In an ST-structure $\STofC{\C}$ generated as in Theorem~\ref{th_configtoSTsteps} the ST-con\-figurations with $S=T$ correspond exactly to the configurations of $\C$. That is to say that $\stintoc(\STofC{\C})\isomorphic\C$.
\end{corollary}

\cp{The above equivalence should also be proven as an adjunction between the two categories.}

One can now check that the map \stintoc\ can be lifted to a functor between \allST\ and \allC\ the same as we did in Proposition~\ref{prop_functorExtension_cintost}.

\begin{proposition}
The new map \cintostSecond\ from Theorem~\ref{th_configtoSTsteps} can be lifted to a functor by defining its application on morphisms to be $\cintostSecond(f)=f$. This is the right adjoint to the functor \stintoc.
\end{proposition}

\begin{proof}
Translating configuration structures into ST-structures does not change the set of events nor the labeling function, therefore it is easy to see that $\cintostSecond(f)$ preserves the labeling.

To show that $\cintostSecond(f)$ preserves the ST-configurations consider some $(S,T)\in\cintostSecond(\C_{1})$ and take two cases cf.\ the definition of \cintostSecond\ from Theorem~\ref{th_configtoSTsteps}.

Case when $S=T$ which means that $S\in\C_{1}$ is a configuration, and since $f$ preserves configurations it means that $f(S)\in\C_{2}$ and thus we have the desired result that $(f(S),f(S))\in\cintostSecond(\C_{2})$, i.e., $f(S,T)\in\cintostSecond(\C_{2})$.

Case when $S\neq T$ means that $(S,T)$ comes from a transition $T\stepTransConfGlabbeek S\in\C_{1}$. This means that $S,T\in\C_{1}$ are configurations preserved by $f$, hence $f(T),f(S)\in\C_{2}$. Since by Lemma~\ref{lemma_morph_preserve_steps} $f$ preserves also asynchronous steps we have the step $f(T)\stepTransConfGlabbeek f(S)\in\C_{2}$ which implies that this is translated into the ST-configuration $(f(S),f(T))\in\cintostSecond(\C_{2})$, i.e., $f(S,T)\in\cintostSecond(\C_{2})$.

It is easy to see that $\cintostSecond(f)$ is local injective.

To show that \cintostSecond\ is right adjoint to \stintoc\ we exhibit the co-unit $\counit:\stintoc\circ\cintostSecond\rightarrow I_{\allC}$ to be the isomorphism from Corollary~\ref{cor_counit_C_ST}.

\newlength{\parindentoutsidemini}
\setlength{\parindentoutsidemini}{\parindent}
\noindent\begin{minipage}[l]{0.70\textwidth}
\setlength{\parindent}{\parindentoutsidemini}
\vspace{0.5ex}
We have to show that for any object \C\ of \allC\ and any morphism $g:\stintoc(\ST)\rightarrow\C$ in \allC\ there exists a unique morphism $g^{\#}\!:\!\ST\!\rightarrow\!\cintostSecond(\C)$ for which the diagram on the right commutes.
The map \stintoc\ preserves events then the events of \ST\ are the same as those of $\stintoc(\ST)$; the same holds for $\cintostSecond$ meaning that the events of \C\ are the same as the events of $\cintostSecond(\C)$. 
Therefore, we can take $g^{\#}$ to be $g$, and the functor returns $\stintoc(g^{\#})=g^{\#}=g$.
\end{minipage}
\hspace{-1ex}\begin{minipage}[r]{0.3\textwidth}
\vspace{-4ex}\begin{diagram}
\stintoc\circ\cintostSecond(\C) & \rTo^{\counit_\C} & \C \\
\uDashto^{\stintoc(g^{\#})} & \ruTo_g &  \\
\stintoc(\ST) &  &  \\
\end{diagram}
\vfill
\end{minipage}

It is easy to see that the diagram commutes: for any $e\in E_{\stintoc(\ST)}$ we have that $g(e)=\counit_{\C}\circ g(e)$ because the isomorphism $\counit_{\C}$ from Corollary~\ref{cor_counit_C_ST} is the identity.

To show the uniqueness of $g^{\#}$ assume the existence of another $f:\ST\rightarrow\cintostSecond(\C)$ for which the diagram commutes but for some event $e\in E_{\stintoc(\ST)}$ it is different $f(e)\neq g^{\#}(e)$. This means to say that $f(e)\neq g(e)$ and that $\stintoc(f)(e)\neq g(e)$. But then the composition with the counit would again result in $\counit_{C}\circ\stintoc(f)(e)\neq g(e)$, i.e., a contradiction.
\end{proof}

%
%
%

\begin{corollary}[filled-in]\label{cor_cubicalProp_configST}
The ST-structure obtained in Th.~\ref{th_configtoSTsteps} is \emph{``filled in''}, in the sense that any cube is filled in. 
By a \textit{``cube''} it is meant an initial ST-configuration $(S,S)$, a final $(S\cup X,S\cup X)$, where $X$ is a nonempty set of events, together with all the ST-configurations $(Y,Y)$ from the subsets $S\subseteq Y\subseteq S\cup X$. 
To be \textit{``filled in''} means that the intermediate ST-configuration $(S\cup X,S)$ exists.
\end{corollary}

\begin{proof}
We call \textit{``corners''} the ST-configurations where $S$ and $T$ are equal. 
By a \textit{``cube''} it is meant an initial corner $(S,S)$ and a final corner $(S\cup X,S\cup X)$ where $X$ is a nonempty set of events that are meant to be executed concurrently; thus the dimension of $X$ makes the \textit{``higher dimension''} of the cube. The subsets $S\subseteq Y\subseteq S\cup X$ are the rest of the corners of the cube. 
To be \textit{``filled in''} means that the intermediate ST-configuration $(S\cup X,S)$ exists, and hence reachable and with all intermediate ST-configurations.
%

The definition of a ``cube'' between some $(S,S)$ and $(S\cup X,S\cup X)$ implies that all the corners of the cube come from configurations $S\subseteq S\cup Y\subseteq S\cup X$. This means that in the configuration structure there exists the asynchronous step $S\stepTransConfGlabbeek S\cup X$. Therefore, by the definition of \cintostSecond\ from Theorem~\ref{th_configtoSTsteps}, this asynchronous step is translated into the ST-configuration $(S\cup X,S)$.
\end{proof}

\begin{proposition}\label{prop_stableSTconf}
For stable and adjacent-closed ST-structures and stable configuration structures there is a one-to-one correspondence. (The adjacency is necessary.)
\end{proposition}

\cp{This result could be proven also as a result of equivalence of categories!? Moreover, the proof below makes heavy use of the condition of ST-structures; does the proof go through with the weaker constraint?}

\begin{proof}
We use the mapping \stintoc\ from Definition~\ref{def_STtoC}.

Since the input \ST\ is stable, by Proposition~\ref{prop_ST_config_connect}, the output $\C(\ST)$ is also stable.

Define a map $\cintostThird:\allC\rightarrow\allST$ by extending that from Definition~\ref{def_STtoC} which associates to each configuration $X\in\C$ an ST-configuration $\cintostThird(X)=(X,X)\in\cintostThird(\C)$, and for each pair of configurations $T$ and $T\cup\{e\}$ add also the intermediate ST-configuration $(T\cup\{e\},T)$. 
We then close the resulting ST-structure under bounded unions and intersections.

\vspace{1ex}
\noindent\textit{Claim:}\hspace{1ex} The ST-structure generated by $\cintostThird(\C)$ is stable if \C\ is stable.
\vspace{0.5ex}

The $\cintostThird(\C)$ is rooted, and also closed under bounded unions and intersections, by its definition. We need to show it is connected.

Also from the definition of \cintostThird\ and the connectedness of the input \C\ structure, we see that all ST-configurations of concurrency degree 0 or 1 are reachable. It remains to show that any ST-configuration of concurrency degree more than 1 is reachable; these are those ST-configurations coming from the closures. Since the generated ST-structure is rooted, then by Proposition~\ref{prop_connectPaths} we can work with paths and show reachability of all ST-configurations.

Since the dimension of a ST-configuration is the same as the length of the rooted paths that reach it, we can use this measure in an inductive reasoning. The inductive hypothesis is that for two reachable ST-configurations $(S,T)$ and $(S',T')$ which enter the requirements of the closure under bounded unions, then their union $(S\cup S',T\cup T')$, which is also an ST-configuration, is also reachable.
At least one of the two ST-configurations is not empty, thus take $(S,T)$ to be reachable from some $(S\setminus e,T)$, i.e., through an s-step (the argument for a t-step is analogous). By the closure it means that also the union $((S\setminus e)\cup S',T\cup T')$ is an ST-configuration of $\cintostThird(\C)$. Moreover this has degree one lower, coming from two ST-configurations reachable through shorter paths. Therefore we may apply the induction hypothesis to obtain a path reaching this smaller $((S\setminus e)\cup S',T\cup T')$. But from this we can make an s-step, with the event $e$ to reach the initial union ST-configuration $(S\cup S',T\cup T')$, thus finding the desired path.

For the closure under bounded intersections a similar inductive reasoning on the length of the path goes through.

\vspace{1ex}
\noindent\textit{Claim:}\hspace{1ex} The ST-structure generated by $\cintostThird(\C)$ respects the constraint from Definition~\ref{def_st_structs}.

\vspace{1ex}
It is not difficult to check that the claim holds for ST-configurations of concurrency degree 1. 

For any two $(S,T)$, $(S',T')$ satisfying the constraint, i.e., $(S,S),(S',S')\in\cintostThird(\C)$ we show that their union ST-configuration also respects the constraint when this is indeed an ST-configuration from $\cintostThird(\C)$. 
Respecting the conditions for closure under bounded unions means that there exists $(S'',T'')$ s.t.\ $(S\cup S',T\cup T')\subseteq (S'',T'')$ and $(S'',T'')$ also satisfies the constraint of Definition~\ref{def_st_structs}, i.e., $(S'',S'')\in\cintostThird(\C)$. But in this case we see that the two ST-configurations $(S,S),(S',S')$ respect too the conditions of closure under bounded unions, which implies that their union is an ST-configuration also $(S\cup S',S\cup S')\in\cintostThird(\C)$, which is our desired result.

\vspace{1ex}
\noindent\textit{Claim:}\hspace{1ex} The map $\cintostThird(\C)$ does not introduce corners $(X,X)\in\cintostThird(\C)$ which do not have correspondent $X\in\C$.
\vspace{0.5ex}

Any new corners can come only from unions or intersections. Assume two $(S,T),(S',T')$ that respect the conditions for closure under bounded unions and their union is a new corner $(S,T)\cup(S',T')=(S'',S'')$. But by the previous claim there exist also the ST-configurations $(S,S)$ and $(S',S')$ which are both smaller than (i.e., included in) $(S'',S'')$, To these the inductive hypothesis says that are not new, but come from \C, i.e., $S,S'\in\C$. Since \C\ is also closed under bounded unions it means that $S\cup S'\in\C$, our desired result.

\vspace{1ex}
To show the one-to-one correspondence we show two results.

One is that for some stable configuration structure \C\ we have that the application of the two association functions above results in an isomorphic configuration structure; i.e., 

\centerline{$\C\isomorphic \stintoc(\cintostThird(\C))$.}

This result is easy to establish because, intuitively, the first map \cintostThird\ adds information which is then forgotten by the application of \C. It is easy to see that \cintostThird\ does not introduce new events; and the same for \stintoc. Therefore exhibiting the isomorphism is done by the identity function between the events of \stintoc\ and $\stintoc(\cintostThird(\C))$. We need to show that it preserves configurations, which means to show that for any configuration $X\in\C$ then the same configuration is found in the right structure, i.e., $Id(X)=X\in\stintoc(\cintostThird(\C))$. Any configuration is translated into a corner $(X,X)\in\cintostThird(\C)$. By the previous claim, no other corners exist. Then each corner is translated into an appropriate configuration $X\in\stintoc(\cintostThird(\C))$.

\vspace{1ex}
More difficult is to establish that for some stable and adjacent-closed ST-structure \ST\ we have that the application of the two association functions above results in an isomorphic ST-structure; i.e., 

\vspace{-1ex}\centerline{$\ST\isomorphic \cintostThird(\C(\ST))$.}

Note that only the requirement of stable is not enough for this result. A counterexample is given by the stable ST-structure from Fig.~\ref{fig_ex_hda}(right-most) which is not adjacent closed and for which the above isomorphism is not the case.

The proof has two parts: first we show that any ST-configuration $(S,T)\in\ST$ has an isomorphic version in $\cintostThird(\C(\ST))$; second is to show that the function applications $\cintostThird(\C(\cdot))$ does not introduce new ST-configurations.

For the first part, if $S=T$ then it is easy to see that $(S,T)\in\cintostThird(\C(\ST))$.

When $S\neq T$ then let $E=S\setminus T$. Because the input \ST\ is stable (hence rooted and connected) and adjacent-closed it means it is closed under single events, cf.~Proposition~\ref{prop_adj_equiv}. Therefore, $(S\setminus e,T)\in\ST$ for all $e\in E$. With a simple inductive argument using the above closure under single events one can easily show that $\forall X:T\subseteq X\subseteq S$ we have $(X,T)\in\ST$. Therefore, together with the requirement on ST-structures that $(X,X)$ exists for any $(X,T)$, it means we have all configurations $X\in\C(\ST)$, for $T\subseteq X\subseteq S$. By the definition of the association function $\cintostThird(\cdot)$ for all pairs of configurations $T\cup e$ and $T$, for all $e\in E$, the function adds an ST-configuration $(T\cup e,T)\in\cintostThird(\C(\ST))$. When closing under bounded unions all these ST-configurations we obtain the desired $(T\cup E,T)\in\cintostThird(\C(\ST))$.

For the second part, assume some $(S,T)\in\cintostThird(\C(\ST))$ then we show that $(S,T)\in\ST$. If $S=T$ then this ST-configuration must come from a configuration $S\in\C(\ST)$ (cf.\ the previous claim about no new corners), which in turn only comes from an ST-configuration $(S,S)\in\ST$.

Assume $(S,T)$ comes from the existence of two configurations $T$ and $T\cup e$ in $\C(\ST)$; i.e., $(S,T)=(T\cup e,T)$. But this means that in \ST\ there exist the ST-configurations $(T,T)$ and $(T\cup e, T\cup e)$. From the fact that \ST\ is stable it means that $(T\cup e, T\cup e)$ is reachable from $(\emptyset,\emptyset)$ through a path of single event steps. Assume we do not remove the event $e$ from the second set of the pair immediately (for otherwise we already have our desired result) and thus there is a series of single steps that remove single events different than $e$ gradually, first removing from the second set. But eventually we must reach a point when we remove $e$ from the second set and not from the first set yet. This means we reach an ST-configuration $(S',T')$ with $S'\subseteq T\cup e$ containing $e$, and $T'\subseteq T$. We can apply the property of closed under bounded unions for $(T,T)$ and $(S',T')$ to obtain $(T\cup e,T)$.

Assume that $(S,T)$ comes from closure under bounded union of two smaller $(S',T')$ and $(S'',T'')$. By induction these are in \ST\ which is closed under bounded unions, hence it also contains $(S,T)$. 
The basis of the induction is essential here. We check it for ST-configurations of concurrency degree 1. Take two $(T'\cup e,T'),(T''\cup f,T'')\in\cintostThird(\C(\ST))$ and show that their union is $(T'\cup e \cup T''\cup f\,,T'\cup T'')\in\ST$. By the previous argument we know that $(T'\cup e,T'),(T''\cup f,T'')\in\ST$ which because $\ST$ is closed under bounder unions delivers the expected result.
\end{proof}

The results in this section also apply to \textit{pure event structures} because these are shown in \cite[Th.2 and Prop.2.2]{GlabbeekP09configStruct} to be equivalent to configuration structures under their respective computational interpretations, i.e., asynchronous steps are preserved through translations.

\cp{
\begin{proposition}
Two stable and adjacent-closed ST-structures are hh-bisimilar iff their corresponding (cf.~Proposition~\ref{prop_stableSTconf}) stable configuration structures are, cf.~\cite[def.9.6]{GlabbeekG01refinement}.
\end{proposition}

\begin{proof}
The proof is tedious.
\end{proof}

}

\subsection{Correspondence with the event structures of Plotkin and van Glabbeek}

We relate the ST-structures with the \textit{(inpure) event structures} of \cite[Def.1.3]{GlabbeekP09configStruct} and the asynchronous transition relation associated to them in \cite[Def.2.3]{GlabbeekP09configStruct}.
An \textit{event structure} (which we call \textit{inpure} since their definition in \cite[Def.1.3]{GlabbeekP09configStruct} is different than standard event structures and also the restriction of being pure is not imposed) is $\E=(E,\enableRelEv)$, a set of events with an \textit{enabling relation} defined between sets of events $\enableRelEv\subseteq 2^{E}\times 2^{E}$. An event structure can be associated with its set of configurations, cf.\ \cite[Def.1.4]{GlabbeekP09configStruct}, $L(E)=\{X\subseteq E \mid \forall Y\subseteq X,\exists Z\subseteq X:Z\enableRelEv Y\}$. Asynchronous transitions between these configurations are then defined in \cite[Def.2.3]{GlabbeekP09configStruct} as $X\stepTransEvGlabbeek Y$ iff $X\subseteq Y$ and $\forall Z\subseteq Y,\exists W\subseteq X: W\enableRelEv Z$.

\begin{theorem}[\allEv\ to\ \allST]\label{th_inpureEv_to_ST}
An inpure event structure can be encoded into an ST-structure s.t.\ any asynchronous concurrent step transition (cf.\ \cite[Def.2.3]{GlabbeekP09configStruct}) $X\stepTransEvGlabbeek Y$ is matched by an appropriate path that passes through the ST-configu\-ration $(Y,X)$.
The encoding is done with the mapping $\stEnc:\allEv\rightarrow\allST$ defined similarly to the one for configuration structures of~Theorem~\ref{th_configtoSTsteps}, considering the set $L(\E)$ of left-closed configurations \cite[Def.1.4]{GlabbeekP09configStruct} of the event structure; i.e., $\stEnc(X)=(X,X)$ for $X\in L(\E)$ and for any transition $X\stepTransEvGlabbeek Y$ add also the ST-configuration $\stEnc(X\stepTransEvGlabbeek Y)=(Y,X)$.
\end{theorem}

\begin{proof}
The above property that the theorem requires on the generated ST-structure captures the concurrency that the event structure transition embodies.

The proof uses induction on the dimension of the asynchronous transitions, i.e., on $|Y\setminus X|$, noting the fact that $X\subseteq Y$.
The proof is similar to what we did in Theorem~\ref{th_configtoSTsteps} and is facilitated by the Corollary~\ref{cor_ev_intermediaryTrans}.

The basis for $|Y\setminus X|=1$ is easy, for $Y=X\cup \{e\}$.

The induction case for $|Y\setminus X|\geq 2$ means we can consider two different events $e\neq e'\in Y\setminus X$. After we prove that $Y\setminus\{e\}$ and $X\cup\{e'\}$ are also part of $L(\E)$ we can use Corollary~\ref{cor_ev_intermediaryTrans} two times, with $X\subseteq X \subseteq Y\setminus\{e\}\subseteq Y$ and with $X\subseteq X\cup\{e'\} \subseteq Y\subseteq Y$, to get asynchronous transitions of shorter length respectively $X\stepTransEvGlabbeek Y\setminus\{e\}$ and $X\cup\{e'\}\stepTransEvGlabbeek Y$, which we can use inductively.

To show that $(Y\setminus\{e\})\in L(\E)$ we must show that $\forall Z\subseteq (Y\setminus\{e\}):\exists W\subseteq(Y\setminus\{e\}):W\enableRelEv Z$. 
This can be shown from the existence of the transition $X\stepTransEvGlabbeek Y$ which says that $\forall Z\subseteq Y$, hence for our $Z\subseteq (Y\setminus\{e\})$ also, there exists $W\subseteq X$ with the property $W\enableRelEv Z$. But since $X\subseteq (Y\setminus\{e\})$ we have found the $W$ that we needed.

A similar argument is carried to prove that $X\cup\{e'\}\in L(\E)$.

Now having the transition $X\stepTransEvGlabbeek Y\setminus\{e\}$ of lower dimension we can apply the inductive hypothesis to obtain that there is a sequence of single steps in the ST-structure $\stEnc(E)$ between the ST-configurations $(X,X)\transition{}^{*}(Y\setminus e,X)\transition{}^{*}(Y\setminus e,Y\setminus e)$. The existence of $(Y,X)$ is guaranteed by the construction, thus having also a single step $(Y\setminus e,X)\transition{}(Y,X)$. Using induction with the other asynchronous transition we get that $(Y,X\cup\{e'\})\transition{}^{*}(Y,Y)$. Thus, we get the sequence of single steps we were looking for because $(Y,X)\transition{}(Y,X\cup\{e'\})$.
\end{proof}

\begin{corollary}[from {\cite[Def.2.3]{GlabbeekP09configStruct}}]\label{cor_ev_intermediaryTrans}
For some transition $X\stepTransEvGlabbeek Y$ all the intermediate smaller transitions exist (i.e., going between any two subsets $X\subseteq S\subseteq S' \subseteq Y$).
\end{corollary}

\begin{proof}
By intermediary smaller transitions we mean the transitions that go between some two subsets $X\subseteq S\subseteq S' \subseteq Y$. Thus, knowing that $X\stepTransEvGlabbeek Y$ we prove that $S\stepTransEvGlabbeek S'$. From \cite[Def.2.3]{GlabbeekP09configStruct} of the step transition relation \stepTransEvGlabbeek\ on inpure event structures we have that: $X\subseteq Y$ and $\forall Z\subseteq Y:\exists W\subseteq X: W\enableRelEv Z$. We prove that $\forall Z'\subseteq S':\exists W'\subseteq S: W'\enableRelEv Z'$. We have that all $Z'\subseteq S'\subseteq Y$ and therefore $\exists W\subseteq X: W\enableRelEv Z'$, and since $X\subseteq S$ we found our $W'\subseteq S$ to be $W$.
\end{proof}

Intuitively, one can view this last corollary as the opposite of the ``filled in'' property that was observed in Corollary~\ref{cor_cubicalProp_configST} for the ST-structures produced from a configuration structure. The ST-structures associated to the inpure event structures are not ``filled in''; i.e., the opposite direction of this last corollary does not hold.

\begin{proposition}\label{prop_evnST_adjacentclosed}
The ST-structures generated from inpure event structures as in Theorem~\ref{th_inpureEv_to_ST} are adjacent-closed (and rooted and connected if the event structure is rooted and connected).
\end{proposition}

\begin{proof}
We show first rootedness and connectedness.

Assume the event structure $\E=(E,\enableRelEv)$ is rooted and prove that the resulting ST-structure $\eintost(\E)$ is also rooted.
The event structure to be rooted means that $\emptyset\enableRelEv\emptyset$, which is equivalent to $\emptyset\in L(\E)$, by the definition of left-closed configurations $L(\E)$ from \cite[Def.1.4]{GlabbeekP09configStruct}.
The translation function \eintost\ from Theorem~\ref{th_inpureEv_to_ST} adds the ST-configuration $(\emptyset,\emptyset)\in\eintost(\E)$, therefore making it also rooted.

Assume now that \E\ is connected, i.e., all $X\in L(\E)$ are reachable from the root $\emptyset$ through a sequence of asynchronous steps $X_{1}\stepTransEvGlabbeek\dots\stepTransEvGlabbeek X_{n}$ with $X_{1}=\emptyset$ and $X_{n}=X$. But Theorem~\ref{th_inpureEv_to_ST} says that for each of these steps there exists a path $\pi_{i}$ that goes from $(X_{i},X_{i})$ to $(X_{i+1},X_{i+1})$, for $i\leq i<n$. These paths can be concatenated (in the right order) to obtain a path from $(\emptyset,\emptyset)$ to $(X,X)$ thus making any ST-configuration from $\eintost(\E)$ that is of the form $(S,S)$ reachable.

Note now that the definition of \eintost\ from the proof of Theorem~\ref{th_inpureEv_to_ST} adds one ST-configuration $(X,X)$ for each configuration $X\in L(\E)$ and one ST-configuration $(Y,X)$ for each asynchronous step from \E. It adds no other ST-configurations than these. This implies that any ST-configuration in $\eintost(\E)$ either it comes from a configuration in $L(\E)$ or it comes from an asynchronous step in \E.
It is therefore, an easy consequence of Theorem~\ref{th_inpureEv_to_ST} that for any ST-configuration $(Y,X)$ there is a path from $(X,X)$ to $(Y,Y)$ that passes through $(Y,X)$ (i.e., reaches it). Since $(X,X)$ is reachable from the root, then also $(Y,X)$ is reachable from $(\emptyset,\emptyset)$. Thus we have connectedness for the rooted $\eintost(\E)$.


We check each restriction for adjacency. 
\begin{enumerate}
\item Assuming $(S,T),(S\cup e,T),(S\cup \{e,e'\},T)$ as ST-configurations it means that these should come from the transitions $T\stepTransEvGlabbeek S$, $T\stepTransEvGlabbeek S\cup e$, $T\stepTransEvGlabbeek S\cup \{e,e'\}$, where $(e\neq e')\not\in S$. To prove the conclusion that $(S\cup e',T)$ is also an ST-configuration we prove that there is a transition $T\stepTransEvGlabbeek S\cup e'$. By the definition we have that $\forall X\subseteq S\cup \{e,e'\}:\exists Y\subseteq T:Y\enableRelEv X$; therefore it also holds that $\forall X\subseteq S\cup e':\exists Y\subseteq T:Y\enableRelEv X$, which is the desired transition.

\item Assuming $(S,T),(S\cup e,T),(S\cup e,T\cup e')$ as ST-configurations it means that these should come from the transitions $T\stepTransEvGlabbeek S$, $T\stepTransEvGlabbeek S\cup e$, $T\cup e'\stepTransEvGlabbeek S\cup e$, where $e'\in S\setminus T$ and $e\not\in S$. To prove the conclusion that $(S,T\cup e')$ is also an ST-configuration we prove that there is a transition $T\cup e'\stepTransEvGlabbeek S$. By the definition we have that $\forall X\subseteq S:\exists Y\subseteq T:Y\enableRelEv X$; therefore it also holds that $\forall X\subseteq S:\exists Y\subseteq T\cup e':Y\enableRelEv X$, which is the desired transition.

\item Assuming $(S,T),(S\cup e,T),(S,T\cup e')$ as ST-configurations it means that these should come from the transitions $T\stepTransEvGlabbeek S$, $T\stepTransEvGlabbeek S\cup e$, $T\cup e'\stepTransEvGlabbeek S$, where $e'\in S\setminus T$ and $e\not\in S$. To prove the conclusion that $(S\cup e,T\cup e')$ is also an ST-configuration we prove that there is a transition $T\cup e'\stepTransEvGlabbeek S\cup e$. 
We may use the Corollary~\ref{cor_ev_intermediaryTrans} to obtain the desired transition.

\item Assuming $(S,T),(S,T\cup e),(S,T\cup \{e,e'\})$ as ST-configurations it means that these should come from the transitions $T\stepTransEvGlabbeek S$, $T\cup e\stepTransEvGlabbeek S$, $T\cup \{e,e'\}\stepTransEvGlabbeek S$, where $e'\in S$. To prove the conclusion that $(S,T\cup e')$ is also an ST-configuration we prove that there is a transition $T\cup e'\stepTransEvGlabbeek S$ using the Corollary~\ref{cor_ev_intermediaryTrans}.
\end{enumerate}
\end{proof}

\begin{proposition}[\allST\ to \allEv]\label{prop_st_to_ev}
Any rooted, connected, and adjacent-closed ST-structure can be translated into an inpure event structure, s.t.\ the transitions of the event structure capture the concurrency embodied by the ST-structure.
\end{proposition}

\begin{proof}
The translation of the ST-structure ensures that all the concurrency is captured in the resulting event structure, in the sense that if the ST-structure expresses that a set of events can be done concurrently, then there is a transition with that set of events in the generated event structure.

The ST-structure expresses that some set of events $X$ are done in parallel whenever we have an ST-configuration where $X=S\setminus T$ (cf.\ Definition~\ref{def_ConcCausal}).

The adjacency constraint ensures that Corollary~\ref{cor_ev_intermediaryTrans} holds.

Define a translation function $\stintoe:\allST\rightarrow\allEv$ similar to what we did in Section~\ref{subsec_configStruct} and Proposition~\ref{prop_ST_config_connect} when embedding ST-structures into configuration structures. Essentially, \stintoe\ keeps only the corners of the ST-structure and uses the rest of the ST-configurations to build the enabling relation of the event structure s.t.\ the transitions that result correspond exactly to those in the ST-structure (i.e., no new transitions are introduced). 

The corners are exactly those ST-configurations where $S=T$.
Since the ST-structure is connected and rooted, for every two immediately close corners $(S,S)$ and $(S\cup e,S\cup e)$ there is also the intermediary ST-configuration $(S\cup e,S)$. Since the ST-structure is adjacent-closed, for every reachable ST-configuration $(S\cup X,S)$ we also find all the intermediary ST-configurations corresponding to the faces of the corresponding cube, cf.~Proposition~\ref{prop_adj_equiv}.

We build the enabling relation \enableRelEv\ recursively starting with the root, which is translated into the empty set. To have the empty set as an admissible left-closed configuration of the event structure we must have $\emptyset\enableRelEv\emptyset$. Since we work with a connected ST-structure then the way we build \enableRelEv\ to ensure that the transitions are respected will also make sure that the reachable sets of events respect the restriction of being left-closed; so all our built sets of events will be left-closed configurations of the event structure.

For a single step transition coming from a sequence of ST-configurations like  $(S,S)\transition{}(S\cup e,S)\transition{}(S\cup e,S\cup e)$ add to \enableRelEv\ the following: $S\enableRelEv e\cup Y$ for all $Y\subseteq S$.

Because of adjacency, and thus because for steps with more events all the intermediary single steps exist, it is enough to extend the above to sets of events $X$ as follows: for every ST-configuration $(S\cup X,S)$ add  $S\enableRelEv X'\cup Y$ for all $Y\subseteq S$ and for all $X'\subseteq X$.

This construction is enough because of connectedness of the ST-structure we started from, which implies that the $S$ was already reached through a transition in the event structure we have built earlier. Therefore, we need to consider only the new subsets. More precisely, to prove that $S\stepTransEvGlabbeek S\cup X$ we need to prove that $\forall Z\subseteq S\cup X.\exists W\subseteq S: W\enableRelEv Z$. From the fact that $S$ is reachable it means that we have that $\forall Y\subseteq S.\exists W\subseteq S'\subseteq S:W\enableRelEv Y$. It means that to get the desired result we need to consider also all the sets formed by adding some part of $X$ to any of these subsets $Y$ of $S$.

We are creating redundancy in \enableRelEv, as the transition relation would require less pairs in the relation \enableRelEv. But this redundancy is artificial four our purpose of capturing the transition relations and the configurations of the two structures (i.e., inpure event structures and adjacent-closed ST-structures).

%
%

One can now check that no transitions are introduced in the event structure that are not present in the original ST-structure. Moreover, since we take all the single steps of the ST-structure, as well as all the ST-configurations of higher dimensions, we are translating all the possible transitions from the ST-structure into the event structure.

\vspace{1ex}
\noindent\textit{Claim:}\hspace{1ex} For any $X$ a left-closed configuration of $\stintoe(\ST)$ then $(X,X)\in\ST$.
\vspace{0.5ex}

This means that $\forall Y\subseteq X.\exists Z\subseteq X:Z\enableRelEv Y$. In particular, for $X$ exists $Z\subseteq X$ s.t.\ $Z\enableRelEv X$. 
By the definition of $\stintoe(\ST)$ this last enabling must come from some $(Z\cup A,Z)\in\ST$ with $X$ divided into $X=B\cup C$ with $C\subseteq Z$ and $B\subseteq A$. This also says that $A\cap Z=\emptyset$, hence $B\cap Z=\emptyset$ and $B\cap C=\emptyset$. From $B\cap Z=\emptyset$, $X=B\cup C$, and $Z\subseteq X$ we have that $Z\subseteq C$, and hence $Z=C$. This means that the above ST-configuration is actually $(C\cup A,C)\in\ST$ for which $X=B\cup C\subseteq C\cup A$. Since \ST\ is closed under single events and connected it means that we can remove the events in $A$ to reach smaller ST-configurations; in particular we remove only the events in $A\setminus B$ thus obtaining $(X,C)\in\ST$. But by the property of \ST\ we then have that also $(X,X)\in\ST$.

\vspace{1ex}
\noindent\textit{Claim:}\hspace{1ex} For any $X\stepTransEvGlabbeek Y$ in $\stintoe(\ST)$ then $(X,Y)\in\ST$.
\vspace{0.5ex}

This means that $X\subseteq Y$ and $\forall Z\subseteq Y.\exists W\subseteq X:W\enableRelEv Z$. In particular, for $Y$ exists $W\subseteq X$ s.t.\ $W\enableRelEv Y$. 
By the definition of $\stintoe(\ST)$ this last enabling must come from some $(W\cup A,W)\in\ST$ with $Y$ divided into $Y=B\cup C$ with $C\subseteq W$ and $B\subseteq A$. This also says that $A\cap W=\emptyset$, hence $B\cap W=\emptyset$ and $B\cap C=\emptyset$. From $B\cap W=\emptyset$, $Y=B\cup C$, and $W\subseteq X\subseteq Y$ we have that $W\subseteq C$, and hence $W=C$. This means that the above ST-configuration is actually $(C\cup A,C)\in\ST$ for which $Y=B\cup C\subseteq C\cup A$ and hence $C=W\subseteq X\subseteq Y\subseteq C\cup A$. This means that $X\setminus C\subseteq A$. Since \ST\ is closed under single events and connected it means that we can add the events from $A$ to $C$ to reach ST-configurations of smaller concurrency degree; in particular we add only the events in $X\setminus C$ thus obtaining $(C\cup A,X)\in\ST$. We can also remove elements from $A$; in particular, removing the elements from $A\setminus B$ we obtain now $(Y,X)\in\ST$, i.e., the desired result.
\end{proof}

\cp{Can prove categorical results like equivalence of categories \allST\ and \allEv; in fact their respective sub-categories of rooted connected and adjacent-closed. Or one can look for adjoints since $E\isomorphic \stintoe(\eintost(E))$ and $ST\isomorphic\eintost(\stintoe(ST))$.}

\subsection{Correspondence with higher dimensional automata}

We recall the definition of higher dimensional automata (\HDA) following the terminology of \cite{Glabbeek06HDA,Pratt03trans_cancel}, defining also additional notions including the restriction to acyclic and non-degenerate \HDAs.

\begin{figure}[tp]
\psfrag{ee}{{\scriptsize  $(\emptyset,\emptyset)$}}
\psfrag{ae}{{\scriptsize $(a,\emptyset)$}}
\psfrag{be}{{\scriptsize $(b,\emptyset)$}}
\psfrag{aa}{{\scriptsize  $(a,a)$}}
\psfrag{bb}{{\scriptsize $(b,b)$}}
\psfrag{bab}{{\scriptsize $(ba,b)$}}
\psfrag{baa}{{\scriptsize $(ba,a)$}}
\psfrag{baba}{{\scriptsize $(ba,ba)$}}
\psfrag{bae}{{\scriptsize $(ba,\emptyset)$}}
\psfrag{q11}{$q_{0}^{1}$}
\psfrag{q12}{$q_{0}^{3}$}
\psfrag{q13}{$q_{0}^{4}$}
\psfrag{q14}{$q_{0}^{2}$}
\psfrag{q21}{$a$}
\psfrag{q22}{$b$}
\psfrag{q23}{$a$}
\psfrag{q24}{$b$}
\psfrag{a}{\small $a$}
\psfrag{b}{\small $b$}
\psfrag{s1}{\small $s_{1}$}
\psfrag{t1}{\small $t_{1}$}
\psfrag{s2}{\small $s_{2}$}
\psfrag{t2}{\small $t_{2}$}
\psfrag{inQ2}{$\in Q_{2}$}
\psfrag{inQ1}{$\in Q_{1}$}
\psfrag{inQ0}{$\in Q_{0}$}
\psfrag{alpha}{\small \color{red}$s_{1}(t_{2}(q_{2}))=t_{1}(s_{1}(q_{2}))$}
\psfrag{A}{$a||b$}
\psfrag{B}{$a;b + b;a$}
\psfrag{C}{$C$}
\psfrag{q3}{$q_{2}$}
\psfrag{s}{{\scriptsize $s$}}
\psfrag{t}{{\scriptsize $t$}}
\begin{center}
\includegraphics[height=3.6cm]{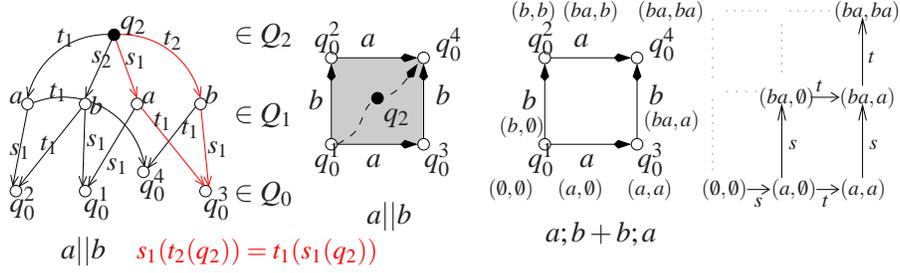}
  \end{center}
\caption{Example of a \HDA\ with two concurrent events labeled by $a$ and $b$: with an instance of cubical laws (left-most), and a more geometrical picturing (middle-left); an ST-structure and its \HDA\ (middle-right) for interleaving, which is not stable but adjacent-closed; and (right-most) a stable ST-structure that is not adjacent-closed.}
\label{fig_ex_hda}
\end{figure}

For an intuitive understanding of the \HDA\ model consider the standard example \cite{Pratt03trans_cancel,Glabbeek06HDA} pictured in Figure~\ref{fig_ex_hda}(middle-left). It represents a \HDA\ that models two concurrent events which are labeled by $a$ and $b$ (we can also have the same label $a$ for both events, giving rise to the notion of \emph{autoconcurrency}). The \HDA\ has four states, $q_{0}^{1}$ to $q_{0}^{4}$, and four transitions between them. This would be the standard picture for interleaving, but in the case of \HDA\ there is also a square $q_{2}$. Traversing through the interior of the square means that both events are executing. When traversing on the lower transition it means that event one is executing but event two has not started yet, whereas, when traversing through the upper transition it means that event one is executing and event two has finished already. In the states there is no event executing; in particular, in state $q_{0}^{4}$ both events have finished, whereas in state $q_{0}^{1}$ no event has started yet.

Similarly, \HDAs\ allow to represent three concurrent events through a cube, or more events through hypercubes. Causality of events is modelled by sticking such hypercubes one after the other. For our example, if we omit the interior of the square (i.e., the grey $q_{2}$ is removed) we are left with a description of a system where there is the choice between two sequences of the same two events, i.e., $a;b+b;a$.
This last \textit{interleaving choice} example can be seen as obtained by sticking together four cubes of dimension 1 by identifying their endpoints; whereas the true concurrency example is just one single cube of dimension 2.

\begin{definition}[higher dimensional automata]\label{def_hda}\ 

A \emph{cubical set} $H=(Q,\overline{s},\overline{t})$ is formed of a family of sets $Q=\mathop{\bigcup}_{n=0}^{\infty}Q_{n}$ with all sets $Q_{n}$ disjoint, and for each $n$, a family of maps $s_{i}, t_{i}:Q_{n}\rightarrow Q_{n-1}$, with $1\leq i\leq n$, which respect the following \emph{cubical laws}:
\begin{equation}\label{eq_cubic_laws}
\hspace{-1ex}\alpha_{i}\circ\beta_{j}=\beta_{j-1}\circ\alpha_{i},\hspace{2ex}1\!\leq\!i\!<\!j\!\leq\!n\mbox{ and }\alpha,\beta\in\!\{s,t\}.
\end{equation}
In $H$, the $\overline{s}$ and $\overline{t}$ denote the collection of all the maps from all the families (i.e., for all $n$).
A \emph{higher dimensional automaton} $(Q,\overline{s},\overline{t},\labelH,I,F)$ over an alphabet $\Sigma $ is a cubical set together with a \emph{labelling function} $\labelH:Q_{1}\rightarrow\Sigma $ which respects $\labelH(s_{i}(q))=\labelH(t_{i}(q))$ for all $q\in Q_{2}$ and $i\in\{1,2\}$; and with $I\in Q_{0}$ \emph{initial} and $F\subseteq Q_{0}$ \emph{final} cells.
\end{definition}

We call the elements of $Q_{0},Q_{1},Q_{2},Q_{3}$ respectively \textit{states}, \textit{transitions}, \textit{squares}, and \textit{cubes}, whereas the general elements of $Q_{n}$ are called \textit{cells} (also known as n-cells, n-dimensional cubes, or hypercubes).
For a transition $q\in Q_{1}$ the $s_{1}(q)$ and $t_{1}(q)$ represent respectively its source and its target cells (which are \textit{states} from $Q_{0}$ in this case). Similarly for a general n-cell $q\in Q_{n}$ there are $n$ source cells and $n$ target cells all of dimension $n-1$. 

Intuitively, an n-dimensional cell $q$ represents a configuration of a concurrent system in which $n$ events are performed at the same time, i.e., concurrently. A source cell $s_{i}(q)$ represents the configuration of the system before the starting of the $i^{th}$ event, whereas a target cell $t_{i}(q)$ represents the configuration of the system immediately after the termination of the $i^{th}$ event. We call all these source and target cells the \emph{faces} of $q$.
A cell of $Q_{1}$ represents a configuration of the system in which a single event is being performed.
The cubical laws account for the \textit{geometry} (concurrency) of the \HDAs, with four kinds of cubical laws depending on the instantiation of $\alpha$ and $\beta$; Figure~\ref{fig_ex_hda}(left) presents one such instantiation.

\begin{definition}[isomorphism of \HDAs]\label{def_isomorphismHDA}
A \emph{morphism} between two \HDAs, $f:H\rightarrow H'$ is a dimension preserving map between their cells $f:Q\rightarrow Q'$, such that:
\begin{enumerate}
\item\label{def_isomorphismHDA_1} the initial cell is preserved: $f(I)=I'$,
\item\label{def_isomorphismHDA_2} the labeling is preserved: $l'(f(q_{1}))=l(q_{1})$ for all $q_{1}\in Q_{1}$,
\item\label{def_isomorphismHDA_3} the mappings are preserved, for any $q_{n}\in Q_{n}$ and $1\leq i\leq n$:
\begin{itemize}
\item $s'_{i}(f(q_{n}))=f(s_{i}(q_{n}))$ and
\item $t'_{i}(f(q_{n}))=f(t_{i}(q_{n}))$.
\end{itemize}
\end{enumerate}
When a morphism is bijective we call it \emph{isomorphism}. Two \HDAs\ are isomorphic, denoted $H\isomorphicHDA H'$, whenever there exists an isomorphism between them.
\end{definition}

The above definition of isomorphism conforms with that in \cite[Def.2]{Glabbeek06HDA} and whereas the definition of morphism conforms with that of \cite[Sec.1.1]{Goubault12Category_Cubical}.

\begin{definition}[paths in \HDAs]\label{def_paths_HDA}
A \emph{single step} in a \HDA\ is either $q_{n-1}\transition{s}q_{n}$ with $s_{i}(q_{n})=q_{n-1}$ or $q_{n}\transition{t}q_{n-1}$ with $t_{i}(q_{n})=q_{n-1}$, where $q_{n}\in Q_{n}$ and $q_{n-1}\in Q_{n-1}$ and $1\leq i\leq n$. A \emph{path} $\pi\defequal q^{0}\transition{\alpha^{1}}q^{1}\transition{\alpha^{2}}q^{2}\transition{\alpha^{3}}\dots$ is a sequence of single steps $q^{j}\transition{\alpha^{j+1}}q^{j+1}$, with $\alpha^{j}\in\{s,t\}$. 
We say that $q\in\pi$ iff $q=q^{j}$ appears in one of the steps in $\pi$. 
The first cell in a path is denoted $\startPath{\pi}$ and the ending cell in a finite path is $\finishPath{\pi}$. 
\end{definition}

Note that the marking of the steps by $s/t$ can be deduced from the fact that the step goes from a lower cell to a higher cell for s-steps (and the opposite for t-steps). It is though useful in many of the proofs to have easily visible the exact map (i.e., the index also) that the step uses, instead of explicitly assuming it every time.

\begin{definition}[histories for \HDA\ -- from {\cite[Sec.7]{Glabbeek06HDA}}]\label{def_history_HDA}\ 

In a \HDA\ two paths are \emph{adjacent}, denoted $\pi\adjacentHDA\pi'$ if one can be obtained from the other by replacing, for $q,q'\in Q$ and $i<j$,
\begin{enumerate}
\item a segment $\transition{s_{i}}q\transition{s_{j}}$ by $\transition{s_{j-1}}q\transition{s_{i}}$, or
\item a segment $\transition{t_{t}}q\transition{t_{i}}$ by $\transition{t_{i}}q\transition{t_{j-1}}$, or
\item a segment $\transition{s_{i}}q\transition{t_{j}}$ by $\transition{t_{j-1}}q\transition{s_{i}}$, or
\item a segment $\transition{s_{j}}q\transition{t_{i}}$ by $\transition{t_{i}}q\transition{s_{j-1}}$.
\end{enumerate}
Two finite paths are \textit{l-adjacent} $\pi\ladjacentHDA{l}\pi'$ when the segment replacement happens at position $l+1$; i.e., $q$ is the $l+1$ cell in the path.
\emph{Homotopy} is the reflexive and transitive closure of adjacency. Two paths that are homotopic, denoted $\pi\homotopicHDA\pi'$, share their respective start and end cells. All homotopic rooted paths that have the same end cell $q$ are said to be a \emph{history of $q$} and is denoted $\homotopyClass{q}$ when this is unique. We use the same notation for the homotopy class of a rooted path, $\homotopyClass{\pi}$, which is also used when a cell has more than one history, as is the case with the interleaving square HDA from Figure~\ref{fig_ex_hda}.
\end{definition}

Above, homotopy is defined for all paths (opposed to the definition in \cite[Sec.1.6]{Goubault12Category_Cubical}) and thus also a cell of higher dimension, like the inside of a square, has a history, not only the state cells of dimension $0$ that form the corners of the square.

Inspired by the definition of history unfolding for process graphs from \cite[Sec.3]{glabbeek96histUnfold} we define the same notion for HDAs (the present author is not aware of this notion being defined for HDAs anywhere else).

\begin{definition}[history unfolding for HDAs]\label{def_unfolding_history}\ 

The \emph{history unfolding} $\unfolding(H)$ of a higher dimensional automaton $H$ is given by:
\begin{itemize}
\item $Q_{n}^{\unfolding(H)}$ is the set of histories that end up in cells on level $Q_n$ of $H$,
\item has the same labeling as $H$ and initial cell the empty rooted history,
\item the $s/t$ maps are built from the corresponding maps between the end cells of the histories; i.e., $s_{i}(\homotopyClass{\pi})=\homotopyClass{\pi'}$ iff $s_{i}(q)=q'\wedge \pi'\transition{s}\pi\wedge \finishPath{\pi'}=q'\wedge \finishPath{\pi}=q$.
\end{itemize}

\end{definition}

\begin{definition}[hh-bisimulation]\label{def_hhbisim}
Two higher dimensional automata $H_{A}$ and $H_{B}$ (with $I_{A}$ and $I_{B}$ the initial cells) are \emph{hereditary history-preserving bisimulation equivalent} (hh-bisimilar), denoted $H_{A} \hhequiv H_{B}$, if there exists a binary relation $R$ between their paths starting at $I_{A}$ respectively $I_{B}$ that respects the following:
\begin{enumerate}
\item if $\pi_{A}R\pi_{B}$ and $\pi_{A}\transition{a^{\pm}}\pi_{A}'$ then $\exists\pi_{B}'$ with $\pi_{B}\transition{a^{\pm}}\pi_{B}'$ and $\pi_{A}'R\pi_{B}'$;
\item if $\pi_{A}R\pi_{B}$ and $\pi_{B}\transition{a^{\pm}}\pi_{B}'$ then $\exists\pi_{A}'$ with $\pi_{A}\transition{a^{\pm}}\pi_{A}'$ and $\pi_{A}'R\pi_{B}'$;
\item\label{hh_HDM_l1} if $\pi_{A}R\pi_{B}$ and $\pi_{A}\homotopic{l}\pi_{A}'$ then $\exists\pi_{B}'$ with $\pi_{B}\homotopic{l}\pi_{B}'$ and $\pi_{A}'R\pi_{B}'$;
\item\label{hh_HDM_l2} if $\pi_{A}R\pi_{B}$ and $\pi_{B}\homotopic{l}\pi_{B}'$ then $\exists\pi_{A}'$ with $\pi_{A}\homotopic{l}\pi_{A}'$ and $\pi_{A}'R\pi_{B}'$;
\item if $\pi_{A}R\pi_{B}$ and $\pi_{A}'\transition{a^{\pm}}\pi_{A}$ then $\exists\pi_{B}'$ with $\pi_{B}'\transition{a^{\pm}}\pi_{B}$ and $\pi_{A}'R\pi_{B}'$;
\item if $\pi_{A}R\pi_{B}$ and $\pi_{B}'\transition{a^{\pm}}\pi_{B}$ then $\exists\pi_{A}'$ with $\pi_{A}'\transition{a^{\pm}}\pi_{A}$ and $\pi_{A}'R\pi_{B}'$.
\end{enumerate}
\end{definition}

A corollary from \cite{Glabbeek06HDA} strengthens the above conditions~\ref{hh_HDM_l1} and \ref{hh_HDM_l2} to \textit{unique} existence.

\begin{corollary}[cf.~{\cite[sec.7.5]{Glabbeek06HDA}}]\label{cor_unique_adjacent}
For a path $\pi$ and a point $l>1$ there exists a unique path $\pi'$ that is $l$-adjacent with $\pi$.
\end{corollary}

Many of the results in this paper work with \emph{acyclic} and \emph{non-degenerate} \HDAs\ in the following sense. Such \HDAs\ are often considered in the literature on concurrent systems and are more general than most of the true concurrency models \cite{Pratt03trans_cancel,Glabbeek06HDA}.

\begin{definition}[acyclic and non-degenerate \HDAs]\label{def_acyclic}\ 

A \HDA\ is called \emph{acyclic} if no path visits a cell twice. 
A \HDA\ is called \emph{non-degenerate} if for any cell $q$ all its faces exist and are different, in the sense of $\forall i\neq j:\alpha_{i}(q)\neq\beta_{j}(q)\wedge\alpha,\beta\in\{s,t\}$, and no two transitions with the same label share both their end states.
\end{definition}

The restriction on \HDAs\ that we call here ``non-degenerate'' is close to that of Cattani and Sassone \cite[Def.2.2]{CattaniSassone96HDTS} and that of van Glabbeek \cite[p.10]{Glabbeek06HDA}. The second constraint of non-degeneracy is close to the notion of strongly labeled of \cite[Def.1.13]{Goubault12Category_Cubical}.
Note that the non-degeneracy still allows for two opposite s and t-maps to be equal, i.e., it is allowed $s_{i}=t_{i}$. But when the \HDA\ is also required to be acyclic then this is also ruled out since it would create a cycle.
In this paper we usually work with non-degenerate \HDAs; and moreover we silently assume all the s/t-maps to be total.

\begin{definition}[\allST\ to \allHDA]\label{def_STtoHDA}
We define a mapping $\stintoh:\allST\rightarrow\allHDA$ from ST-structures into HDAs which for a $\ST=(E,ST,l)$ with the events linearly ordered as a list $\evlist{E}$ (i.e., each event being indexed by a natural number) returns the HDA $\stintoh(\ST)$ which
\begin{itemize}
\item has cells $Q=\{q^{(S,T)}\in Q_{n} \mid (S,T)\in ST \mbox{ and } |S\setminus T|=n\}$;

\item for any two cells $q^{(S,T)}$ and $q^{(S\setminus e,T)}$ add the map entry $s_{i}(q^{(S,T)})=q^{(S\setminus e,T)}$ where $i$ is the index of the event $e$ in the listing $\evlist{E}\!\!\downarrow_{(S\setminus T)}$;

\item for any two cells $q^{(S,T)}$ and $q^{(S,T\cup e)}$ add the map entry $t_{i}(q^{(S,T)})=q^{(S,T\cup e)}$ where $i$ is the index of the event $e$ in the listing $\evlist{E}\!\!\downarrow_{(S\setminus T)}$;

\item has labeling $l(q^{(T\cup e,T)})=l(e)$ for any $q^{(T\cup e,T)}\in Q_{1}$.
\end{itemize}
More precisely, by $\evlist{E}\!\!\downarrow_{(S\setminus T)}$ we represent the listing of the events in $S\setminus T$, i.e., a list of dimension $|S\setminus T|$ obtained from the original listing $\evlist{E}$ by removing all other events. This new listing has the events of $S\setminus T$ in the same original order but with new indexes attached (ranging from $1$ to $|S\setminus T|$).
\end{definition}

\begin{theorem}\label{th_stintohda}
For a rooted, connected, and adjacent-closed ST-structure \ST\ the mapping \stintoh\ associates a $\stintoh(\ST)$ which is a higher dimensional automaton respecting all cubical laws and is acyclic and non-degenerate.
\end{theorem}

\begin{proof}
We first show that $\stintoh(\ST)$ is a HDA in the sense of Definition~\ref{def_hda}.
For any cell $q^{(S,T)}$ all immediately lower cells $q^{(S\setminus e,T)}$ and $q^{(S,T\cup e)}$, with $e\in S\setminus T$ exist because the ST-structure is rooted, connected, and adjacent-closed, and by Proposition~\ref{prop_adj_equiv} is closed under single events, therefore all ST-configurations $(S\setminus e,T)$ and $(S,T\cup e)$ exist and thus have the above associated cells. 
Consider for now that each immediately lower cell $q^{(S\setminus e,T)}$ is linked through $s_{e}((S,T))=(S\setminus e,T)$. Note that the s-maps are not indexed as in the definition, but are indexed by an event. We will replace these event indexes by numbers. Link these cells also to $q^{(S,T\cup e)}$ through $t_{e}(S,T)=(S,T\cup e)$. 

To get the cubical laws right we must use a discipline in replacing the event indexes for the s and t maps by numbers. This is what the Definition~\ref{def_STtoHDA} does (inspired from \cite{Glabbeek06HDA}). 
The listing of the events that the ST-structure comes with provides a bijective indexing map $i(\cdot)$ from $E$ to $\mathbb{N}$. For a specific ST-configuration $(S,T)$ this indexing map becomes a map from $S\setminus T$ to $I=\{1,\dots,n\}$, with $n=|S\setminus T|$, that respects the original ordering of the events from the listing of $E$. Call this indexing $i\downarrow_{(S\setminus T)}(\cdot)$.
For the cell $q^{(S,T)}$ replace $s_{e}$ by $s_{i\downarrow_{(S\setminus T)}(e)}$ and $t_{e}$ by $t_{i\downarrow_{(S\setminus T)}(e)}$. For each immediately lower cell, like $q^{(S\setminus e,T)}$, which is linked as $s_{i\downarrow_{(S\setminus T)}(e)}(S,T)$, their corresponding indexing maps look like $i\downarrow_{(S\setminus e)\setminus T}$.
The relationship between these two maps $i\downarrow_{(S\setminus T)}$ and $i\downarrow_{(S\setminus e)\setminus T}$ is easy to see; for simplicity of notation denote the two maps respectively by $i$ and $i_{e}$. The indexing map $i_{e}(\cdot)$ is defined on $(S\setminus e)\setminus T$ as $i_{e}(f)=i(f)$ if $i(f)< i(e)$, and $i_{e}(f)=i(f)-1$ if $i(f) > i(e)$.  The same holds for $q^{(S,T\cup e)}$.

One can check that the cubical laws hold. This is easier done by keeping in mind an intuitive association between each cubical law and the corresponding adjacent-closure constraint.
As an example: $t_{i\downarrow_{(S\setminus e)\setminus T}(f)}(s_{i\downarrow_{(S\setminus T)}(e)}(q^{(S,T)}))=s_{i\downarrow_{S\setminus(T\cup f)}(e)}(t_{i\downarrow_{(S\setminus T)}(f)}(q^{(S,T)}))$ for two events $e,f\in S\setminus T$ under the assumption that $i(f)<i(e)$.

The labeling of the \HDA\ is obtained from the labeling of the ST-structure. Each $q^{(S,T)}\in Q_{1}$ is labeled with $l(e)$ where $\{e\}=S\setminus T$ is the single event that is concurrent in $(S,T)$.

\vspace{1ex}
\noindent\textit{Claim:}\hspace{1ex} The $\stintoh(\ST)$ is acyclic and non-degenerate.
\vspace{0.5ex}

To prove \textit{non-degeneracy} one can notice that for showing that any cell has all its faces distinct it is enough to recall how the faces of some cell $q^{(S,T)}$ have been built. One s-face is a cell $q^{(S\setminus e,T)}$ that is obtained from an ST-configuration that can immediately reach $(S,T)$ through an s-step in \ST, and which adds the event $e$; hence the labeling of the corresponding s-map by $s_{e}$. Since we added one such map and face for each distinct event from $S\setminus T$, then all the resulting cells are distinct. The same for the t-maps.

A note is in order. In the definition and the argumentation above, two generated cells $q^{(S,T)}$ and $q^{(S',T')}$ are considered equal (respectively different) iff $(S,T)=(S',T')$ (respectively $(S,T)\neq(S',T')$).

We now finis proving non-degeneracy. For two transitions, i.e., cells of dimension one, hence obtained as $q^{(S,T)}$ and $q^{(S',T')}$ with $S\setminus T=\{e\}$ (respectively $S'\setminus T'=\{f\}$) with $l(e)=l(f)$ assume they have the same source. This implies that $T=T'$ and thus the two transitions are $q^{(Te,T)}$ and $q^{(Tf,T)}$. Since these two transitions are assumed different then it implies that $e\neq f$. The target of the first transition thus becomes $q^{(Te,Te)}$ and of the second $q^{(Tf,Tf)}$ which are different.

To prove that the obtained \HDA\ is acyclic note first that each step in the ST-structure is matched precisely by a corresponding single step of the same type in the \HDA. Moreover, one step in the ST-structure increases strictly the dimension of the ST-configuration (since it adds one new event to one of the two sets).
Because each cell in the resulting \HDA\ is labeled by an ST-configuration to which it corresponds, we can define a weight for each cell to be the dimension of the ST-configuration that it is labeled with. With this we can define a weight for each finite path to be the weight of the cell it ends in.

Each path in the \HDA\ is matched by one path in the ST-structure. Since each extension of a path reaches an ST-configuration of strictly larger dimension, it means that each extension of a path in the \HDA\ will have strictly larger weight. To have a cycle, the \HDA\ must have one path that visits the same cell twice; say $q^{(S,T)}$. This means that an initial segment of this path that ends in $q^{(S,T)}$, which has weight $|(S,T)|$, is extended to another path that ends in the same cell, hence having he same weight. But this is a contradiction, since any extension strictly increases the weight.
\vspace{1ex}
\end{proof}

The next lemma ensures that it is immaterial which listing of the events is picked in the definition of the mapping \stintoh.

\begin{lemma}\label{lemma_listings_different}
For some \ST\ and two listings $\evlist{E_{1}},\evlist{E_{2}}$ of the events $E$, the \HDAs\ resulting from the application of \stintoh\ with each listing are isomorphic up to reindexing of the maps.
\end{lemma}

\begin{proof}
Take the two generated \HDAs\ to be respectively  $H_{1}$ and $H_{2}$. Since $\evlist{E_{1}}$ and $\evlist{E_{2}}$ are two listings of the same set then we get a permutation $p$ of their indexes, in the sense that if $e$ is on position $i$ in $\evlist{E_{1}}$ then the same event is on position $p(i)$ in $\evlist{E_{2}}$.

The two generated \HDAs\ are isomorphic through the identity morphism $F(q^{(S,T)})=q^{(S,T)}$. The only thing to check is that it preserves the mappings up to the reindexing of the maps according to the above permutation; i.e., instead of showing that $s_{i}(F(q^{(S,T)}))=F(s_{i}(q^{(S,T)}))$ we show that $s_{p\downarrow_{(S\setminus T)}(i)}(F(q^{(S,T)}))=F(s_{i}(q^{(S,T)}))$.

In $H_{1}$ we have that $s_{i}(q^{(S,T)})=q^{(S\setminus e,T)}$ with $e$ having index $i$ in the listing $\evlist{E_{1}}\!\downarrow_{(S\setminus T)}$, thus making the whole right-hand side of the equality $q^{(S\setminus e,T)}$. On the left side, $F(q^{(S,T)})$ returns the same $q^{(S,T)}$ in $H_{2}$; and $s_{p\downarrow_{(S\setminus T)}(i)}(q^{(S,T)})=q^{(S\setminus g,T)}$ where $g$ is the event on index $p\downarrow_{(S\setminus T)}(i)$ in the listing $\evlist{E_{2}}\!\downarrow_{(S\setminus T)}$. By the notation $p\downarrow_{(S\setminus T)}(i)$ we mean the restriction of the permutation $p$ to $S\setminus T$ in the following sense. We have $p\downarrow_{(S\setminus T)}:|\evlist{E_{1}}\!\downarrow_{(S\setminus T)}|\rightarrow \evlist{E_{2}}\!\downarrow_{(S\setminus T)}$ defined as $p\downarrow_{(S\setminus T)}(i)=l$ if $\evlist{E_{2}}\!\downarrow_{(S\setminus T)}[l]=\evlist{E_{2}}[p(k)]$ with $\evlist{E_{1}}[k]=\evlist{E_{1}}\!\downarrow_{(S\setminus T)}[i]$.
It is easy to see that $g=e$ and hence the desired result.
\end{proof}

\begin{figure}[tp]
\psfrag{e}{\scriptsize $s$}
\psfrag{d}{\scriptsize $b$}
\psfrag{f}{\scriptsize $f$}
\psfrag{00}{{\scriptsize $(\emptyset,\emptyset)$}}
\psfrag{d0}{{\scriptsize $(b,\emptyset)$}}
\psfrag{e0}{{\scriptsize $(s,\emptyset)$}}
\psfrag{ee}{{\scriptsize $(s,s)$}}
\psfrag{dd}{{\scriptsize $(b,b)$}}
\psfrag{ded}{{\scriptsize $(bs,b)$}}
\psfrag{dede}{{\scriptsize $(bs,bs)$}}
\psfrag{dfd}{{\scriptsize $(bf,b)$}}
\psfrag{dfdf}{{\scriptsize $(bf,bf)$}}
\psfrag{0}{{\scriptsize $\emptyset$}}
\psfrag{0d}{{\scriptsize $\{b\}$}}
\psfrag{0e}{{\scriptsize $\{s\}$}}
\psfrag{0ed}{{\scriptsize $\{b,s\}$}}
\psfrag{sd0}{{\small \begin{rotate}{125}$\enableRelEv$\end{rotate}}}
\psfrag{se0}{{\small \begin{rotate}{45}$\enableRelEv$\end{rotate}}}
\psfrag{sed}{{\small \begin{rotate}{45}$\enableRelEv$\end{rotate}}}
\psfrag{hh}{{\scriptsize $\hhequiv$}}
\psfrag{iso}{{\scriptsize $\isomorphic$}}
\psfrag{niso}{{\scriptsize $\not\isomorphic$}}
\psfrag{ST}{{\scriptsize on STs}}
\psfrag{HDA}{{\scriptsize on \HDAs}}
  \begin{center}
    \includegraphics[height=2.5cm]{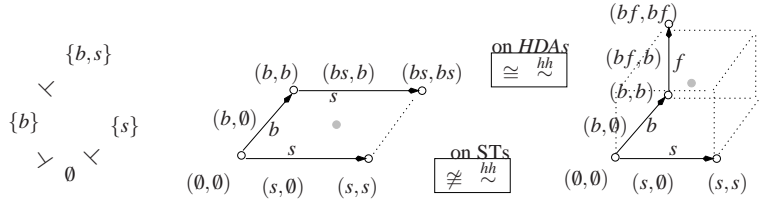}
  \end{center}
\caption{\textit{Strong asymmetric conflict} $s+b;s$ as inpure event structure (left), and ST-structure (middle). Related isomorphic \HDAs\ give rise to non-isomorphic ST-structures (middle and right).}
\label{fig_ex_asymconflict1}
\end{figure}

\begin{example}[Strong asymmetric conflict]\label{example_asym_confl}
This example, taken from \cite[Ex.3]{GlabbeekP09configStruct} (called \textit{strong} in \cite[p.22]{Pratt03trans_cancel}), shows the gain in expressive power of the ST-structures. Asymmetric conflict cannot be captured in the \emph{pure event structures} of \cite[Def.1.5]{GlabbeekP09configStruct}, hence not by the configuration structures. Asymmetric conflict can be captured by the inpure event structures of \cite{GlabbeekP09configStruct}, and thus, also by the adjacent-closed ST-structure of Fig.~\ref{fig_ex_asymconflict1}(middle).

The example has no concurrency and involves two events, imposing the only restriction that once event $s$ happens, event $b$ cannot happen any more.

Within \HDAs\ it is more cumbersome to represent this example because \HDAs\ are not good at identifying the particular events. \HDAs\ abstract from the concrete events and concentrate only on the labels. 
One way of identifying events is by equivalence classes of transitions, where two transitions are equivalent when they are parallel in the border of a filled square (i.e., what we assumed until now in our \HDAs\ examples).

Applying this technique to the \HDA\ in Fig.~\ref{fig_ex_asymconflict1}(middle) would not result in the corresponding 2-events ST-structure (which is what we want), but would result in the 3-events ST-structure of Fig.~\ref{fig_ex_asymconflict1}(right), and these two ST-structures are not isomorphic.
On the other hand, the two representations of \HDA\ from Fig.~\ref{fig_ex_asymconflict1}(middle and right) are isomorphic. 
Nevertheless, if we are interested in representing systems only up to hh-bisimulation, then both \HDAs\ and ST-structures are as good, because the different representations of ST-structures (with 2 or 3 events) would be equated by the hh-bisimulation.
\end{example}

\begin{proposition}\label{prop_stintoh_iso}
\ 

\begin{enumerate}
\item\label{prop_stintoh_iso_1} The mapping \stintoh\ from Definition~\ref{def_STtoHDA} preserves isomorphism; i.e., for $\ST\isomorphic\ST'$ then $\stintoh(\ST)\isomorphic\stintoh(\ST')$.

\item\label{prop_stintoh_iso_collapse} The mapping \stintoh\ may collapse non-isomorphic ST-structures into isomorphic \HDAs.
\end{enumerate}
\end{proposition}

\begin{proof}
For the second part of the proposition consider the two ST-structures from Figure~\ref{fig_ex_asymconflict1} which are not isomorphic as the left one is defined on two events whereas the right one is on three events. But the \HDAs\ that the mapping \stintoh\ associates are isomorphic.

\vspace{1ex}
For the first part of the proposition consider two isomorphic ST-structures $\ST\isomorphic\ST'$ with $f:E\rightarrow E'$ their respective isomorphism. To show that $\stintoh(\ST)\isomorphic\stintoh(\ST')$ we build an isomorphism between the two generated \HDAs\ as $F:Q\rightarrow Q'$ given by $F(q^{(S,T)})=q^{(f(S),f(T))}$. We prove that $F$ is a dimension preserving isomorphism of the two $\stintoh(\ST)$ and $\stintoh(\ST')$ as in Definition~\ref{def_isomorphismHDA}.

Because of Lemma~\ref{lemma_listings_different}, if for the translation of \ST\ we pick some listing of $E$, then for the translation of $\ST'$ we pick the listing of $f(E)$ such that the order of the events is preserved; i.e., if $e_{i}<e_{j}$ in the listing of $E$ then also $f(e_{i})<f(e_{j})$ in the listing of $f(E)$. Since $f$ is a bijection this means that if $e_{i}$ is on position $i$ in the listing of $E$ then  we find $f(e_{i})$ on the same position $i$ in the listing of $f(E)$.

For $q^{(S,T)}$ which was generated from the $(S,T)\in\ST$, the isomorphism of ST-structures ensures that $(f(S),f(T))\in\ST'$, which means that the mapping will associate the cell $q^{(f(S),f(T))}\in\stintoh(\ST')$. This makes $F$ well defined. Moreover, because the isomorphism $f$ preserves the concurrency degree of the ST-configurations, i.e., $|S\setminus T|=|f(S)\setminus f(T)|$, then $F$ preserves the dimension of the cells.
It is easy to see that $F$ preserves the initial cell.
$F$ also preserves the labeling since $l'(F(q^{(Te,T)}))=l'(q^{(f(T)f(e),f(T))})\stackrel{(1)}{=}l'(f(e))\stackrel{(2)}{=}l(e)$, where the equality (1) comes from the Definition~\ref{def_STtoHDA} of the \stintoh\ map and (2) comes from the Definition~\ref{def_isomorphism} of isomorphism for ST-structures; and since $l(q^{(Te,T)})\stackrel{(1)}{=}l(e)$, we obtain the requirement \ref{def_isomorphismHDA_2} of Definition~\ref{def_isomorphismHDA}.

It remains to show that $F$ preserves the mappings, i.e., $s_{i}(F(q^{(S,T)}))=F(s_{i}(q^{(S,T)}))$. (The case for t-maps is analogous.) By the Definition~\ref{def_STtoHDA} of the \stintoh, $s_{i}(q^{(S,T)})=q^{(S\setminus e,T)}$ for $e$ with index $i$ in the listing of the events $E\!\downarrow_{S\setminus T}$, and thus $F(s_{i}(q^{(S,T)}))=q^{(f(S\setminus e),f(T))}$. Since $F(q^{(S,T)})=q^{(f(S),f(T))}$, by Definition~\ref{def_STtoHDA}, $s_{i}(q^{(f(S),f(T))})=q^{(f(S)\setminus g,f(T))}$ for $g$  the event with index $i$ in the listing of the events $f(E)\!\downarrow_{f(S)\setminus f(T)}$. If we choose the listing of the events of $\ST'$ such that the isomorphisms $f$ preserves their order, as we explained before, then the event $g$ is exactly $f(e)$. Therefore, we have the equality we are looking for; $s_{i}(F(q^{(S,T)}))=q^{(f(S)\setminus f(e),f(T))}$.
\end{proof}

A corollary of Proposition~\ref{prop_stintoh_iso}\refeq{prop_stintoh_iso_collapse} is that \stintoh\ in not an embedding from \allST\ to \allHDA\ since it looses information, i.e., the events.

Nevertheless, the mapping \stintoh\ preserves hh-bisimulation.

\begin{definition}[\allHDA\ to\ \allST]\label{def_hdaTOst}\ 

Consider a non-degenerate \HDA\ $H=(Q,\overline{s},\overline{t},l,I)$.
Define a relation $\eventEquivHDAs\subseteq Q_{1}\times Q_{1}$ on transitions as 
\[
q_{1}\eventEquivHDAs q_{1}' \mbox{\hspace{2ex} iff \hspace{2ex}} \exists q_{2}\in Q_{2}:\alpha_{i}(q_{2})=q_{1}\wedge \beta_{i}(q_{2})=q_{1}'
\]
for some $i\leq 2$ and $\alpha,\beta\in\{s,t\}$.
Consider the reflexive and transitive closure of the above relation, and denote it the same. This is now an equivalence relation on $Q_{1}$. Consider an equivalence class $\equivClass{q_{1}}$ to be all $q_{1}'$ equivalent with $q_{1}$.
Such an equivalence class is called \emph{an event}.

Define a map $\hintost:\allHDA\rightarrow\allST$ which builds an ST-structure $\hintost(H)$ by associating to each rooted path $\pi\in H$ an ST-configuration as follows.
\begin{enumerate}
\item\label{hintost_1} for the minimal rooted path which ends in $I$ associate $(\emptyset,\emptyset)$;

\item\label{hintost_2} for any path $\pi$ which ends in a transition $\finishPath{\pi}=q_{1}\in Q_{1}$ then 
\begin{enumerate}
\item\label{hintost_21} add the ST-configuration $\hintost(\pi)=\hintost(\pi_{s})\cup(\equivClass{q_{1}},\emptyset)$ with $\pi_{s}\transition{s}q_{1}\in\homotopyClass{\pi}$;

\item\label{hintost_22} add the ST-configuration $\hintost(\pi\transition{t}q_{0})=\hintost(\pi)\cup(\emptyset,\equivClass{q_{1}})$;
\end{enumerate}

\item\label{hintost_3} for any path $\pi$ which ends in a higher cell $\finishPath{\pi}=q_{n}\in Q_{n}$, with $n\geq 2$, then add the ST-configuration $\hintost(\pi)=\hintost(\pi^{i})\cup\hintost(\pi^{j})$, with $\pi^{i}\neq\pi^{j}$, $\pi^{i}\transition{s}q_{n}\in\homotopyClass{\pi}$, and $\pi^{j}\transition{s}q_{n}\in\homotopyClass{\pi}$.
\end{enumerate}
\end{definition}

Note that in the case~\refeq{hintost_3} above the paths $\pi^{i},\pi^{j}$ always exist because we work with non-degenerate \HDAs. The same goes for the path $\pi_{s}$ used in \refeq{hintost_21}.

\cp{
The definition above is based on the following lemma.

\begin{lemma}
Two homotopic paths $\pi\homotopic{hom}\pi'$ are translated in the same ST-configuration:

\centerline{$\hintost(\pi)=\hintost(\pi')$.}
\end{lemma}

\begin{proof}

\end{proof}
}

\begin{proposition}\label{prop_hdaintost}
For an acyclic and non-degenerate \HDA\ the resulting ST-structure $\hintost(H)$ is rooted, connected, and adjacent-closed.
\end{proposition}

\begin{proof}
%
%
Rootedness is easy because it corresponds to the minimal rooted path of the \HDA, i.e., the initial cell.

Connectedness is satisfied when the \HDA\ that we translate is connected. This is the case because we work with \HDAs\ that are closed under reachable parts. This means that we consider only those cells that are reachable from the initial cell, as we stated before.

In a non-degenerate \HDA\ every $Q_{1}$ cell has exactly one $s$ and one $t$ map. This means that every path $\pi$ with $\finishPath{\pi}=q_{1}$ there exists exactly one path $\pi_{s}$ which can reach $\pi$ through an s-step. Moreover, for $\pi$ there is exactly one continuation by a t-step, and this reaches a $q_{0}$ state cell; this motivates in the definition the consideration of the paths $\pi\transition{t}q_{0}$. In fact any state cell $q_{0}\in Q_{0}$ can be reached through such a path.

In a non-degenerate \HDA, every higher dimensional cell $q_{n}\in Q_{n}$, with $n\geq 2$, has at least two source maps which enter a cubical law $s_{i}(s_{j}(q_{n}))=s_{j-1}(s_{i}(q_{n}))=q_{n-2}\in Q_{n-2}$, and with $s_{j}(q_{n})\neq s_{i}(q_{n}) \in Q_{n-1}$. There are uniquely corresponding paths for the original $\pi$ and the cells involved and for this cubical law, these paths being connected through the corresponding s-steps. This motivates the last point in the definition.

Connectedness is proven using induction on the length of the paths that are used to generate the ST-configurations.
It is easy to see that the ST-configuration introduced in \ref{hintost_21} is connected to the ST-configuration of the immediately shorter path through adding one new event to the $S$ set. This event is new because the \HDA\ is acyclic, and thus the path $\pi$ never goes through a cell twice; in particular it has never been through the $q_{1}$ that is used in the definition.
Similarly, the ST-configuration introduced in \ref{hintost_22} is connected to the the ST-configuration of the immediately shorter path through a t-step, i.e., terminating the event $\equivClass{q_{1}}$.

For a higher dimensional cell $q_{n}$ we show that the associated ST-configuration differs from each immediately lower paths reaching it through an s-step by only one event in the $S$ set; thus ensuring connectedness.
For this we show that for any two transitions $q_{1},q_{1}'$ entering the cubical law $s_{1}(s_{1}(q_{2}))=s_{1}(s_{2}(q_{2}))$, with $s_{1}(q_{2})=q_{1}$ and $s_{2}(q_{2})=q_{1}'$ they cannot be equivalent, i.e., not denote the same event.

%

To show adjacent-closure we use Proposition~\ref{prop_adj_equiv} because the ST-structure is rooted and connected. Therefore it is enough to show closure under single events. 
\cp{
We base the rest of the proof on the following.

\vspace{1ex}
\noindent\textit{Claim:}\hspace{1ex} $\forall\pi: \finishPath{\pi}\in Q_{n} \Rightarrow \hintost(\pi) \mbox{ has concurrency degree }n$.
\vspace{0.5ex}

The second property of Definition~\ref{def_closeSingleEv} is obtained from the above fact that in the extension step the ST-configuration that we introduce differs in exactly one event; and since we have chosen two arbitrary s-maps, we obtain the result for all the concurrent events. Showing the first constraint of Definition~\ref{def_closeSingleEv} is simple from the completion step and the above, since this step considers each and all the s-maps, where each such map corresponds to one of the concurrent events.
\cp{Redo this last paragraph!!!}
}
\end{proof}

\cp{
\begin{proposition}\label{prop_Hpreserveshh}
For two acyclic and non-degenerate \HDAs, $\modelH$ and $\modelH'$, their corresponding rooted, connected and adjacent-closed ST-structures $\mathsf{ST}(\modelH)$ and $\mathsf{ST}(\modelH')$ are hh-bisimilar (cf.~Def.~\ref{def_hh_ST}) iff the original higher dimensional automata are hh-bisimilar.
\end{proposition}

\begin{proof}[sketch]
Intuitively, the $l$-adjacency steps for HDA correspond to the restriction 1 in Def.~\ref{def_hh_ST} of $f$-isomorphism together with the adjacent-closure properties for ST-structures.
\end{proof}
}

\begin{proposition}\label{prop_hintost_iso}
\ 

\begin{enumerate}
\item\label{prop_hintost_iso_1} The mapping \hintost\ from Definition~\ref{def_hdaTOst} preserves isomorphism of reachable parts; i.e., for $H\isomorphic H'$ then $\hintost(H)\isomorphic\hintost(H')$.

\item\label{prop_hintost_collapse} The mapping \hintost\ may collapse non-isomorphic \HDAs\ into isomorphic ST-structures.
\end{enumerate}
\end{proposition}

\begin{figure}[tp]
\psfrag{a}{\scriptsize $a$}
\psfrag{b}{\scriptsize $b$}
\psfrag{c}{\scriptsize $c$}
\psfrag{d}{\scriptsize $d$}
\psfrag{e}{\scriptsize $e$}
\psfrag{I}{{\scriptsize $(\emptyset,\emptyset)$}}
\psfrag{ab}{{\scriptsize $(ab,ab)$}}
\psfrag{dfd}{{\scriptsize $(bf,b)$}}
\psfrag{F}{{\scriptsize $(abc,abc)$}}
\psfrag{hh}{{\scriptsize $\hhequiv$}}
\psfrag{nhh}{{\scriptsize $\not\hhequiv$}}
\psfrag{iso}{{\scriptsize $\isomorphic$}}
\psfrag{niso}{{\scriptsize $\not\isomorphic$}}
\psfrag{ST}{{\scriptsize on STs}}
\psfrag{HDA}{{\scriptsize on \HDAs}}
\psfrag{with}{{\scriptsize with $d,e$}}
  \begin{center}
    \includegraphics[height=3.5cm]{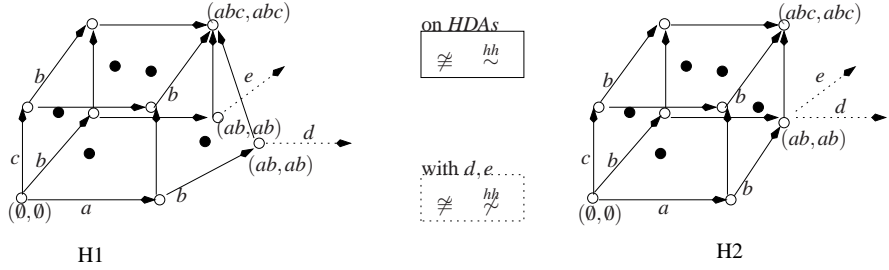}
  \end{center}
\caption{Identifying of non-isomorphic and non-hh-bisimilar \HDAs\ through \hintost.}
\label{fig_ex_Glabbeek}
\end{figure}

\begin{proof}
For the second part of the proposition consider the two \HDAs\ from Figure~\ref{fig_ex_Glabbeek} without the two dotted transitions, which are translated into the same ST-structure by \hintost. Even when the dotted transitions are added, they are mapped to the same ST-structure.

For the first part we build an isomorphism $I_{ST}$ between  $\hintost(H)$ and $\hintost(H')$ starting from the isomorphism $I_{H}$ between $H\isomorphic H'$ as follows. Take 
\[
\equivClass{q_{1}} I_{ST} \equivClass{q'_{1}}\mbox{ iff }\exists q\in\equivClass{q_{1}},q'\in\equivClass{q'_{1}}:q I_{H} q'.
\]
\cp{Finish proof for part 1!!}
\end{proof}

We can define the \emph{category} \categoryHDA\ to have objects \HDAs\ and morphisms defined as in Definition~\ref{def_isomorphismHDA}.

\begin{proposition}
The mapping \stintoh, from Definition~\ref{def_STtoHDA}, can be lifted to a functor between the categories \categoryST\ and \categoryHDA\ by defining its application to morphisms as follows: for $f:\ST\rightarrow\ST'$ have $\stintoh(f)=F$ with $F(q^{(S,T)})=q^{(f(S,T))}$ for any $q^{(S,T)}\in\stintoh(\ST)$.
\end{proposition}

\begin{proof}
Since the morphism $f$ preserves ST-configurations, we have $f(S,T)\in\ST'$ and thus $q^{(f(S,T))}\in\stintoh(\ST')$ making $F$ well defined as a function from the cells of $\stintoh(\ST)$ to $\stintoh(\ST')$. The rest of the proof that $F$ is a morphism of \HDAs\ goes as in the proof of Proposition~\ref{prop_stintoh_iso}(\ref{prop_stintoh_iso_1}).

It is easy to see that \stintoh\ preserves the identity morphisms.

We are left to show that \stintoh\ preserves composition; i.e., $\stintoh(f\circ g)=\stintoh(f)\circ\stintoh(g)$.
\cp{Finish!!}
\end{proof}

\begin{proposition}
The mapping \hintost, from Definition~\ref{def_hdaTOst}, can be lifted to a functor between the categories \categoryHDA\ and \categoryST\ by defining its application to morphisms as follows: for some morphism of \HDAs\ $F:H\rightarrow H'$ have $\hintost(F)=f$ with $f(\equivClass{q_{1}})=\equivClass{F(q_{1})}$ for any $\equivClass{q_{1}}$ an event generated by $\hintost(H)$.
\end{proposition}

\begin{proof}
From the statement is clear that $\hintost(F)$ is defined over all events of $\hintost(H)$. Since for any $q_{1}\in\hintost(H)$ the $F(q_{1})$ is reachable by some path $\pi'\in H'$ it means that the equivalence class $\equivClass{F(q_{1})}$ will be used in building some ST-configuration of $\hintost(H')$, in the case \ref{hintost_21} of Definition~\ref{def_hdaTOst}, and therefore $\equivClass{F(q_{1})}$ will be an event of $\hintost(H')$, thus making the codomain of $f$ to be $\hintost(H')$; i.e., $f:E_{\hintost(H)}\rightarrow E_{\hintost(H')}$.

To show that $f$ is well defined we show the following:

\vspace{1ex}
\noindent\textit{Claim:}\hspace{1ex} For two \HDAs\ related by a morphism $F:H\rightarrow H'$ and for any two cells $q_{1}^{a},q_{1}^{b}\in Q_{1}\in H$ we have 
\[
q_{1}^{a}\eventEquivHDAs q_{1}^{b} \mbox{ then } F(q_{1}^{a})\eventEquivHDAs F(q_{1}^{b}).
\]
\vspace{0.5ex}
Since $q_{1}^{a}\eventEquivHDAs q_{1}^{b}$ then it means that $\exists q_{2}\in Q_{2}:\alpha_{i}(q_{2})=q_{1}^{a}\wedge\beta_{i}(q_{2})=q_{1}^{b}$, with $\alpha,\beta\in\{s,t\}$ and $i\in\{1,2\}$. Since $F$ preserves the mappings we have $\alpha_{i}(F(q_{2}))=F(\alpha_{i}(q_{2}))$ and $\beta_{i}(F(q_{2}))=F(\beta_{i}(q_{2}))$. This means we have found $F(q_{2})$ that witnesses that $F(q_{1}^{a})\eventEquivHDAs F(q_{1}^{b})$.

To show that $f$ preserves ST-configuration consider an $(S,T)\in\hintost(H)$ which means that it was obtained from some path $\hintost(\pi)=(S,T)$. But since $F$ preserves the maps ti means that the path $\pi$ corresponds to a path $F(\pi)\in\hintost(H')$ with the same maps and indexes and where each cell on this path is the $F$-image of the corresponding cell on $\pi$. This means that if $\finishPath{\pi}=q_{1}$ then $\finishPath{F(\pi)}=F(q_{1})$ and that $F(s_{i}(\pi))=s_{i}(F(\pi))$. We show by induction on the length of the paths that $f(\hintost(\pi))=\hintost(F(\pi))$, therefore implying that it is an ST-configuration from $\hintost(H')$. The base case for the empty path is easy. We then take cases depending on the cell that the path ends in, according to Definition~\ref{def_hdaTOst}.
When $\finishPath{\pi}=q_{1}\in Q_{1}$ then case~\ref{hintost_21} applies, which means that $\hintost(\pi)=\hintost(s_{1}(\pi))\cup(\equivClass{q_{1}},\emptyset)$. When $f$ is applied to this ST-configuration we have $f(\hintost(\pi))=f(\hintost(s_{1}(\pi)))\cup(f(\equivClass{q_{1}}),\emptyset)$. By the induction hypothesis $f(\hintost(s_{1}(\pi)))=\hintost(F(s_{1}(\pi)))$, and by the definition of $f$ we have $(f(\equivClass{q_{1}}),\emptyset)=(\equivClass{F(q_{1})},\emptyset)$. These imply, by the same case~\ref{hintost_21} of Definition~\ref{def_hdaTOst}, the desired result that $f(\hintost(\pi))=\hintost(F(\pi))$.
When $\finishPath{\pi}=q_{0}\in Q_{0}$ then case~\ref{hintost_22} applies and a similar argument as before is used.
When $\finishPath{\pi}=q_{n}\in Q_{n}$ then case~\ref{hintost_3} applies, and thus $\hintost(\pi)=\hintost(\pi^{i})\cup\hintost(\pi^{j})$. By the induction hypothesis, since the twp paths are of shorter length, we have $f(\hintost(\pi^{i}))=\hintost(F(\pi^{i}))$ and $\finishPath{F(\pi^{i})}=F(\finishPath{\pi^{i}})$ which is $F(s_{i}(\finishPath{\pi}))$ which is equal to $s_{i}(F(\finishPath{\pi}))=s_{i}(\finishPath{F(\pi)})$; analogous $f(\hintost(\pi^{j}))=\hintost(F(\pi^{j}))$ with $\finishPath{F(\pi^{j})}=s_{j}(\finishPath{F(\pi)})$. Therefore, by the same case of Definition~\ref{def_hdaTOst} we have $f(\hintost(\pi))=\hintost(F(\pi^{i}))\cup\hintost(F(\pi^{j}))=\hintost(F(\pi))$.

To show that $f$ is locally injective we use induction on the length of the path, since the ST-configura\-tions are build in Definition~\ref{def_hdaTOst} from paths shorter with one step. We show that for any ST-configuration $(S,T)$ obtained from some path as $\hintost(\pi)$ and for any two equivalence classes $\equivClass{q_{1}^{a}}\neq \equivClass{q_{1}^{b}}$ we have $f(\equivClass{q_{1}^{a}})\neq f(\equivClass{q_{1}^{b}})$ which is the same as $\equivClass{F(q_{1}^{a})}\neq\equivClass{F(q_{1}^{b})}$. In other words, if for two transition cells that have been added by \hintost\ on some path $\pi$ which are not equivalent in the sense of $q_{1}^{a}\not\eventEquivHDAs q_{1}^{b}$ then their corresponding cells through $F$ are also not equivalent, $F(q_{1}^{a})\not\eventEquivHDAs F(q_{1}^{b})$.
When we treat the case~\ref{hintost_22} for when $\finishPath{\pi}\in Q_{0}$ it is trivial because it follows from the inductive hypothesis since we only add to the $T$ set of the ST-configuration.
The case~\ref{hintost_21} for when $\finishPath{\pi}=q_{1}\in Q_{1}$ assumes that $f\restrictedToSet{\hintost(\pi_{0})}$ is locally injective as $\pi_{0}\transition{}q_{1}=\pi$ and from the definition $\hintost(\pi)=\hintost(\pi_{0})\cup(\equivClass{q_{1}},\emptyset)$. Therefore we look only at the situation when $\equivClass{q_{1}}\not\in \hintost(\pi_{0})$.
The case~\ref{hintost_3} needs some more care and it goes through using the cubical law that the two shorter paths enter into.

It remains to show that \hintost\ preserves identity morphisms and respects composition of morphisms.
It is not difficult to show that $\hintost(Id_{H})=Id_{\hintost(H)}$, since $Id_{\hintost(H)}(\equivClass{q_{1}})\stackrel{def}{=}\equivClass{Id_{H}(q_{1})}=\equivClass{q_{1}}$.
We show that $\hintost(F_{1}\circ F_{2})(\equivClass{q_{1}})=\hintost(F_{1})\circ \hintost(F_{2})(\equivClass{q_{1}})$. We know that $\hintost(F_{1}\circ F_{2})(\equivClass{q_{1}})=\equivClass{F_{1}\circ F_{2}(q_{1})}$. On the other side, $\hintost(F_{1})\circ \hintost(F_{2})(\equivClass{q_{1}}) = \hintost(F_{1})(\equivClass{F_{2}(q_{1})}) = \equivClass{F_{1}(F_{2}(q_{1}))}$, which is the desired result.
\end{proof}

\begin{remark}[no adjoint]
For the two functors \stintoh\ and \hintost\ between the categories \categoryST\ and \categoryHDA\ we cannot find a unit to make \stintoh\ the left adjoint of \hintost\ because of the example of the ST-structure of Figure~\ref{fig_ex_asymconflict1}(middle). For this ST-structure there is no way to associate a morphism to its translation through $\hintost\circ\stintoh$, which is the ST-structure from Figure~\ref{fig_ex_asymconflict1}(right). There is also not possible to get the adjunction the other way, because of the example of Figure~\ref{fig_ex_interleaving_triangle}(right) showing unfolding of the triangle \HDA. For this \HDA\ there is no way to associate a morphism to its translation through $\stintoh\circ\hintost$, which is unfolded.
\end{remark}

\cp{What is the category of sculptures? And what is its relation with the category of \HDAs?
Sculptures are special morphisms (together with their respective objects) from the category of \HDAs. If we take such morphisms as objects, then what are the morphisms between sculptures?

Note that a bulk is a sculpture wrt.\ all other bulks of higher dimension. So there are many such morphism with the origin in a bulk. We are interested only in those sculpture morphisms which are \textit{simplistic}. What is their categorical characterization in the category \categoryHDA?
}

\section{Sculpting}\label{subsec_sculpting}

\begin{figure}[tp]
\psfrag{ee}{{\scriptsize  $(\emptyset,\emptyset)$}}
\psfrag{ae}{{\scriptsize $(\equivClass{q_{1}^{1}},\emptyset)$}}
\psfrag{be}{{\scriptsize $(b,\emptyset)$}}
\psfrag{aa}{{\scriptsize  $(\equivClass{q_{1}^{1}},\equivClass{q_{1}^{1}})$}}
\psfrag{bb}{{\scriptsize $(b,b)$}}
\psfrag{bab}{{\scriptsize $(ba,b)$}}
\psfrag{baa}{{\scriptsize $(b\equivClass{q_{1}^{1}},\equivClass{q_{1}^{1}})$}}
\psfrag{baba}{{\scriptsize $(ba,ba)$}}
\psfrag{bae}{{\scriptsize $(ba,\emptyset)$}}
\psfrag{q11}{\small $q_{0}^{1}$}
\psfrag{q12}{\small $q_{0}^{3}$}
\psfrag{q13}{\small $q_{0}^{4}$}
\psfrag{q15}{\small $q_{0}^{5}$}
\psfrag{q14}{\small $q_{0}^{2}$}
\psfrag{q21}{$q_{1}^{1}$}
\psfrag{q22}{$q_{1}^{2}$}
\psfrag{q23}{$q_{1}^{3}$}
\psfrag{q24}{$q_{1}^{4}$}
\psfrag{a}{\small $a$}
\psfrag{b}{\small $b$}
\psfrag{c}{\small $c$}
\psfrag{d}{\small $d$}
\psfrag{H1}{$H_{1}$}
\psfrag{H2}{$H_{2}$}
\psfrag{H3}{$H_{3}$}
\psfrag{H4}{$H_{4}$}
\psfrag{different}{$\equivClass{q_{1}^{1}}\neq\equivClass{q_{1}^{2}}\neq\equivClass{q_{1}^{3}}\neq\equivClass{q_{1}^{4}}$}
\psfrag{hh}{{\scriptsize $\hhequiv$}}
\psfrag{nhh}{{\scriptsize $\not\hhequiv$}}
\psfrag{iso}{{\scriptsize $\isomorphic$}}
\psfrag{niso}{{\scriptsize $\not\isomorphic$}}
\psfrag{ST}{{\scriptsize on STs}}
\psfrag{HDA}{{\scriptsize on \HDAs}}
\psfrag{STH1}{{\scriptsize $\hintost(\!H_{1}\!)\!\isomorphic\!\hintost(\!H_{2}\!)$}}
  \begin{center}
    \includegraphics[height=3.9cm]{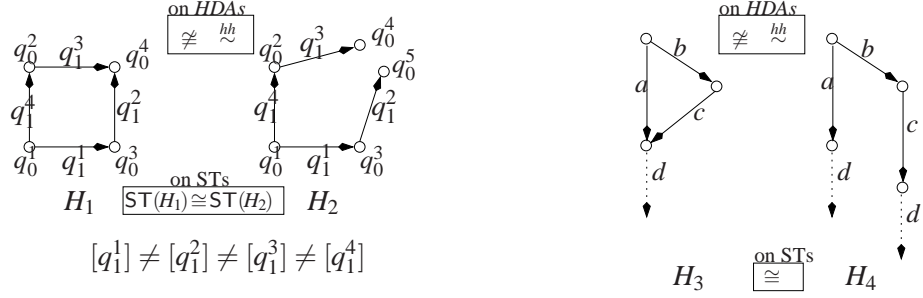}
  \end{center}
\caption{Unfoldings of \HDAs\ through \hintost.}
\label{fig_ex_interleaving_triangle}
\end{figure}

There are several issues with the above mappings that we want to address in this section using the method of \textit{sculpting}, which is much like what Pratt has used in \cite{Pratt96reconcilingevent,Pratt00HDArev}.
The mapping \hintost\ works like an unfolding since it works with paths; in fact it is more close to the \textit{history unfolding} of the \HDA\ that it manipulates (cf.\ Definition~\ref{def_unfolding_history}). This is obvious from the example in Figure~\ref{fig_ex_interleaving_triangle}(right) where the right structure is the unfolding of the left triangle-like \HDA. But history unfolding is hh-bisimilar to the original structure, so we could try to check if \hintost\ is good up to hh-bisimulation. The example of Figure~\ref{fig_ex_Glabbeek}, disregarding the two dotted transitions, also shows two hh-bisimilar \HDAs\ the left being the history-unfolding of the right one, and which are mapped into the same ST-structure. So we could try to show that for \HDAs\ that are not hh-bisimilar the \hintost\ would map them to not hh-bisimilar ST-structures. But this is dismissed by the example of Figure~\ref{fig_ex_Glabbeek}, this time considering also the dotted transitions. These two \HDAs\ are not hh-bisimilar, but they are mapped to isomorphic ST-structures, hence hh-bisimilar.

The above issues are related to the fact that it is not clear how to identify the events in a \HDA. The best example for this is the fact that the mapping \hintost\ destroys the interleaving square; which is exactly due to the fact that our method of identifying events in a \HDA\ by equivalent transitions opposite in a filled square fails for this \textit{unfilled} square; see Figure~\ref{fig_ex_interleaving_triangle}(left). This same issue about events is also the one that causes the problem for the other mapping \stintoh\ where we could not say in the \HDA\ that was generated whether this was representing two or three events.

These issues are solved through sculpting since this will allow us to identify the events in a \HDA\ in the same way as ST-structures work with events. We will see that ST-structures capture \HDAs\ which can be seen as \textit{sculptures}. At first the sculpting method seams orthogonal to the history-unfolding aspect because:

\begin{enumerate}
\item There are \HDAs\ which are sculptures but for which their history unfolding is not a sculpture. This is the example of van Glabbeek~\cite[Fig.11]{Glabbeek06HDA}, pictured here in Figure~\ref{fig_ex_Glabbeek}, with the cube with one face missing, and its strange looking unfolding where the corner is split into two.

\item There are \HDAs\ which are not sculptures, but for which their history unfolding is a sculpture. 
Consider the example from Figure~\ref{fig_ex_interleaving_triangle}(right) with the triangle where the end state is reached either through one event or through a sequence of two events.

\item There are \HDAs\ which are sculptures, and also their history-unfolding is a sculpture (of a different dimension though). 
Consider the example from Figure~\ref{fig_ex_interleaving_triangle}(left) of the interleaving square.

\item There are \HDAs\ which are not sculptures and also their history-unfoldings are not sculptures either. This is the example of the game of the angelic vs.\ demonic choice from Example~\ref{ex_game_angelic_vs_demonic}, on page~\pageref{ex_game_angelic_vs_demonic}, that depends on the speed of the players.
\end{enumerate}

\begin{definition}[bulks]\label{def_bulks}
We call a \HDA\ \emph{a bulk} iff there exists a unique cell of highest concurrency (i.e., $\exists q_{n}\in Q_{n}: |Q_{n}|=1\wedge|Q_{n+i}|=0$ for all $i\in\mathbb{N}^{+}$) and all other cells from $Q$ are just the faces of this cell $q_{n}$. A bulk is simply a single (possibly infinitely-dimensional) cube, and thus denoted $B_{n}$ or just $q_{n}$ when no confusion can appear.
We work with acyclic and non-degenerate bulks only.
\end{definition}

\begin{definition}[sculptures]\label{def_sculptures}
We say that a $H=(Q,\overline{s},\overline{t},\labelH,I)$ can be \emph{sculpted} from a bulk $B_{n}$ whenever there exists an embedding morphism, i.e., injective morphism, $\embedMorphism:H\rightarrow B_{n}$, which means that the cells of this \HDA\ are just a subset of the cells of the bulk and the rest of the notions are just restrictions to this subset. 
We call a \emph{sculpture}, a \HDA\ together with a bulk and an embedding morphism. Since a bulk can be equated with its dimension, we will denote a sculpture as $\sculpture{H}{n}=(H,B_{n},\embedMorphism)$.
\end{definition}

It is easy to see that if a \HDA\ can be sculpted from a bulk $B_{n}$ of dimension $n$ then it can also be sculpted from any bulk of dimension higher than $n$. This is because any bulk $B_{n}$ can be seen as a sculpture from a bulk $B_{m}$, with $n<m$, by taking the trivial embedding $\embedMorphism^{m}_{n}:B_{n}\rightarrow B_{m}$ which associates to $q_{n}$ any of the cells of $B_{m}$ of dimension $n$ and to all the other cells just the corresponding faces. We say that a sculpture $\sculpture{H}{n}=(H,B_{n},\embedMorphism)$ can be over-complicated to become the sculpture $\sculpture{H}{m}=(H,B_{m},\embedMorphism')$ of higher dimension by taking the $\embedMorphism'=\embedMorphism^{m}_{n}\circ\embedMorphism$.
Therefore, we are interested in the minimal bulks, when they exist. We call a sculpture \emph{simplistic} if it cannot be simplified, i.e., has a minimal bulk. A sculpture $\sculpture{H}{n}=(H,B_{n},\embedMorphism)$ can be \emph{simplified} when $\exists i\leq n: s_{i}(q_{n})=q_{n-1}$ and $\embedMorphism$ is also an embedding into the smaller bulk, i.e., $\embedMorphism:H\rightarrow q_{n-1}$.

\begin{definition}[isomorphism of sculptures]\label{def_iso_sculptures}
Two sculptures $\sculpture{H}{n}$ and $\sculpture{H'}{m}$ are isomorphic iff their respective simplistic versions have the same dimension and $H$ and $H'$ are isomorphic as \HDAs.
\end{definition}

We can give an equivalent definition without the notion of simplistic versions. We need to over-complicate one sculpture to have the same dimensions. After this we take the obvious isomorphism of the two bulks. If this isomorphism when restricted becomes an isomorphism of the two underlying \HDAs\ then the sculptures are considered isomorphic.

\begin{remark}
One \HDA\ can be seen as two different sculptures, e.g., from two different dimensional bulks, in both cases being a simplistic sculpture (i.e., it all depends on the embedding morphism). Then this \HDA\ object enters as source object of several sculpture morphisms (as seen in Figure~\ref{fig_ex_asymconflict1}). Because of this we cannot determine from a \HDA\ alone in which sculpture it enters.

Working with history unfoldings is not not particularly good either. The interleaving square from Figure~\ref{fig_ex_interleaving_triangle}(left) can be seen as a sculpture from 2, but its history unfolding can be seen as a sculpture from 3 or from 4; we cannot decide which.
\end{remark}

\begin{proposition}\label{prop_HgivesSculptures}
The mapping \stintoh\ from Definition~\ref{def_STtoHDA} generates \HDAs\ which can be seen as sculptures.
\end{proposition}

But the mapping \stintoh\ from Definition~\ref{def_STtoHDA} does not tell exactly which sculpture it generates, i.e., which is the dimension of the sculpture and the embedding.
The next definition shows how this can be done, thus also giving the proof for Proposition~\ref{prop_HgivesSculptures}.

\begin{definition}[\allST\ to sculptures]\label{def_stintosculptures}
Define the mapping \stintosculpture\ the same as in Definition~\ref{def_STtoHDA} to generate the \HDA\ $H$, only that it also returns a bulk and an embedding, thus a sculpture, as follows.

For $\ST=(E,ST,l)$ build the bulk $B_{n}$, with $n=|E|$, by adding one cell $b^{(E,\emptyset)}$ and all its faces are added as in Definition~\ref{def_STtoHDA} for the \HDA\ part using the same listing of the events as used to generate the \HDA. Each cell of the bulk corresponds to a pair of subsets of $E$, i.e., $b^{(S,T)}$ with $S,T\subseteq E$, corresponds to a possible ST-configuration. 
The embedding $\embedMorphism:H\rightarrow B_{n}$ is defined as $\embedMorphism(q^{(S,T)})=b^{(S,T)}$.
We obtain $\stintosculpture(\ST)=(H,B_{n},\embedMorphism)$.
\end{definition}

\begin{proposition}
The mapping \stintosculpture\ from Definition~\ref{def_stintosculptures} does not collapse non-isomorphic ST-structures; i.e., for $\ST\not\isomorphic\ST'$ then $\stintosculpture(\ST)\not\isomorphic\stintosculpture(\ST')$ in the sense of Definition~\ref{def_iso_sculptures}.
\end{proposition}

\begin{proof}[sketch]
We prove the contrapositive, i.e., if $\stintosculpture(\ST)\isomorphic\stintosculpture(\ST')$ then $\ST\isomorphic\ST'$. We show the existence of an isomorphism over the ST-structures, knowing the isomorphism over their translations as sculptures, making use of a fixed listing of events that the maps \stintosculpture\ and \stintoh\ work with.
The rest is tedious details.
\end{proof}

\begin{definition}[sculptures to \allST]\label{def_sculptures_to_ST}
Define a mapping $\sculpintost:\allHDA\rightarrow\allST$ only over sculptures, which for a sculpture $\sculpture{H}{n}=(H,B_{n},\embedMorphism)$ associates the ST-structure $\sculpintost(\sculpture{H}{n})$ as follows.
Take a linearly ordered set $E$ (of events) of cardinality as the dimension of the bulk cell $q_{n}$. The ST-configurations of $\sculpintost(\sculpture{H}{n})$ are obtained from the cells of $H$, i.e., $ST_{s}=\{\sculpintost(q)|q\in H\}$.
One ST-configuration $\sculpintost(q)$ is obtained as $\sculpintost(q)=(S,T)=\alphachain{k}(E,\emptyset)$ for $\alphachain{k}(q_{n})=\embedMorphism(q)$. The \emph{$\alpha$-chain} $\alphachain{k}$ is a sequence of s/t-maps applications (with correct indexes) $\alphachain{k}=\alpha_{i_{1}}\circ\alpha_{i_{2}}\circ\dots\circ\alpha_{i_{k}}$ where $\alpha_{i_{k}}$ is either an s-map or a t-map of some correct index $i_{k}$. The application of the $\alpha$-chain to a cell returns another cell. We abuse the notation in $\alphachain{k}(E,\emptyset)$ and apply an $\alpha$-chain to an ST-configuration w.r.t.\ a predefined listing $E$, which is defined as: $\alphachain{k}(S,T)=\alphachain{k-1}(\alpha_{i_{k}}(S,T))$ with 
\[
\alpha_{i_{k}}(S,T)=\begin{cases} (S\setminus e,T), & \mbox{if } \alpha=s\wedge E\!\!\downarrow_{S\setminus\!T}\!\![i_{k}]=e \\ 
(S,T\cup e), & \mbox{if } \alpha=t\wedge E\!\!\downarrow_{S\setminus\!T}\!\![i_{k}]=e \end{cases}
\]
thus returning an ST-configuration.

\end{definition}

Intuitively, for the bulk cell $q_{n}$ associate the ST-configuration $(E,\emptyset)$. For any other cell $q$ of the bulk take a descending chain of s/t maps that reach it from the bulk cell. Build a ST-configuration $(S_{q},T_{q})$ starting with $(E,\emptyset)$ by removing from $E$, if an $s$, respectively adding to the $\emptyset$, if a $t$, the corresponding event $e_{i}$, where at each level of descent a renumbering of the events is made by removing the $e_{i}$ from the list. In the beginning, i.e., at the highest level of the bulk cell and the ST-configuration $(E,\emptyset)$, the listing of events is in correspondence with the s/t mappings of the bulk cell, i.e., $e_{i}$ corresponds to $s_{i}/t_{i}$ (this is kept all throughout the descent).
The mapping \sculpintost\ keeps only those ST-configurations that correspond to cells from the sculpture.

Since a sculpture has finite concurrency then \sculpintost\ returns only finite ST-configurations, even if the bulk may be of infinite concurrency.

\cp{Prove the next!}
\begin{enumerate}
\item The main property to use to prove that the example of van Glabbeek cannot be a sculpture is: In a bulk, for a cell and its unique history, for any two paths in this history and two different states, each on one of these paths, then the two states have associated different sets of events.

\item If a \HDA\ is a sculpture, then the mapping \hintost\ associates to every cell exactly one ST-configuration; in particular, it associates to every state exactly one set of events.

\end{enumerate}

\begin{remark}
Since the ST-structures and Chu spaces over 3 are isomorphic, cf.~Proposition~\ref{prop_STstructChu3}, and since ST-structures capture sculptures \HDAs, which are a strict subset of the \HDAs, we can conclude that the Chu spaces over 3 is not enough to capture all \HDAs, not even the acyclic ones. This observation is supplementing the results of Pratt, and is motivating our investigation in Section~\ref{sec_STCstruct} where we ask how STC-structures and Chu spaces over 4 are capturing non-sculptures and acyclic \HDAs.
\end{remark}

When applied to a sculpture $\sculpture{H}{n}$ the mapping \hintost\ from Definition~\ref{def_hdaTOst} can use the extra information that the sculpture gives to determine the events in the \HDA\ correctly.

\begin{definition}[\hintost\ from sculptures]\label{def_hdaTOst_sculptures}
For sculptures $\sculpture{H}{n}$ extend the mapping \hintost\ from Definition~\ref{def_hdaTOst} such that instead of using as events the equivalence classes of $Q_{1}$ cells as defined in Definition~\ref{def_hdaTOst}, it uses the following equivalence classes (as coming from the bulk); denote this mapping \hintostScultures.

Define the relation $\eventEquivHDAsculpture\subseteq Q_{1}\times Q_{1}$ with $Q_{1}\in H$ which is from the sculpture $\sculpture{H}{n}=(H,B_{n},\embedMorphism)$, as 
\[
q_{1}\eventEquivHDAsculpture q_{1}' \mbox{\hspace{2ex} iff \hspace{2ex}} \exists q_{2}\in Q^{B}_{2}:\alpha_{i}(q_{2})=\embedMorphism(q_{1})\wedge\beta_{i}(q_{2})=\embedMorphism(q_{1}')
\]
for some $i\leq 2$ and $\alpha,\beta\in\{s,t\}$, and $Q_{2}^{B}\in B_{n}$ part of the bulk. The reflexive and transitive closure of this relation generates equivalence classes denoted as before $\equivClass{q_{1}}$ and representing the \emph{events}.
\end{definition}

After giving some results on \textit{$\alpha$-chains} we will be able to prove the following theorem.
This theorem intuitively says that if we use the bulk to determine correctly all the cells that are equivalent, i.e., determine the events correctly, then both ways of translating sculptures are correct, i.e., either by using the unfolding method, $\hintostScultures$, or by sculpting out from the bulk directly, \sculpintost.

\begin{theorem}\label{th_on_scultures}
For a sculpture $\sculpture{H}{n}=(H,B_{n},\embedMorphism)$ we have
\[
\hintostScultures(\sculpture{H}{n})\isomorphic \sculpintost(\sculpture{H}{n}).
\]
\end{theorem}

An $\alpha$-chain is a sequence of applications of s/t-maps $\alpha=\alpha_{i_{1}}\circ\dots\alpha_{i_{k}}$ of correct indexes, where one member of the chain can be identified with its position in the chain as $\alpha^{k}=\alpha_{i_{k}}$. This example of $\alpha$-chain has \emph{length} $k$. The $\alpha$-chains are thus meant to be applied to cells, and depending on which cell these are applied to the indexes $i_{k}$ must be correctly bounded wrt.\ the dimension of the cell. To an $\alpha$-chain it is natural to associate the list of its indexes as $[i_{1},\dots,i_{k}]_{\alpha}$. We define an equivalence relation on $\alpha$-chains that is motivated by the equivalence relation in a bulk on the transition cells, as we gave in Definition~\ref{def_hdaTOst_sculptures}, and also motivated by the cubical laws which apply to such $\alpha$-chains and determine that different chains applied to the same cell give the same resulting cell. We denote this relation on $\alpha$-chains as \chainEquivHDAsculpture.

\begin{definition}
Define \chainEquivHDAsculpture\ on $\alpha$-chains to be reflexive and transitive and respecting the following:
\begin{itemize}
\item[] $\alpha\chainEquivHDAsculpture\alpha'$ iff both have the same length and either
\begin{enumerate}
\item their list of indexes is the same, i.e., $[i_{1},\dots,i_{k}]_{\alpha}=[i'_{1},\dots,i'_{k}]_{\alpha'}$; or
%

\item they are different by two consecutive maps which satisfy a cubical law: i.e., for some $l$ have $\alpha^{l}\circ\alpha^{l+1}=\alpha_{i}\circ\beta_{j}=\beta_{j-1}\circ\alpha_{i}=\alpha'^{l}\circ\alpha'^{l+1}$.
\end{enumerate}
\end{itemize}
\end{definition}

\begin{definition}
Inside a bulk $B_{n}$ define a relation on any two cells $q\cellEquivBulk q'$ in the following coinductive manner:
\begin{enumerate}
\item $\forall q\in B_{n}:q\cellEquivBulk q$;

\item $q_{k}\cellEquivBulk q'_{k}$ iff $q_{k},q'_{k}\in Q_{k}$ for some $1\leq k\leq n$ (i.e., have the same dimension) and $\exists q_{k+1},q'_{k+1}\in Q_{k+1}:q_{k+1}\cellEquivBulk q'_{k+1}\wedge \alpha_{i}(q_{k+1})=q_{k}\wedge \beta_{i}(q'_{k+1})=q'_{k}$ for some $i\leq k+1$.
\end{enumerate}
\end{definition}

\begin{lemma}
Inside a bulk $B_{n}$ if two transition cells are $q_{1}\cellEquivBulk q'_{1}$ then the cells are also $q_{1}\eventEquivHDAsculpture q'_{1}$.
\end{lemma}

\cp{Give Proof!!}

\begin{lemma}\label{lemma_chains_equiv}
Inside a bulk $B_{n}$ we have the following double implication:
\[
q_{k}\cellEquivBulk q'_{k} \ \ \Leftrightarrow\ \ \exists q_{l+k},\exists\alpha,\alpha':\alpha\chainEquivHDAsculpture\alpha'\wedge\alpha(q_{l+k})=q_{k}\wedge\alpha'(q_{l+k})=q'_{k}
\]
\end{lemma}

\begin{proof}
We assume $q_{k}\neq q'_{k}$ for otherwise it is trivial.

We prove the \textit{left-to-right} implication.
From the definition we have that $q_{k}\cellEquivBulk q'_{k}$ iff 
$\exists q_{k+1},q'_{k+1}\in Q_{k+1}:q_{k+1}\cellEquivBulk q'_{k+1}\wedge \alpha_{i}^{1}(q_{k+1})=q_{k}\wedge \beta_{i}^{1}(q'_{k+1})=q'_{k}$.
If $q_{k+1}=q'_{k+1}$ are the same then we have found the requirements for the right-side of the implication of the claim, i.e., take $q_{l+k}=q_{k+1}$ and $\alpha=\alpha_{i}^{1}$, $\alpha'=\beta_{i}^{1}$, i.e., of length $1$, and we have the desired $\alpha\chainEquivHDAsculpture\alpha'$ since their list of indexes is the same.
Otherwise, when $q_{k+1}\neq q'_{k+1}$, we apply the definition again to obtain that $q_{k}\cellEquivBulk q'_{k}$ iff  
$\exists q_{k+1},q'_{k+1}\in Q_{k+1}:\alpha_{i}^{1}(q_{k+1})=q_{k}\wedge \beta_{i}^{1}(q'_{k+1})=q'_{k}$ and 
$\exists q_{k+2},q'_{k+2}\in Q_{k+2}:q_{k+2}\cellEquivBulk q'_{k+2}\wedge \alpha_{j}^{2}(q_{k+2})=q_{k+1}\wedge \beta_{j}^{2}(q'_{k+2})=q'_{k+1}$.
Again, if the two cells are the same, $q_{k+2}=q'_{k+2}$ then we can stop the recursive reasoning and exhibit the required elements for the right-side of the implication; i.e., $q_{n}=q_{k+2}$ and $\alpha=\alpha_{i}^{1}\circ\alpha_{j}^{2}$, $\alpha'=\beta_{i}^{1}\circ\beta_{j}^{2}$ of length $2$ being equivalent $\alpha\chainEquivHDAsculpture\alpha'$ since their list of indexes is the same.
This recursive reasoning always eventually stops in the unique cell $q_{n}$ of the bulk.

We prove the \textit{right-to-left} implication.
The two $\alpha$-chains being equivalent have the same length: $\alpha_{i_{1}}\circ\dots\circ\alpha_{i_{l}}=\alpha\chainEquivHDAsculpture\alpha'=\alpha'_{j_{1}}\circ\dots\circ\alpha'_{j_{l}}$. From this, two cases are distinguished: one when the lists of indexes are the same; and another when there is a sequence of $\alpha$-chains each pair in the sequence having chains different by a cubical law.

(1) Having $\alpha_{i_{1}}\circ\dots\circ\alpha_{i_{l}}(q_{l+k})=q_{k}$ and $\alpha'_{j_{1}}\circ\dots\circ\alpha'_{j_{l}}(q_{l+k})=q'_{k}$ using the definition for \cellEquivBulk\ for the base case (i.e., when we work with the same cell) we have $\alpha_{i_{l}}(q_{l+k})\cellEquivBulk\alpha'_{j_{l}}(q_{l+k})$. We then use the definition with the standard case and obtain after applying it $l-1$ times the expected $q_{k}\cellEquivBulk q'_{k}$.

(2) There exists a sequence of $\alpha$-chains $\alpha\chainEquivHDAsculpture\alpha^{0}\chainEquivHDAsculpture\dots\chainEquivHDAsculpture\alpha^{m}\chainEquivHDAsculpture\alpha'$ each differing from the other by a cubical law instance; in particular, for $\alpha=\alpha_{i_{1}}\circ\dots\circ\alpha_{i_{l}}$ and $\alpha^{0}=\alpha_{j_{1}}^{0}\circ\dots\circ\alpha_{j_{l}}^{0}$ there exist two indexes $i_{z},i_{z+1}$ s.t.\ for all other indexes we have $i_{y}=j_{y}$ for $y\neq z,z+1$, and $\alpha_{i_{z}}\circ\alpha_{i_{z+1}}=\alpha_{j_{z}}\circ\alpha_{j_{z+1}}$ as a cubical law. Since it does not matter for the equivalence which exact maps the $\alpha$-s are, we assume all these to be the same s-map.
This means that $\alpha_{i_{z+2}}\circ\dots\circ\alpha_{i_{l}}(q_{l+k})=\alpha_{j_{z+2}}^{0}\circ\dots\circ\alpha_{j_{l}}^{0}(q_{l+k})=q_{z+2}$ which by the cubical law means that $\alpha_{i_{z}}\circ\alpha_{i_{z+1}}(q_{z+2})=\alpha_{j_{z}}\circ\alpha_{j_{z+1}}(q_{z+2})=q_{z}$ to which we apply the rest of the maps to obtain the same cell. This is done for all the $\alpha$-chains in the sequence, obtaining $\alpha(q_{l+k})=\alpha'(q_{l=k})$ and by reflexivity we have the result.
\end{proof}

From Lemma~\ref{lemma_chains_equiv} we obtain the following.

\begin{corollary}
Any two chains which are equivalent, $\alpha\chainEquivHDAsculpture\alpha'$, when applied to the bulk cell $q_{n}$ result in equivalent cells. 
\end{corollary}

In particular, for a bulk of dimension $n$, any two $\alpha$-chains of dimension $n-1$ that are equivalent reach the same equivalence class of $Q_{1}$ transition cells; and also all $\alpha$-chains that reach such an equivalence class are equivalent. Therefore, there is a one-to-one correlation of the equivalence classes of transition cells (i.e., the events of the \hintostScultures) and the equivalence classes of $\alpha$-chains of dimension $n-1$.
The question is how many equivalence classes of $\alpha$-chains of dimension $n-1$ are? For our purposes these should be as many as the number of events of \sculpintost.

\begin{lemma}\label{lemma_on_scultures_equalEventSets}
For a listing $E$ of cardinality $n$ and two $\alpha$-chains $\alpha,\alpha'$ of length less than $n$ we have
\[
\alpha\chainEquivHDAsculpture\alpha' \mbox{\ \ iff\ \ }\applyChainList{\alpha}{E}=\applyChainList{\alpha'}{E}.
\]
\end{lemma}

\begin{proof}
The notation $\applyChainList{\alpha}{E}$ is close to what we used in Definition~\ref{def_sculptures_to_ST} but here we are only interested in the lists of events obtained by removing from the initial list $E$ the events at the indexes corresponding to the chain steps. More precisely, for $\alpha=\alpha_{i_{1}}\circ\dots\circ\alpha_{i_{k}}$ of length $k<n$, define $\applyChainList{\alpha}{E}$ as $\applyChainList{\alpha_{i_{1}}\circ\dots\circ\alpha_{i_{k}}}{E}\defequal\applyChainList{\alpha_{i_{1}}\circ\dots\circ\alpha_{i_{k-1}}}{\applyChainList{\alpha_{i_{k}}}{E}}$ where for one element of the chain the application is defined as $\applyChainList{\alpha_{i_{k}}}{E}\defequal [E\setminus i_{k}]$ to be the list of length one less than $E$ which is the same as $E$ but with the element on index $i_{k}$ removed (in other notation: $E\!\downarrow_{(E\setminus E[i_{k}])}$).

We prove the \textit{left-to-right} implication.
Having $\alpha\chainEquivHDAsculpture\alpha'$ we assume there is a sequence of $\alpha$-chains each different than the other by a cubical law, i.e., $\alpha\chainEquivHDAsculpture\alpha^{0}\chainEquivHDAsculpture\dots\chainEquivHDAsculpture\alpha^{m}\chainEquivHDAsculpture\alpha'$ with the indexes of $\alpha$ the same as with those of $\alpha^{0}$ with the exception of two consecutive ones which enter a cubical law.
(The case when the indexes are the same is trivial.)

Take the first equivalence in the chain $\alpha\chainEquivHDAsculpture\alpha^{0}$, i.e., $\alpha_{i_{1}}\circ\dots\circ\alpha_{i_{k}}\chainEquivHDAsculpture\alpha_{j_{1}}^{0}\circ\dots\circ\alpha_{j_{k}}^{0}$ with two consecutive indexes for $l,l+1$, with $1\leq l<k$, s.t.\ $i_{m}=j_{m}$ for $m\neq l,l+1$ and $1\leq m\leq k$. Because of this it means that these applied to some list $F$ return the same list, i.e., $\applyChainList{\alpha_{i_{m}}}{F}=\applyChainList{\alpha_{j_{m}}}{F}$, and therefore we have $\applyChainList{\alpha_{i_{l+2}}\circ\dots\alpha_{i_{k}}}{E}=\applyChainList{\alpha_{j_{l+2}}^{0}\circ\dots\alpha_{j_{k}}^{0}}{E}=E_{l+2}$.

Now for the $\alpha_{i_{l}}\circ\alpha_{i_{l+1}}$ and $\alpha_{j_{l}}^{0}\circ\alpha_{j_{l+1}}^{0}$ we know that they enter a cubical law, which means that $i_{l}=o<p=i_{l+1}$ and $j_{l}=p-1,j_{l+1}=o$ (the case for when $o\geq p$ has the same argument). It is easy to see that for a list $E_{l+2}$ when removing first an element on a position $p$ higher than a position $o$ which we remove afterwards, it is the same as first removing the element from this lower position $o$ and then removing the element on the position one-lower than $p$, since all the before higher positions than $o$ have been decreased by one after the first removal; i.e., $[[E_{l+2}\setminus p]\setminus o]=[[E_{l+2}\setminus o]\setminus p-1]$. Therefore, $\applyChainList{\alpha_{i_{l}}\circ\alpha_{i_{l+1}}}{E_{l+2}}=\applyChainList{\alpha_{j_{l}}^{0}\circ\alpha_{j_{l+1}}^{0}}{E_{l+2}}=E_{l}$. 
To this last one when we apply the remainder of the two $\alpha$-chains that have the same indexes we obtain the same list, thus the desired result.

We prove the \textit{right-to-left} implication.
Given two $\alpha$-chains of some dimension $k<n$ their application to the list $E$ is defined as before and results in the same list $\applyChainList{\alpha}{E}=\applyChainList{\alpha'}{E}=E_{n-k}$. Knowing that from an $n$-dimensional list $E$ we obtain the list $E_{n-k}$ of dimension $k$ lower we can view the two $\alpha$-chains and how they are applied to the initial list as a table. Take the rows of the table to correspond to the indexes of the initial list $E$ and take the columns to stand for the element in the $\alpha$-chain; i.e., we have a $n\times k$ matrix. Each application of one element of the $\alpha$-chain removes one different element from the list. The matrix represents the operation of removing the elements by the particular $\alpha$-chain by putting a $1$ value on the position where the element of the $\alpha$-chain is removing the respective element of the list; the rest of the matrix is filled with $0$. In consequence we have a single $1$  on each column (since a chain element is applied only once, and all chain elements remove some element of the list) and on each row (since an element is removed only once) that corresponds to a removed event; all rows corresponding to the events in the remaining list $E_{n-k}$ are completely filled with $0$. 

Each $\alpha$-chain that is applied to the list $E$ and returns the list $E_{n-k}$ is represented by one such matrix with the rows corresponding to the events of $E_{n-k}$ completely filled with $0$. If we remove these empty rows we are left with a square matrix of $n-k\times n-k$ which is a \textit{permutation matrix}.

We want to show that $\alpha\chainEquivHDAsculpture\alpha'$. If their corresponding matrices are the same it means that they have the same indexes and the result is trivial.
Otherwise we show that there is a sequence of $\alpha$-chains each different than the previous by a cubical law. Let us understand what it means a cubical law difference in terms of the corresponding matrices. All the indexes are the same with the exception of two adjacent ones. This means that the matrices are the same with the exception of two columns that correspond to the two indexes. The cubical law on these indexes then means the swapping of the two rows of the matrix that have the value $1$ on the two corresponding columns.
See this through the example in Figure~\ref{fig_ex_matrix}.


\begin{figure}[tp]
\begin{tabular}{lll}
$$
\bordermatrix{ \alpha= & \overset{1}{\overset{=}{i_1}} & \overset{2}{\overset{=}{i_2}} & \overset{4}{\overset{=}{i_{3}}} & \overset{\mathbf{5}}{\overset{=}{i_{4}}} & \overset{\mathbf{2}}{\overset{=}{i_5}}  \cr
                E_{6} & 0 & 0 & 0 & 1 & 0\cr
                E_{5} & 0 & 0 & 1 & 0 & 0\cr
                \mathbf{E_{4}} & 0 & 0 & 0 & 0 & 0\cr
                E_{3} & 0 & 1 & 0 & 0 & 0\cr
                E_{2} & 0 & 0 & 0 & 0 & 1\cr
                E_{1} & 1 & 0 & 0 & 0 & 0
}
$$
&
$\transition{2\circ 6=5\circ 2}$
&
$$
\bordermatrix{ \alpha'= & \overset{1}{\overset{=}{i_1}} & \overset{2}{\overset{=}{i_2}} & \overset{4}{\overset{=}{i_{3}}} & \overset{\mathbf{2}}{\overset{=}{i_{4}}} & \overset{\mathbf{6}}{\overset{=}{i_5}}  \cr
                E_{6} & 0 & 0 & 0 & 0 & \mathbf{1}\cr
                E_{5} & 0 & 0 & 1 & 0 & 0\cr
                \mathbf{E_{4}} & 0 & 0 & 0 & 0 & 0\cr
                E_{3} & 0 & 1 & 0 & 0 & 0\cr
                E_{2} & 0 & 0 & 0 & \mathbf{1} & 0\cr
                E_{1} & 1 & 0 & 0 & 0 & 0
}
$$\\
%
%
%
%
\end{tabular} 

\caption{}
\label{fig_ex_matrix}
\end{figure}

This cubical law change corresponds on the matrices to a \textit{adjacent transposition} on the corresponding permutations. In other words, if two $\alpha$-chains are different by a cubical law then their matrices are different in a adjacent transposition of their corresponding permutation; and each adjacent transposition corresponds to a cubical law change.

Take now the tables of the two given $\alpha$-chains $\alpha,\alpha'$. Since these $\alpha$-chains return the same list then we are in the situation above where we associate to each $\alpha$-chain a square matrix that corresponds to one permutation. It is a known result that any two permutations can be obtained one from the other by a sequence of adjacent transpositions. This means that there is a sequence of matrices that differ only in a cubical law manner, hence the equivalence of the initial $\alpha$-chains.
\end{proof}

\begin{proof}[of Theorem~\ref{th_on_scultures}]
We must exhibit a bijective map $f$ between the events generated by \sculpintost\ and those generated by \hintostScultures, and show that it respects Definition~\ref{def_isomorphism} of being a morphism. 

The set of events generated by \sculpintost\ is $E$, having the cardinality the same as the dimension of the bulk $B_{n}$, i.e., $|E|=n$, and being linearly ordered. The linear order coincides with the indexes of the mappings from the bulk cell $q_{n}$. 

The map \hintostScultures\ generates events which are the equivalence classes given by the relation \eventEquivHDAsculpture\ from Definition~\ref{def_hdaTOst_sculptures} which comes from the bulk $B_{n}$.
%
%

Intuitively, we define $f$ to associate to each event in $E$, which can be seen as obtained from an application of an $\alpha$-chain, i.e., $f(\alpha_{n-1}(E))$, an equivalence class which is obtained using the same $\alpha$-chain from the bulk cell $\equivClass{\alpha_{n-1}(q_{n})}$. Note that the length of the $\alpha$-chain is related to the cardinality of $E$.
The Lemma~\ref{lemma_on_scultures_equalEventSets} ensures that the definition of the function on events as $f(\alpha_{n-1}(E))=\equivClass{\alpha_{n-1}(q_{n})}$ is a bijection.

We show that $f$ preserves ST-configurations, i.e., for $(S,T)\in\sculpintost$ then $f(S,T)\in\hintostScultures$.
By Definition~\ref{def_sculptures_to_ST}, $(S,T)\in\sculpintost$ means that $\exists q\in H$ (with $H$ from the sculpture $\sculpture{H}{n}=(H,B_{n},\embedMorphism)$) reachable in the bulk from the bulk cell by some $\alpha$-chain, $\embedMorphism(q)=\alpha_{k}(q_{n})$, which this $\alpha$-chain determines the ST-configuration, $(S,T)=\alpha_{k}(E,\emptyset)$.
Therefore, we want to show that $f(\alpha_{k}(E,\emptyset))\in\hintostScultures$, which amounts to showing that $\exists\pi:\hintostScultures(\pi)=f(\alpha_{k}(E,\emptyset))$.

We show the following claim, which implies the above expected result.

\vspace{1ex}
\noindent\textit{Claim:}\hspace{1ex} For any $\pi$ with $\finishPath{\pi}=q$ and any $\alpha$-chain with $\alpha(q_{n})=q$ then 
\[
f(\alpha(E,\emptyset))=\hintostScultures(\pi).
\]
\vspace{0.5ex}

There are three cases to consider depending on $q$, which correspond to the three ways of generating ST-configurations by \hintostScultures\ in Definition~\ref{def_hdaTOst}.

Note that the order on the events in $E$ that \sculpintost\ is considering in Definition~\ref{def_sculptures_to_ST} is the same as the indexes of the maps of the $q_{n}$ bulk cell. This is the same as the order of the events the \hintostScultures\ generates since these are the equivalence classes on the $Q_{1}$ cells of the bulk, which are the same as the equivalence classes of $\alpha$-chains, cf.\ the above results on the correspondence of equivalence of $\alpha$-chains and events. An equivalence class of $\alpha$-chains $\equivClass{\alpha_{n-1}}$ is on position $i_{k}$ in the listing of the events generated by \hintostScultures\ iff no $\alpha$-chain in this equivalence ends in index $i_{k}$.

Since any $\alpha$-chain in an equivalence class generates equivalent cells (when applied to the bulk cell) all these cells will be associated to ST-configurations exhibiting the same set of concurrent events. In particular, any two $\alpha$-chains from an equivalence class that have the same number of s- and t-maps but differ only through cubical laws interchange of their indexes, reach the same cell. All such $\alpha$-chains when applied to the $(E,\emptyset)$ generate the same ST-configuration. Therefore, we are free to work with any such $\alpha$-chain (not necessarily with the one from the claim); in particular the $\alpha$-chains that are of the form $\alpha^{s}\circ\alpha^{t}$ are of interest, where $\alpha^{s}$ and $\alpha^{t}$ are, possibly empty, sequences of only s- respectively t-maps.

In a bulk every cell has only one history. 
Nevertheless, in the sculpture a cell may have several histories.
In a history all paths have the same length.
For a cell in the sculpture and for any of its histories, then all the paths in this history are included in the paths of the uniques history corresponding to this cell in the bulk.
The mapping \hintostScultures\ works with the paths in the sculpture, and two such paths there may be from two different histories of the same cell. Nevertheless, in the bulk both paths are homotopic, therefore they will involve the same equivalence classes of $Q_{1}$ cells, as given by \cellEquivBulk. We are thus free to work in this proof with any path leading to a cell, which may very well not be from the sculpture since the same ST-configuration would be generated. This is not the case for a non-sculpture and the mapping \hintost; all the above are due to the fact that \hintostScultures\ works inside a bulk.
In particular we are interested in paths of the form $\pi^{s}\pi^{t}$ where $\pi^{t}$ may be empty, but not $\pi^{s}$.

With the definition of events that \hintostScultures\ uses, if two transitions denote the same event in the sculpture, then they denote the same event in the bulk also. 
The bulk may equate more transitions as the same event; in other words, two events which are considered different in the sculpture may be collapsed in the bulk.

We use induction on the length of the path that reaches the cell that is used to build the current ST-configuration.
The base case is for the initial cell which \sculpintost\ translates into $(\emptyset,\emptyset)$; to which when $f$ is applied results in the same ST-configuration which is also one of the configurations generated by \hintostScultures. It is easy to see this because in $H$ the initial cell is reachable only through $\alpha$-chains of length $n$ and of only s-maps. The application of such a chain to $(E,\emptyset)$ results in the root ST-configuration $(\emptyset,\emptyset)$.

Recall that $f(S,T)=(f(S),f(T))=(\{f(e)\mid e\in S\},\{f(e)\mid e\in T\})$ and that we work with $\alpha$-chains in the form $\alpha^{s}\circ\alpha^{t}$. For the case when we look at a cell of dimension $k\geq 2$ then we work with an $\alpha$-chain of at most dimension $n-2$ (with $n$ the dimension of the bulk). So, for the cell $q=\alpha^{s}\circ\alpha^{t}(q_{n})$ we look at $f(\alpha^{s}\circ\alpha^{t}(E,\emptyset))$. Taking the application of $\alpha$-chains from Definition~\ref{def_sculptures_to_ST} we have that $\alpha^{s}\circ\alpha^{t}(E,\emptyset)=(S,T)$ with $T=E\setminus\applyChainList{\alpha^{t}}{E}$ and $S=T\cup\applyChainList{\alpha^{s}\circ\alpha^{t}}{E}$; we thus have $f(\alpha^{s}\circ\alpha^{t}(E,\emptyset))=(f(E\setminus\applyChainList{\alpha^{t}}{E})\cup f(\applyChainList{\alpha^{s}\circ\alpha^{t}}{E}),f(E\setminus\applyChainList{\alpha^{t}}{E}))$ where we denote the two sets involved as $(T\cup B,T)$. Note that we have used the application of the $\alpha$-chains to lists, as we defined earlier.

Since $q$ has at least two concurrent events, then it also has at least two s-maps; take any two different $s_{1},s_{2}$. The two cells reachable through these two maps are closer to the initial cell and are reachable through paths of length one shorter than the path reaching $q$. Take the two non-equivalent chains that reach these cells to be $q^{1}=s_{1}\circ\alpha^{s}\circ\alpha^{t}(q_{n})$ and $q^{2}=s_{2}\circ\alpha^{s}\circ\alpha^{t}(q_{n})$. The ST-configurations that \sculpintost\ associates are $\sculpintost(q^{1})=s_{1}\circ\alpha^{s}\circ\alpha^{t}(E,\emptyset)$ and $\sculpintost(q^{2})=s_{2}\circ\alpha^{s}\circ\alpha^{t}(E,\emptyset)$ and the function $f$ is applied to these in a previous induction step. The application results in: $f(s_{1}\circ\alpha^{s}\circ\alpha^{t}(E,\emptyset))=(T\cup f(\applyChainList{s_{1}\circ\alpha^{s}\circ\alpha^{t}}{E}),T)$ and $f(s_{2}\circ\alpha^{s}\circ\alpha^{t}(E,\emptyset))=(T\cup f(\applyChainList{s_{2}\circ\alpha^{s}\circ\alpha^{t}}{E}),T)$; denote the two new sets as $B_{1}$ and $B_{2}$.

But $B_{1}$ is the same as $B$ but with one event missing, i.e., the one removed by the $s_{1}$ map; call this event $e_{1}$ and $B_{1}=B\setminus e_{1}$. The same for $B_{2}=B\setminus e_{2}$ where $e_{1}\neq e_{2}$ but with $e_{1}\in B_{2}$ and $e_{2}\in B_{1}$. This means that $B_{1}\cup B_{2}=B$. Therefore, $(T\cup f(\applyChainList{s_{1}\circ\alpha^{s}\circ\alpha^{t}}{E}),T)\cup(T\cup f(\applyChainList{s_{2}\circ\alpha^{s}\circ\alpha^{t}}{E}),T)=(T\cup B,T)$; i.e., $f(s_{1}\circ\alpha^{s}\circ\alpha^{t}(E,\emptyset))\cup f(s_{2}\circ\alpha^{s}\circ\alpha^{t}(E,\emptyset))=(f(\applyChainList{\alpha^{t}}{E})\cup f(\applyChainList{\alpha^{s}\circ\alpha^{t}}{E}),f(E\setminus\applyChainList{\alpha^{t}}{E}))=f(\alpha^{s}\circ\alpha^{t}(E,\emptyset))$.

By the induction hypothesis we have $f(s_{1}\circ\alpha^{s}\circ\alpha^{t}(E,\emptyset))=\sculpintost(\pi_{1})$, for any $\pi_{1}$ reaching $q_{1}=s_{1}(q)$, and $f(s_{2}\circ\alpha^{s}\circ\alpha^{t}(E,\emptyset))=\sculpintost(\pi_{2})$ for $\pi_{2}$ reaching $q_{2}=s_{2}(q)$. Since $s_{1},s_{2}$ were two arbitrary s-maps of $q$, we have the desired result $f(\alpha^{s}\circ\alpha^{t}(E,\emptyset))=\sculpintost(\pi)$ for any $\pi$ reaching $q$.

The case for cells $q_{1}\in Q_{1}$ works with $\alpha$-chains of length $n-1$.
There is a base case for the cells $q_{1}$ reachable in one step from the initial cells; these are also reachable only through $\alpha$-chains of only s-maps, denoted $\alpha_{n-1}^{s}(q_{n})=q_{1}$. Then $f(\alpha_{n-1}^{s}(E,\emptyset))=f((\applyChainList{\alpha_{n-1}^{s}}{E},\emptyset))=(f(\applyChainList{\alpha_{n-1}^{s}}{E}),\emptyset)$ which by the definition of $f$ is $(\equivClass{\alpha_{n-1}^{s}(q_{n})},\emptyset)$.
\end{proof}

\cp{
Can I give results relating the sculptures with configuration deterministic and configuration unique and configuration preserving properties on \HDAs\ as given by van Glabbeek?

Looks like the following:
\begin{enumerate}
\item All sculptures are configuration preserving.
\item All sculptures are configuration deterministic.
Are there \HDAs\ that are configuration deterministic, or configuration preserving but are not sculptures?
\item The example of the interleaving square as a \HDA\ is not configuration unique. But if we see it as a sculpture and take events as equivalence classes from the bulk then it becomes configuration unique.
\end{enumerate}

}

The above proof indicates the following definition and result.

\begin{definition}
Define $\equiatingEvents{\sim}$ to take an ST-structure $\ST=(E,ST,l)$ and an equivalence relation on its events $\sim\subseteq E\times E$, and return the ST-structure $\ST'=(E',ST',l')$ with $E'=\{\equivClass[\sim]{e}\mid e\in E\}$, $ST'=\{\quotientofwrt{(S,T)}{\sim} \mid (S,T)\in ST\}$ where $\quotientofwrt{(S,T)}{\sim}=(\quotientofwrt{S}{\sim},\quotientofwrt{T}{\sim})$ is the quotient of $(S,T)$ wrt.\ $\sim$, and $l'(\equivClass[\sim]{e})=l(e')$ for some $e'\in\equivClass[\sim]{e}$.

For a sculpture $\sculpture{H}{n}=(H,B_{n},\embedMorphism)$ the function $\equiatingEvents{\eventEquivFromBulk{n}}$ applies the equivalence relation \eventEquivFromBulk{n} which is defined over equivalence classes $\equivClass[\eventEquivHDAs]{q_{1}}$, coming from the Definition~\ref{def_hdaTOst}, as: 
\[
\equivClass[\eventEquivHDAs]{q_{1}} \eventEquivFromBulk{n} \equivClass[\eventEquivHDAs]{q'_{1}} \mbox{\ iff \ }
q_{1} \eventEquivHDAsculpture  q'_{1}
\]
with \eventEquivHDAsculpture\ defined for \sculpture{H}{n} as in Definition~\ref{def_hdaTOst_sculptures}.
\end{definition}

\begin{proposition}\label{prop_eqcircst}
For a sculpture $\sculpture{H}{n}=(H,B_{n},\embedMorphism)$ we have
\[
\equiatingEvents{\eventEquivFromBulk{n}}\circ\hintost(H)\isomorphic \hintostScultures(\sculpture{H}{n})
\]
\end{proposition}

\begin{proof}
Note that in the example of Figure~\ref{fig_ex_asymconflict1} the two \HDAs\ are isomorphic but their sculpture versions are not. Because it preserves isomorphism, the \hintost\ will map the left \HDA\ into the ST-structure on the right over three events, instead of the desired left ST-structure only over two events. But when seen as sculptures, even if \hintost\ maps to isomorphic ST-structures, the equating of the events that $\equiatingEvents{\eventEquivFromBulk{n}}$ does will transform the left ST-structure into the corresponding two events ST-structure we are expecting, thus breaking the artificial isomorphism induced by \hintost.

\cp{Finish proof!!}
\end{proof}

The following results show a one-to-one correspondence between the ST-structures and sculptures.

\begin{proposition}\label{prop_stSculptst}
For an arbitrary ST-structure \ST\ we have
\[
\sculpintost(\stintosculpture(\ST))\isomorphic \ST
\]
\end{proposition}

\begin{proposition}
For a sculpture \sculpture{H}{n} we have
\[
\stintosculpture(\sculpintost(\sculpture{H}{n}))\isomorphic \sculpture{H}{n}
\]
\end{proposition}

\begin{corollary}
For an arbitrary ST-structure \ST\ we have
\[
\equiatingEvents{\eventEquivFromBulk{n}}\circ\hintost(\stintosculpture(\ST))\isomorphic \ST
\]
\end{corollary}

\begin{proof}
Proposition~\ref{prop_eqcircst} gives us $\equiatingEvents{\eventEquivFromBulk{n}}\circ\hintost(H)\isomorphic \hintostScultures(\sculpture{H}{n})$ which by Theorem~\ref{th_on_scultures} becomes $\equiatingEvents{\eventEquivFromBulk{n}}\circ\hintost(H)\isomorphic \sculpintost(\sculpture{H}{n})$, where the sculpture $\sculpture{H}{n}$ that we work with is given by $\stintosculpture(\ST)$. Finally, the Proposition~\ref{prop_stSculptst} gives the desired result $\equiatingEvents{\eventEquivFromBulk{n}}\circ\hintost(H)\isomorphic \sculpintost(\stintosculpture(\ST))\isomorphic \ST$, where the $H$ is the one from the sculpture $\stintosculpture(\ST)$ obtained from the ST-structure.
\end{proof}

\begin{corollary}
\ 

\begin{enumerate}
\item For any two \HDAs\ where one $H$ cannot be a sculpture and $H'$ can be sculpted, then $H\not\isomorphic H'$.

\item For any acyclic and non-degenerate $H'$ that cannot be sculpted there exists some sculpture $\sculpture{H}{n}$ s.t.\ $\hintost(H')\isomorphic\sculpintost(\sculpture{H}{n})$.
\end{enumerate}
\end{corollary}

\begin{proof}
For part 2 the sculpture that we are looking for is given by the application of the mapping \stintosculpture; i.e., take $\sculpture{H}{n}$ to be $\stintosculpture(\hintost(H'))$.
\end{proof}

We end this section by making two conjectures related to history unfolding of a HDA. 

For any non-degenerate \HDA\ $H$ we have
\[
H \hhequiv \unfolding(H) \mbox{\ and\ }
\]
\[
\hintost(H)\isomorphic \hintost(\unfolding(H)).
\]
%
%

\notforsubmission{
}


\section{Action refinement for ST-structures}\label{subsec_actref}

We define the notion of \textit{action refinement} \cite{GlabbeekG01refinement} for ST-structures  using a \textit{refinement function} $\reffun:\Sigma\rightarrow \allST$. The $\reffun(a)$ is a non-empty ST-structure which is to replace each and all events that are labeled with $a$. In this way, what before was abstracted away into a single event, now using action refinement, can be given more (concurrent) structure. The definition of \reffun\ is over all action labels of $\Sigma$, but normally only a (small) subset of actions is refined, whereas the rest should remain the same. For all such actions just refine with a singleton ST-structure labeled the same 
(i.e., $\reffun(b)=(\{(\emptyset,\emptyset),(f,\emptyset),(f,f)\},l(f)=b)$ with $f$ a new event).

\begin{definition}[action refinement]\label{def_actref}
Consider $\ST=(ST,l)$ over $\Sigma$ an \emph{ST-structure to be refined} by a \emph{refinement function} $\reffun:\Sigma\rightarrow \allST$. We call the pair of sets $(\refinement{S},\refinement{T})$ \emph{a refinement of an ST-config\-uration $(S,T)\in\ST$ by \reffun} iff:
\[
(\refinement{S},\refinement{T})=\bigcup_{e\in S\setminus T}(\{e\}\times S_{e},\{e\}\times T_{e})\cup\bigcup_{f\in T}(\{f\}\times S_{f},\{f\}\times S_{f})
\]
where each $(S_{e},T_{e})$ is a non-empty and non-maximal ST-configuration from $\reffun(l(e))$ and each $(S_{f},S_{f})$ is a maximal ST-configuration from $\reffun(l(f))$.
The refinement of\,\ \ST\ is defined as $\reffun(\ST)=(ST_{\reffun},l_{\reffun})$ with
\begin{itemize}
\item $ST_{\reffun}=\{(\refinement{S},\refinement{T})\mid (\refinement{S},\refinement{T})\mbox{ is a refinement of some }(S,T)\in\ST\mbox{ by }\reffun\}$;
\item $l_{\reffun}((e,e'))=l_{\reffun(l(e))}(e')$.
\end{itemize}
\end{definition}

Note that because of the closure restriction in the definition of ST-structures, any maximal ST-configuration, wrt.\ set inclusion, must have both $S$ and $T$ equal.

\begin{proposition}[refinement is well defined]\label{prop_ref1}\ 

For two isomorphic ST-structures $\ST\isomorphic\ST'$ and two isomorphic refinement functions, i.e.\ $\forall a\in\Sigma:\reffun(a)\isomorphic\reffun'(a)$, we have: 

\begin{enumerate}
\item $\reffun(\ST)$ is also an ST-structure;
\item $\reffun(\ST)\isomorphic\reffun'(\ST)$; and
\item $\reffun(\ST)\isomorphic\reffun(\ST')$.
\end{enumerate}
\end{proposition}

\begin{proof}
We first have to show that every set $(\refinement{S},\refinement{T})\in ST_{\reffun}$ is a well defined ST-configuration. This is easy to see because we use unions in the definition of $(\refinement{S},\refinement{T})$ and in the right-most unions the resulting $S$ and $T$ sets are the same, whereas in the left unions all events in the $T$ sets are also in the $S$ sets because they are build from ST-configurations $(S_{e},T_{e})$, i.e., $\{e\}\times T_{e}\subseteq\{e\}\times S_{e}$. Union of sets preserves set inclusion.

We need to show now that for any set $(\refinement{S},\refinement{T})\in ST_{\reffun}$ there is the case that $(\refinement{S},\refinement{S})\in ST_{\reffun}$. Assume that $(\refinement{S},\refinement{T})$ is a refinement of some $(S,T)$ and assume there are some events $e\in S\setminus T$ and that we have used some ST-configuration $(S_{e},T_{e})$ from $\reffun(l(e))$.
(Otherwise, when $S=T$ it implies, by construction, that also $\refinement{S}=\refinement{T}$.)
Because the refinement function uses only ST-structures then it means that in $\reffun(l(e))$ there is also an ST-configuration $(S_{e},S_{e})$. 
If $(S_{e},S_{e})$ is maximal, then the refinement of $(S,T)$ that uses it will result in the desired $(\refinement{S},\refinement{S})$.
Otherwise, when $(S_{e},S_{e})$ is not maximal, by the Definition~\ref{def_actref} it means we eventually build another refinement of $(S,T)$ using this $(S_{e},S_{e})$. Therefore, we would have both the $S$ and $T$ sets the same. These arguments must be carried for all events $e\in S\setminus T$ to obtain the desired $(\refinement{S},\refinement{S})$.

The final step is to show that the new labeling function $l_{\reffun}$ is correctly build in the sense that for each new event it assigns some label, in fact the correct label coming from the refining ST-structures given by \reffun. The definition of $l_{\reffun}$ is for every new event $(e,e')$ where $e$ is from the old ST-structure and $e'$ is from the new refining structures. The new label is the same as the corresponding label in the ST-structure where the $e'$ comes from.

The refinement operation is well defined also wrt.\ isomorphic structures.
\begin{itemize}
\item For a \ST\ and two isomorphic refinement functions $\reffun$ and $\reffun'$, i.e.\ $\forall a\in\Sigma:\reffun(a)\isomorphic\reffun'(a)$, then $\reffun(\ST)\isomorphic\reffun'(\ST)$.
\item For two isomorphic ST-structures $\ST\isomorphic\ST'$ and some refinement function $\reffun$ we have that \linebreak $\reffun(\ST)\isomorphic\reffun(\ST')$.
\end{itemize}

The proof uses the fact that isomorphic ST-structures agree on the labeling functions, and uses similar arguments as above.
\end{proof}

\begin{proposition}[preserving properties]\label{def_ref3}\ 

The refinement operation preserves the properties of the refined structure, i.e., for some $\ST$ and $\reffun$:
\begin{itemize}
\item If\,\ \ST\ is rooted then $\reffun(\ST)$ is rooted.
\item If\,\ \ST\ is connected and all $\reffun(a)$ are connected then $\reffun(\ST)$ is connected.
\item If\,\ \ST\ and all $\reffun(a)$ are adjacent-closed then $\reffun(\ST)$ is adjacent-closed.
\item If\,\ \ST\ and all $\reffun(a)$ are closed under bounded unions (or intersections) then $\reffun(\ST)$ is closed under bounded unions (rsp.\ intersections).
\end{itemize}
\end{proposition}

\begin{proof}
The property of being rooted is easy. 

For connectedness consider some non-empty $(\refinement{S},\refinement{T})\in\reffun(\ST)$ which is a refinement of some (non-empty) $(S,T)\in\ST$. The fact that is non-empty it implies that $\exists (e,e')\in\refinement{S}$ an event, which by the definition of refinement it means that $e$ comes from $S$. Therefore $(S,T)$ is also non-empty. By the connectedness of the original $\ST$ it means that there exists some $f\in S$ s.t.\ either (1) $(S\setminus f,T)\in\ST$ or (2) $(S,T\setminus f)\in\ST$.

Assume that (1) $(S\setminus f,T)\in\ST$ and take a refinement of it where for all events different than $f$ we take the same ST-configurations $(S_{e},T_{e})\in\reffun(l(e))$ as we did in the refinement $(\refinement{S},\refinement{T})$. This means that $(\refinement{S},\refinement{T})$ has all the events of $(\refinement{S\setminus f},\refinement{T})$ and on top it has some more events coming from the refinement of $f$. If there is only one such event $(f,f')$, as for example coming from refinement with a $(f',\emptyset)\in\reffun(l(f))$, then the proof is finished since we found this event which if removed from $\refinement{S}$ we obtain a new ST-configuration which is also in $\reffun(\ST)$. Therefore, consider the case when there are more events $(f,f'_{i})$ as coming from some chosen ST-configuration $(S_{f},T_{f})\in\reffun(l(f))$. Since $\reffun(l(f))$ is also connected it mean that we can find some event $f'_{k}$ s.t.\ either $(S_{f}\setminus f'_{k},T_{f})\in\reffun(l(f))$ or $(S_{f},T_{f}\setminus f'_{k})\in\reffun(l(f))$. No matter which is the case we have the following: one refinement of $(S,T)$ is $(\refinement{S},\refinement{T})\in\reffun(\ST)$ using $(S_{f},T_{f})\in\reffun(l(f))$ and another refinement of the same $(S,T)$ uses all the same ST-configurations except for $(S_{f},T_{f})$, in place of which another ST-configuration is used which has exactly one less event. This concludes this case, as we found the single event $(f,f'_{k})$ which can be removed to obtain another refinement.

Assume (2) $(S,T\setminus f)\in\ST$ and take the refinement of $(S,T\setminus f)$ the same as that for $(S,T)$. This is possible because before, in $(S,T)$, $f$ was part of $T$ and thus it was refined using some maximal configuration $(S_{f},S_{f})\in\reffun(l(f))$. But since now $f\in S\setminus(T\setminus f)$ we can refine with any configuration from $\reffun(l(f))$, and hence also with the maximal one $(S_{f},S_{f})\in\reffun(l(f))$. Because $\reffun(l(f))$ is connected then there exists some event $f'_{k}$ in the maximal configuration $(S_{f},S_{f})$ s.t.\ $(S_{f},S_{f}\setminus f'_{k})\in\reffun(l(f))$. Use this configuration to refine $f$ instead of the one $(S_{f},S_{f})$ used to obtain $(\refinement{S},\refinement{T})$. In this way we obtain a refinement configuration in $\reffun(\ST)$ that differs from $(\refinement{S},\refinement{T})$ by one single event $(f,f'_{k})$.

For bounded unions and intersections the proof should be similar to that in \cite[Prop.5.6]{GlabbeekG01refinement} and should use argument specific to ST-structures as we used above.

We concentrate on the new property of adjacent closure. 
One can prove this directly using Definition~\ref{def_adj_ST} of adjacent-closure and take (more tedious) cases.
We will take the alternative route through Proposition~\ref{prop_adj_equiv} and use the closure under single events. This results in fewer cases to consider.

Therefore, we want to show that for some arbitrary $(\refinement{S},\refinement{T})\in\reffun(\ST)$, being a refinement of some $(S,T)\in\ST$ we have
\begin{enumerate}
\item $\forall (e,e')\in\refinement{S}\setminus\refinement{T}:(\refinement{S},\refinement{T}\cup(e,e'))\in\reffun(\ST)$ and
\item $\forall (e,e')\in\refinement{S}\setminus\refinement{T}:(\refinement{S}\setminus(e,e'),\refinement{T})\in\reffun(\ST)$.
\end{enumerate}
Since $(e,e')\in\refinement{S}\setminus\refinement{T}$ it means that $e\in S\setminus T$ and $e'\in(S_{e},T_{e})$, where $(S_{e},T_{e})\in\reffun(l(e))$ is the chosen ST-configuration. Because $\reffun(l(e))$ is closed under single events, the first requirement then says that also $(S_{e},T_{e}\cup e')\in\reffun(l(e))$. Take now another refinement of $(S,T)$ that is the same as before only that in place of $(S_{e},T_{e})\in\reffun(l(e))$ uses $(S_{e},T_{e}\cup e')$. Clearly this new refinement has all the event of the old refinement with the exception that now the event $(e,e')$ is also contained in $\refinement{T}$. This proves the first requirement.

To show the second requirement consider the existence of $(S_{e}\setminus e',T_{e})\in\reffun(l(e))$ and the same argument applies as before. The only difference is that the new $(S_{e}\setminus e',T_{e})$ may in fact be $(\emptyset,\emptyset)$. In this case we use the fact that the original $\ST$ is closed under single events and thus, for the $e\in S\setminus T$ we can find $(S\setminus e,T)\in\ST$. Take the refinement of this which will have all the event of the old one, except the one $(e,e')$ which is not in $\refinement{S}$ anymore.
\end{proof}

Single steps in the new structure relate to steps in the old structure, before the refinement, and the refining structures given by \reffun. A single s-step in the refined structure comes either from a single s-step in the old structure, and thus coupled with an initial step in \reffun\ (i.e., one like $(\emptyset,\emptyset)\transition{s}(e,\emptyset)$) or only from an s-step in the \reffun\ (with the ST-configuration unchanged). \cp{What about t-steps??}

\begin{proposition}\label{prop_hhPreserved}
The hh-bisimulation is preserved under action refinement; i.e., 

\centerline{for $\ST\hhequiv\ST'$\ \ then\ \ $\reffun(\ST)\hhequiv\reffun(\ST')$.}
\end{proposition}

\begin{proof}
We consider given a hereditary history preserving bisimulation (hh-bisimulation) $R$ that relates the two initial ST-configurations of $\ST$ and $\ST'$. We construct a relation $\refinement{R}$ between the refinements $\reffun(\ST)$ and $\reffun(ST')$ which will also equate their initial empty configurations, and we show that it respects the restrictions of Definition~\ref{def_hh_ST} of being a hh-bisimulation.

We will also show that the proof works also when we consider $R$ to be only history preserving bisimulation. Moreover, we point out how the proof can be changed to show a result where we do not refine with the same refinement function, but with refinement functions that are also hh-bisimilar (or only h-bisimilar in the other case).

Define $\refinement{R}$ as:

$((\refinement{S},\refinement{T}),(\refinement{S'},\refinement{T'}),\refinement{f})\in\refinement{R}$ iff $\exists ((S,T),(S',T'),f)\in R$ s.t.
\begin{enumerate}
\item $(\refinement{S},\refinement{T})$ is a refinement of $(S,T)$;

\item $(\refinement{S'},\refinement{T'})$ is a refinement of $(S',T')$ that uses the same choices as doe for $(\refinement{S},\refinement{T})$;

\item $\refinement{f}:\refinement{S}\rightarrow \refinement{S'}$ is defined as $\refinement{f}(e,e')=(f(e),e')$.
\end{enumerate}

When we want to prove the result for hh-bisimilar refinement functions $\reffun\hhequiv\reffun'$ then we need to complicate the definition of $\refinement{R}$ by adding one more requirement:
\begin{enumerate}
\setcounter{enumi}{3}
\item $\forall e\in S$, with $(S_{e},T_{e})\in\reffun(l(e))$ the refining configuration that makes $(\refinement{S},\refinement{T})$, $\exists f_{e}:S_{e}\rightarrow S'_{f(e)}$ s.t.\ $((S_{e},T_{e}),(S'_{f(e)},T'_{f(e)}),f_{e})\in R_{l(e)}$, with $R_{l(e)}$ being the hh-bisimulation relating the refinement choices for the label $l(e)$ for $(\refinement{S'},\refinement{T'})$;
\end{enumerate}
and we also need to change the requirement 3 above to satisfy $\refinement{f}(e,e')=(f(e),f_{e}(e'))$.

It is not difficult to see from the definition above that because $(\emptyset,\emptyset,\emptyset)\in R$ then also  $(\emptyset,\emptyset,\emptyset)\in \refinement{R}$.
It remains to prove the restrictions of Definition~\ref{def_hh_ST}.

\begin{enumerate} 
\item We prove that $\refinement{f}$ is an isomorphism of $(\refinement{S},\refinement{T})$ and $(\refinement{S'},\refinement{T'})$. Since $f$ is an isomorphism between $(S,T)$ and $(S',T')$ then, by Definition~\ref{def_isomorphism}, $f$ is an isomorphism of $S$ and $S'$ that agrees on the $T$ and that preserves the labeling. In consequence, when the refinement $(\refinement{S},\refinement{T})$ involves some event $e\in S$ then the isomorphic image $f(e)$ has the same label, hence it is refined with the same ST-structure $\reffun(l(f(e)))=\reffun(l(e))$. Here is where the constraint 2.\ in the definition of $\refinement{R}$ says to make the same choice of $(S_{e},T_{e})$.
By the definition of $\refinement{f}$, this is also an isomorphism of $\refinement{S}$ and $\refinement{S'}$ since we use the same events $e'$. It is easy to see that $\refinement{f}$ agrees on $\refinement{T}$. The $\refinement{f}$ also preserves the labeling function of the new refinements because the new events get the label of the second component $e'$ which is related to the label of either $e$ of $f(e)$, which are the same.

When proving the proposition for two bisimilar refining functions then the argument above works because of the extra requirement 4. This gives an equivalent configuration to pick when obtaining $(\refinement{S'},\refinement{T'})$, i.e., pick $(S'_{f(e)},T'_{f(e)})$. Then in the definition of $\refinement{f}$ we use not the same $e'$ but an isomorphism $f_{e}$, therefore the $\refinement{f}$ is also an isomorphism.

\item We prove the second requirement of Definition~\ref{def_hh_ST} and assume there is a step $(\refinement{S},\refinement{T})\transition{a}(\refinement{S_{a}},\refinement{T_{a}})$ in the refinement $\reffun(\ST)$, for $((\refinement{S},\refinement{T}),(\refinement{S'},\refinement{T'}),\refinement{f})\in\refinement{R}$. This is equivalent to saying that we have $((S,T),(S',T'),f)\in R$ satisfying the three requirements from before, i.e., that $(\refinement{S},\refinement{T})$ is a refinement of $(S,T)$, $(\refinement{S'},\refinement{T'})$ is a refinement of $(S',T')$ with the same choices, and that $\refinement{f}(e,e')=(f(e),e')$.
We take cases depending on what kind and how did this step get formed.
\begin{enumerate}
\item When we work with an s-step which is formed from an s-step in the refinement and the same $(S,T)$ in the original \ST.
The s-step comes from a configuration $(S_{e},T_{e})\transition{a}(S_{e}\cup g,T_{e})$ corresponding to refining some $e\in S$. More precisely, if $(\refinement{S},\refinement{T})=(\{e\}\times S_{e},\{e\}\times T_{e})\cup\bigcup_{e'\neq e\in S\setminus T}(\{e'\}\times S_{e'},\{e'\}\times T_{e'})\cup\bigcup_{f\in T}(\{f\}\times S_{f},\{f\}\times S_{f})$ and knowing that the above s-step is in $\reffun(l(e))$ and therefore also in $\reffun(l(f(e)))$, since the isomorphism $f$ preserves labeling, the the s-step we are assuming, i.e., $(\refinement{S},\refinement{T})\transition{a}(\refinement{S_{a}},\refinement{T_{a}})$ is obtained by having $(\refinement{S_{a}},\refinement{T_{a}})$ also a refinement of $(S,T)$ with the same choices as for $(\refinement{S},\refinement{T})$ with one difference: $(\refinement{S_{a}},\refinement{T_{a}})=(\{e\}\times (S_{e}\cup g),\{e\}\times T_{e})\cup\bigcup_{e'\neq e\in S\setminus T}\dots$. Which means that one new event is added to $\refinement{S}$, and that is $(e,g)$.
Knowing that $(\refinement{S'},\refinement{T'})$ is a refinement of $(S',T')$ which is in relation $R$ with $(S,T)$ and $f$, then we take another refinement of $(S',T')$, the same as $(\refinement{S'},\refinement{T'})$ in all respects except that for the event $f(e)$ we take the configuration $(S_{e}\cup g,T_{e})$ (which we know from before that it exists, because the above step exists in $\reffun(l(f(e)))$). Denote this new refinement as $(\refinement{S'_{a}},\refinement{T'_{a}})$ which has the difference in the ST-configuration $(f(e)\times(S_{e}\cup g),f(e)\times T_{e})$, i.e., the single step that we are looking for adds the new event $(f(e),g)$. Clearly there is a single s-step $(\refinement{S'},\refinement{T'})\transition{a}(\refinement{S'_{a}},\refinement{T'_{a}})$. Moreover, the new configurations are in the relation we built $((\refinement{S_{a}},\refinement{T_{a}}),(\refinement{S'_{a}},\refinement{T'_{a}}),\refinement{f_{a}})\in\refinement{R}$ where $\refinement{f_{a}}$ extends $\refinement{f}$ with $\refinement{f_{a}}(e,g)=(f(e),g)$.
It is easy to show this last statement, using the definition for $\refinement{R}$ above. Just take the same $((S,T),(S',T'),f)\in R$, and thus have that $(\refinement{S_{a}},\refinement{T_{a}})$ is a refinement of $(S,T)$ by construction, and $(\refinement{S'_{a}},\refinement{T'_{a}})$ a refinement of $(S',T')$ using the same choices (in particular choosing $(S_{e}\cup g,T_{e})$ for refining $e$). It is easy to see that $\refinement{f_{a}}$ respects the condition 3 since is extends $\refinement{f}$ which does.

\vspace{1ex}We are interested how the proof changes when working with two refinement functions. Because the refining configurations are hh-bisimilar it means that instead of the same configuration as before, we find a bisimilar one which comes as an s-step extension of the old one; i.e., we find $(S_{e}\cup f_{e}(g),T_{e})$. In consequence the isomorphism is extended with $\refinement{f_{a}}(e,g)=(f(e),f_{e}(g))$.

\vspace{1ex}The rest of the three cases are similar and we skip their details.\vspace{1ex}

\item When we work with an s-step which is formed from an s-step in the original \ST, and a minimal s-step in the refinement.
More precisely, this step comes from the step $(S,T)\transition{a}(S\cup e,T)$ and some initial step $(\emptyset,\emptyset)\transition{a}(g,\emptyset)$ in $(S_{e},T_{e})\in\reffun(l(e))$ in the following way. Take $(\refinement{S_{a}},\refinement{T_{a}})$ to be the refinement of $(S\cup e,T)$ which is exactly like $(\refinement{S},\refinement{T})$ on the sub-configuration $(S,T)$ and for the new event $e$ it uses the above ST-configuration $(g,\emptyset)$, which is non-empty. This new ST-configuration refinement has extra to $\refinement{S}$ the event $((e,g),\emptyset)$.
Because $((S,T),(S',T'),f)\in R$ then a matching step exists  $(S',T')\transition{a}(S'\cup f'(e),T')$, where $f'$ extends $f$ and is also an isomorphism, hence preserving the label of $e$.
In consequence, we can find the refinement $(\refinement{S'_{a}},\refinement{T'_{a}})$ of $(S'\cup f'(e),T')$ to be the same as $(\refinement{S'},\refinement{T'})$ and for the new event $f'(e)$ choose the same non-empty minimal ST-configuration $(g,\emptyset)$.
One can show that $((\refinement{S_{a}},\refinement{T_{a}}),(\refinement{S'_{a}},\refinement{T'_{a}}),\refinement{f'})\in\refinement{R}$ and $\refinement{f'}$ extends $\refinement{f}$, from the unrefined case. All the restrictions from the definition of $\refinement{R}$ are satisfied and the extension of the isomorphism is the case because $\refinement{f'}((e,g))=(f'(e),g)$.

\item When we work with a t-step which is formed from a t-step in the refinement and the same $(S,T)$ in the original \ST.

\item When we work with a t-step which is formed from a t-step in the original \ST, and a maximal t-step in the refinement.
\end{enumerate}

\item The third requirement of Definition~\ref{def_hh_ST} is symmetric to the one above.

\vspace{1ex}By this point we have proved that the history preserving bisimulation alone is preserved under action refinement, because in the proof we did not make reference to the backwards step requirements on the original $R$. Moreover, in when working with two bisimilar refinement functions we again did not make reference to the backwards steps.

\item Proving the forth requirement of Definition~\ref{def_hh_ST}, i.e., for backwards steps.
Similar arguments as before are used only that we remove events, instead of adding. The same cases need to be considered depending on what kind of steps we are working with.
\end{enumerate}
\end{proof}

From the proof of the above proposition one gets also that the history preserving bisimulation is preserved under refinement. The proof can also be extended to show that these bisimulations are congruences for action refinement.

\cp{
\begin{proposition}
ST-trace equivalence and cc-equivalence are preserved under action refinement.
\end{proposition}

\begin{proposition}
Action refinement for higher dimensional automata is defined by replacing transitions by new \HDAs. This result should be deducible from the refinement on adjacent-closed ST-structures.
\end{proposition}
}

\section{STC-structures}\label{sec_STCstruct}

We extend ST-structures (following Pratt \cite{Pratt03trans_cancel}) to include the notion of \textit{cancellation}, and call this extension \textit{STC-structures}. These are richer than ST-structures, acyclic \HDAs, or the inpure event structures of \cite{GlabbeekP09configStruct}. 
STC-structures overcome several shortcomings of ST-structures: the inability to associate a natural \textit{termination predicate}, as Example~\ref{ex_termination} illustrates;  they do not properly capture the behavior of \textit{\HDAs\ with cycles}, cf.~Example~\ref{ex_shutdown}; and cannot distinguish the angelic vs.\ demonic choices, cf.~Example~\ref{ex_agelicdemon}. Some kind of cycles can be captured by ST-structures alone, but some more convoluted cyclic \HDAs\ require notions of cancellation also, as Example~\ref{ex_shutdown} illustrates. 
Note also that STC-structures model well the two examples of \HDAs\ from Figure~\ref{fig_ex_Glabbeek} when the dotted transitions are added. But without these transitions, i.e., only over the three events, the STC-structures also equate the two example \HDAs. Therefore STC-structures do not manage to distinguish that one is a sculpture and the other is not.

\begin{definition}[STC-structures]\label{def_STCstructures}\ 

An \emph{STC-configuration} over $E$ is a set triple $(S,T,C)$, with $S\subseteq E$ finite, respecting the following restrictions:
\begin{center}
\begin{tabular}{rl}
(start before terminate) & $T\subseteq S$;\\
(cancellation) & $S\cap C=\emptyset$.\\
\end{tabular} 
\end{center}
An \emph{STC-structure} is a tuple $\STC=(E,STC,l)$, with $STC$ a set of \emph{STC-configurations} over $E$, the labeling function $l$ defined as for ST-structures, and satisfying:
\[
\forall (S,T,C)\in\STC:\exists (S,S,C')\in\STC \mbox{ with }C\subseteq C'.
\]
\end{definition}

\begin{proposition}
ST-structures are strictly included in the STC-structures.
\end{proposition}

\begin{proof}
For an arbitrary ST-configuration associate the STC-configuration with the same S and T sets and $C=\emptyset$.
The simple Example~\ref{ex_agelicdemon} shows the strictness.
\end{proof}

\begin{example}[angelic vs.\ demonic choice]\label{ex_agelicdemon}
The simple example, used by Pratt \cite[sec.3.3]{Pratt03trans_cancel}, of angelic vs.\ demonic choice at the level of the events (not actions) cannot be captured in the general event structures of \cite{GlabbeekP09configStruct}. 
This example involves three events, with $e$ and $f$ conflicting (i.e., a choice between them is made) and are causally depending on $d$. Branching semantics normally distinguish two systems, depending on when the choice is made. Their STC-structures are given in Figure~\ref{fig_ex_angelicdemonic} where the (left) makes a late choice and the (right) an early choice. But when removing the third component $C$ of all the STC-configurations, the two resulting ST-structures are the same. No ST-structure over three events can make this distinction between the two kinds of choices.
As \HDAs, the angelic choice can be seen as a sculpture from a 3-bulk, but the demonic choice cannot be seen as a sculpture.
\begin{figure}[tp]
\psfrag{000}{{\scriptsize $(\emptyset,\!\emptyset,\!\emptyset)$}}
\psfrag{a00}{{\scriptsize $(d,\!\emptyset,\!\emptyset)$}}
\psfrag{aa0}{{\scriptsize $(d,\!d,\!\emptyset)$}}
\psfrag{a0b}{{\scriptsize $(d,\!\emptyset,\!e)$}}
\psfrag{a0c}{{\scriptsize $(d,\!\emptyset,\!f)$}}
\psfrag{aab}{{\scriptsize $(d,\!d,\!e)$}}
\psfrag{aac}{{\scriptsize $(d,\!d,\!f)$}}
\psfrag{acab}{{\scriptsize $(df,\!d,\!e)$}}
\psfrag{abac}{{\scriptsize $(de,\!d,\!f)$}}
\psfrag{acacb}{{\scriptsize $(df,\!df,\!e)$}}
\psfrag{ababc}{{\scriptsize $(de,\!de,\!f)$}}
  \begin{center}
    \hspace{-2ex}\includegraphics[height=1.2cm]{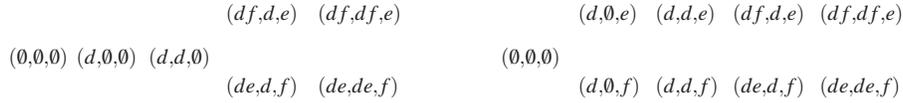}
  \end{center}
\caption{Angelic vs.\ demonic choice at the level of the events $e,f$, not actions.}
\label{fig_ex_angelicdemonic}
\end{figure}
\end{example}


For STC-structures the computational interpretation simply extends the s- and t-steps from ST-structures by not changing the $C$ component. Moreover, two \textit{cancellation steps} are added. There are several kinds of cancellation steps, but in all of them progress must be made, hence the start or termination of a single event must occur. A cancellation step may cancel only one or more events, or may both cancel and enable events.

\begin{definition}[cancellation steps]\label{def_c_steps}\ 

A \emph{cancellation step} between two STC-configurations is denoted $(S,T,C)\transitionUpDown{a}{c}(S',T',C')$, with $c\in\{\,^{+}_{\mathbf{1}}\mathbf{cs}\,,\,^{+}_{\mathbf{1}}\mathbf{ct}\,,\,^{+}_{\mathbf{n}}\mathbf{cs}\,,\,^{+}_{\mathbf{n}}\mathbf{ct}\,,\,^{\pm}_{\mathbf{n}}\mathbf{cs}\,,\,^{\pm}_{\mathbf{n}}\mathbf{ct}\}$, and defined as:
\begin{enumerate}
\item Single event and canceling only:
\begin{description}
\item[$^{+}_{\mathbf{1}}$cs-step:] $T=T'$, $S\subset S'$, $C\subset C'$, $S'\setminus S=\{e\}$, $l(e)=a$ and $|C'\setminus C|=1$;
\item[$^{+}_{\mathbf{1}}$ct-step:] $S=S'$, $T\subset T'$, $C\subset C'$, $T'\setminus T=\{e\}$, $l(e)=a$ and $|C'\setminus C|=1$.
\end{description}

\item Multiple events and canceling only:
\begin{description}
\item[$^{+}_{\mathbf{n}}$cs-step:] $T=T'$, $S\subset S'$, $C\subset C'$, $S'\setminus S=\{e\}$, $l(e)=a$ and $C'\setminus C\neq\emptyset$;
\item[$^{+}_{\mathbf{n}}$ct-step:] $S=S'$, $T\subset T'$, $C\subset C'$, $T'\setminus T=\{e\}$, $l(e)=a$ and $C'\setminus C\neq\emptyset$.
\end{description}

\item Multiple events and both canceling and enabling:
\begin{description}
\item[$^{\pm}_{\mathbf{n}}$cs-step:] $T=T'$, $S\subset S'$, $S'\!\setminus\!S=\!\{e\}$, $l(e)=a$, $e\not\in C$, and $C\subset C' \vee C'\subset C$;
\item[$^{\pm}_{\mathbf{n}}$ct-step:] $S=S'$, $T\subset T'$, $T'\setminus T=\{e\}$, $l(e)=a$ and $C\subset C' \vee C'\subset C$.
\end{description}
\end{enumerate}
\end{definition}

The single event cancellation steps are not enough to model the cyclic structures, where an event cancels all remaining events of a repetition, like in Example~\ref{ex_shutdown}. This would motivate the cancellation steps where more events may be canceled at once. But the Example~\ref{ex_game_angelic_vs_demonic} suggests that such cancellation steps may not be enough, and the preferred would be those that also enable events.

Note also the extra condition in the \textbf{$^{\pm}_{\mathbf{n}}$cs-step} which ensures that currently canceled events cannot be started. This condition is implicit (deducible) in the other kinds of steps, but for this particular kind of step it may be that the same event that is started belongs to the canceled events and is removed (enabled) in the current step; this would result in a correct STC-configuration. We disallow such steps.

Moreover, if we choose to work with \textbf{$^{\pm}_{\mathbf{n}}$cs-step} we need to relax the constraint on STC-structures so that it no longer requires $C\subseteq C'$, the set of canceled event to increase.

We could even give a very general form of steps, to include s/t steps and cancellation steps. We just need to relax the first two kinds of cancellation steps to allow for the set of canceled events to not increase (i.e., be $0$ respectively $\emptyset$). Or we could just combine these first two into one single s/t step that can cancel $0,1$, or several events. We can generalize this even more by enabling events, and thus include the third kind of cancellation too.

For pedagogical reasons we prefer to stick with the clearly split definition from above.

\begin{example}[game of angelic vs.\ demonic speed]\label{ex_game_angelic_vs_demonic}
Consider the following game of two players, where the ``demon'' player and the ``angel'' player each are having a task to do, i.e., the respective events labeled by $d$ respectively $a$. If the angel finishes first she gets to choose the next action to do, whereas if the demon finishes first he gets his initial choice. Both the demon and the angel are starting at the same time and their tasks are going in parallel. The two choice that are to be made are $g$ (good) and $e$ (evil); with $g$ being the winning preference of the angel and $e$ the preference of the demon.

This example is nicely depicted as a \HDA\ which is not a sculpture and is non-degenerate and acyclic (see Figure~\ref{fig_ex_angelicdemonic_game}(left)). Moreover, this \HDA\ is also its own history-unfolding. This \HDA\ puts together the angelic and demonic choice patterns in a single system. This example is not representable as ST-structures.
But it is representable as an STC-structure.
The computational steps in this STC-structure involve cancellation of both angelic and demonic kind. It is not enough to use simple cancellation steps because when the $a$ event finishes then the canceled events must be removed so to give the appropriate angelic choice. Events must be enabled again in such a step.

\begin{figure}[tp]
\psfrag{000}{{\scriptsize $(\emptyset,\!\emptyset,\!\emptyset)$}}
\psfrag{a00}{{\scriptsize $(d,\!\emptyset,\!\emptyset)$}}
\psfrag{aa0}{{\scriptsize $(d,\!d,\!\emptyset)$}}
\psfrag{a0b}{{\scriptsize $(d,\!\emptyset,\!e)$}}
\psfrag{a0c}{{\scriptsize $(d,\!\emptyset,\!f)$}}
\psfrag{aab}{{\scriptsize $(d,\!d,\!e)$}}
\psfrag{aac}{{\scriptsize $(d,\!d,\!f)$}}
\psfrag{acab}{{\scriptsize $(df,\!d,\!e)$}}
\psfrag{abac}{{\scriptsize $(de,\!d,\!f)$}}
\psfrag{acacb}{{\scriptsize $(df,\!df,\!e)$}}
\psfrag{ababc}{{\scriptsize $(de,\!de,\!f)$}}
  \begin{center}
    \hspace{-2ex}\includegraphics[height=3.2cm]{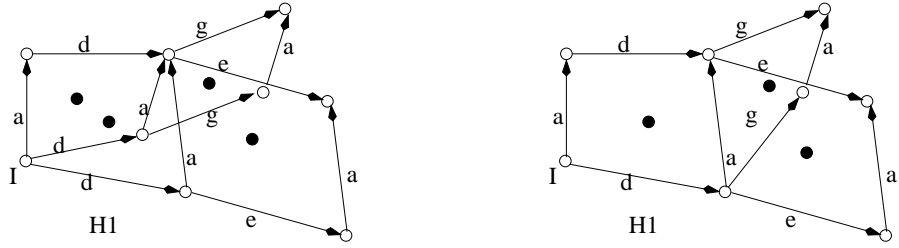}
  \end{center}
\caption{The game of angelic vs.\ demonic speed (on the left) as a non-sculpture. \newline On the right is a related sculpture which does not capture the example, not even when spiting the events.
}
\label{fig_ex_angelicdemonic_game}
\end{figure}
\end{example}

\begin{example}[termination]\label{ex_termination}
Instead of considering any maximal configuration to be terminal we want to have a more general \textit{termination predicate} over ST-configurations, as is done in \cite[Def.4.1\&Def.5.1]{GlabbeekG01refinement} for event structures and configuration structures. But there a configuration can be terminal only if it is maximal, which is a natural requirement that we want to stick with.
Take now the ST-structure for asymmetric conflict from Fig.~\ref{fig_ex_asymconflict1}(middle) where both ST-configurations $(s,s)$ and $(bs,bs)$ should be final, but only $(bs,bs)$ is maximal. The same issue appears for the ST-structures that we discuss below for the two cyclic \HDAs\ from Fig.~\ref{fig_ex_STC_structs}(middle-right and right). But note though that it is very natural to assign \textit{final cells} to these two \HDAs: they are $q_{0}^{4}$ respectively $q_{0}^{2}$.

STC-structures make it natural for maximal STC-configurations to be terminal. Take the STC-structure for asymmetric conflict from Fig.~\ref{fig_ex_STC_structs}(left) where both $(s,s,b)$ and $(bs,bs,\emptyset)$ are maximal.
\begin{figure}[tp]
\psfrag{e}{\scriptsize $s$}
\psfrag{d}{\scriptsize $b$}
\psfrag{f}{\scriptsize $f$}
\psfrag{00}{{\scriptsize $(\emptyset,\emptyset)$}}
\psfrag{d0}{{\scriptsize $(b,\emptyset)$}}
\psfrag{e0}{{\scriptsize $(s,\emptyset)$}}
\psfrag{de0}{{\scriptsize $(bs,\emptyset)$}}
\psfrag{ee}{{\scriptsize $(s,s)$}}
\psfrag{dd}{{\scriptsize $(b,b)$}}
\psfrag{ded}{{\scriptsize $(bs,b)$}}
\psfrag{dede}{{\scriptsize $\mathbf{\boldsymbol{(}bs,bs\boldsymbol{)}}$}}
\psfrag{dfd}{{\scriptsize $(bf,b)$}}
\psfrag{dfdf}{{\scriptsize $(bf,bf)$}}
\psfrag{001}{{\scriptsize $(\emptyset,\!\emptyset,\!\emptyset)$}}
\psfrag{d01}{{\scriptsize $(b,\!\emptyset,\!\emptyset)$}}
\psfrag{e01}{{\scriptsize $(\!s,\!\emptyset,\!b)$}}
\psfrag{ee1}{{\scriptsize $\mathbf{\boldsymbol{(}\!s,\!s,\!b\boldsymbol{)}}$}}
\psfrag{dd1}{{\scriptsize $(b,\!b,\!\emptyset)$}}
\psfrag{ded1}{{\scriptsize $(bs,\!b,\!\emptyset)$}}
\psfrag{dede1}{{\scriptsize $\mathbf{\boldsymbol{(}bs,\!bs,\!\boldsymbol{\emptyset)}}$}}
\psfrag{q11}{$q_{0}^{1}$}
\psfrag{q12}{$q_{0}^{3}\!\!=$}
\psfrag{q13}{$\mathbf{q_{0}^{4}}$}
\psfrag{q131}{$=q_{0}^{4}$}
\psfrag{q14}{$q_{0}^{2}$}
\psfrag{q141}{$\mathbf{q_{0}^{2}}$}
\psfrag{q21}{\scriptsize $b$}
\psfrag{q22}{\scriptsize $s$}
\psfrag{q23}{\scriptsize $b$}
\psfrag{q24}{\scriptsize $s$}
\psfrag{q3}{$q_{2}$}
\psfrag{0}{{\scriptsize $\emptyset$}}
\psfrag{0d}{{\scriptsize $\{b\}$}}
\psfrag{0e}{{\scriptsize $\{s\}$}}
\psfrag{0ed}{{\scriptsize $\{b,s\}$}}
\psfrag{sd0}{{\small \begin{rotate}{125}$\enableRelEv$\end{rotate}}}
\psfrag{se0}{{\small \begin{rotate}{45}$\enableRelEv$\end{rotate}}}
\psfrag{sed}{{\small \begin{rotate}{45}$\enableRelEv$\end{rotate}}}
\psfrag{hh}{{\scriptsize $\hhequiv$}}
\psfrag{iso}{{\scriptsize $\isomorphic$}}
\psfrag{niso}{{\scriptsize $\not\isomorphic$}}
\psfrag{i}{i}
\psfrag{ii}{ii}
\psfrag{iii}{iii}
\psfrag{iv}{iv}
\psfrag{v}{v}
  \begin{center}
    \includegraphics[height=2.9cm]{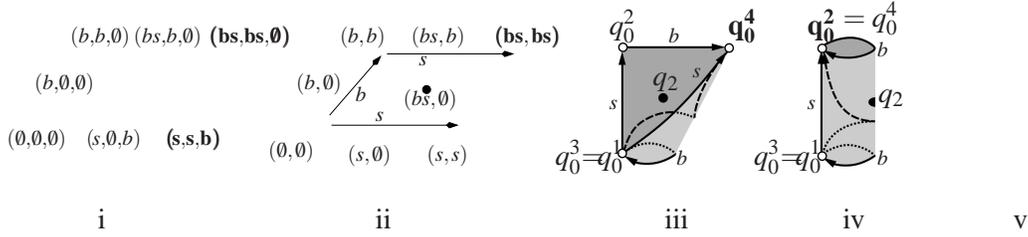}
  \end{center}
\caption{STC-structure for \textit{asymmetric conflict} $s+b;s$ (i); ST-structure for shutdown-backup Ex.~\ref{ex_shutdown} (ii), and a cyclic \HDA\ for its cyclic extension (iii); a cyclic \HDA\ for $s||b^{*}$ (iv); a cyclic and degenerate \HDA\ for $b^{*}\seqSTC(s||b)\seqSTC b^{*}$. Maximal configurations and final cells (i.e., from the set $F$) are in bold.}
\label{fig_ex_STC_structs}
\end{figure}
\end{example}

\cp{
The \HDA\ from Figure~\ref{fig_ex_STC_structs}(v) can be expressed with STC-structure but under the multiple-cancellation steps since the $s$ temporarily cancels all the $b$ events except the current $b$ until both $s$ and $b$  (which can run in parallel) are finished; after this all the $b$ events are enabled again to run as a Kleene-star.
}

\begin{example}[shutdown-backup]\label{ex_shutdown}
Initially one may model a linux-like system abstractly using two events labeled ${s}$ for \textit{shutdown} and ${b}$ for \textit{backup}. At later, more concrete stages these actions may be refined in processes with more structure, eg., part of a shutdown various actions are performed, like closing web or database services. 
The ${b}$ and ${s}$ are considered as being done concurrently, but at this abstract level of modeling the only clear constraint that we have is that ${s}$ must wait for ${b}$ to finish, before itself may finish. 
This does not mean that we first perform the ${b}$ and after it is finished we start the ${s}$. 

This example is modeled in Fig.~\ref{fig_ex_STC_structs}(middle-left) as the three sides of the square for the asymmetric conflict but with the inside filled in, to model the fact that the two actions can happen concurrently. 
%
This example cannot be captured in the event structures of \cite{GlabbeekP09configStruct}, nor configuration structures, nor adjacent-closed ST-structures, nor non-degenerate \HDAs.
This example is not adjacent-closed. As a \HDA\ this is \textit{degenerate} because one of the $t$ maps of the inside cell is missing. Moreover, this example is not closed under unions nor intersections, but it is rooted and connected.


This example is naturally extended to one involving cycles in \HDAs.
We now say that the system performs \textit{backup on a constant basis}, in a loop. But the shutdown may be issued at any time point. Therefore, we model the shutdown as happening concurrently with all the backup events. As soon as a shutdown is started, the currently running backup (if any) is allowed to finish, but no other backups may start before the system shuts down. (Naturally, no more backup can be performed after a shutdown).

This is not a simple parallel composition $s\, ||\, b^{*}$, which is modeled by the cyclic \HDA\ of Fig.~\ref{fig_ex_STC_structs}(right), but it is modeled as the cyclic \HDA\ from Fig.~\ref{fig_ex_STC_structs}(middle-right) which is like the parallel square but with the two lower corners  $q_{0}^{1}=q_{0}^{3}$ equated.
Both these \HDAs\ are non-degenerate. The (right) one can be encoded as an ST-structure over the set of events $\{s\}\cup\{b_{i}\mid i\in\mathbb{N}^{+}\}$ by thinking of unfolding the cylinder \HDA\ into infinitely many copies of the parallel square attached one after the other.
An unfolding for the (middle-right) example is not easy to see. But it is naturally encoded as an STC-structure over the same set of events, but where the $C$ component of the STC-configurations takes care of the cancellation of infinitely many copies of the backup events. The complete description is given in the Appendix, but intuitively, whenever the $s$ is executed it cancels all the remaining backup events, i.e., those that do not appear already in the $S$ or $T$ sets; until then $b_{i}$ events can happen in sequence.

Essential is that when removing the $C$ component from the above STC-structure we obtain an ST-structure isomorphic to the one for the $s\, ||\, b^{*}$. Therefore, the two cyclic \HDAs\ cannot be distinguished using ST-structures, even before thinking of termination.
 

We describe the set of STC-configurations that form the STC-structure over the events $E=\{s\}\cup\{b_{i}\mid i\in\mathbb{N}^{+}\}$ which describes the shutdown-backup example pictured as the cyclic \HDA\ of Figure~\ref{fig_ex_STC_structs}(middle-right). We define this set as a union of sets $S_{k}^{m}$ with $m\in\mathbf{N}$ used just for notation purposes, and $k\in\mathbb{N}^{+}$ used as an index correlated with the index of the $b_{i}$ events, as we see further.

\[
S_{k}^{0}\defequal\{(\{b_{i}\mid i<k\},\{b_{i}\mid i<k\},\emptyset)\mid k\in\mathbb{N}^{+}\}
\]
\[
S_{k}^{1}\defequal\{(s\cup\{b_{i}\mid i<k\},\{b_{i}\mid i<k\},\{b_{i}\mid i\geq k\})\mid k\in\mathbb{N}^{+}\}
\]
\[
S_{k}^{2}\defequal\{(s\cup\{b_{i}\mid i<k\},s\cup\{b_{i}\mid i<k\},\{b_{i}\mid i\geq k\})\mid k\in\mathbb{N}^{+}\}
\]
\[
S_{k}^{3}\defequal\{(s\cup\{b_{i}\mid i<k\},\{b_{i}\mid i<k\},\emptyset)\mid k\in\mathbb{N}^{+}\}
\]
\[
S_{k}^{4}\defequal\{(\{b_{i}\mid i<k+1\},\{b_{i}\mid i<k\},\emptyset)\mid k\in\mathbb{N}^{+}\}
\]
\[
S_{k}^{5}\defequal\{(s\cup\{b_{i}\mid i<k+1\},\{b_{i}\mid i<k+1\},\{b_{i}\mid i\geq k\})\mid k\in\mathbb{N}^{+}\}
\]
\[
S_{k}^{6}\defequal\{(s\cup\{b_{i}\mid i<k\},s\cup\{b_{i}\mid i<k\},\emptyset)\mid k\in\mathbb{N}^{+}\}
\]
\[
S_{k}^{7}\defequal\{(s\cup\{b_{i}\mid i<k+1\},s\cup\{b_{i}\mid i<k\},\{b_{i}\mid i\geq k+1\})\mid k\in\mathbb{N}^{+}\}
\]
\[
S_{k}^{8}\defequal\{(s\cup\{b_{i}\mid i<k+1\},\{b_{i}\mid i<k+1\},\{b_{i}\mid i\geq k+1\})\mid k\in\mathbb{N}^{+}\}
\]

But note that $S_{k}^{8}\subset S_{k}^{1}$ since $S_{k}^{8}$ is the same as $S_{k+1}^{1}$.
Each set from above contains STC-configurations that have a pattern in the sens that in an unfolding of the cyclic \HDA\ these STC-configurations would always represent the same corner. For example, the set $S_{k}^{0}$ contains the root of the STC-structure, when $k=1$ since the STC-configuration is just $(\emptyset,\emptyset,\emptyset)$, but it also contains another starting corner $(b_{1},b_{1},\emptyset)$ which is the STC-configuration after one backup has happened and at which point the same parallel behavior of $s$ with the next backup event can start.

There are steps between these STC-configurations that can be viewed as between $S_{k}$ sets, i.e., as between appropriate STC-configurations from those sets. These are \textbf{cs-steps} labeled by $s$ between STC-configurations of $S_{k}^{0}$ and $S_{k}^{1}$; \textbf{s-steps} labeled by $s$ between $S_{k}^{0}$ and $S_{k}^{3}$; and \textbf{s-steps} labeled $b$ between $S_{k}^{0}$ and $S_{k}^{4}$.
There are now \textbf{t-steps} between $S_{k}^{4}$ and $S_{k+1}^{0}$; these would contribute to forming the fix-point. Note that this last step goes to the next iteration level since one backup event has terminated and we may now start all the process again, but with the other remaining backups, hence working from $k+1$.
As we can see cancellation happen in the step above on the start of the shutdown event, cancellation can also happen on the termination of $s$ as with \textbf{ct-steps} labeled by $s$ between $S_{k}^{3}$ and $S_{k}^{2}$.

The \textit{final} cells of the \HDA\ now correspond to \textit{maximal} STC-configurations, and these are exactly those from $S_{k}^{2}$.

The example with the cyclic \HDA\ the is pictured as a cylinder in Figure~\ref{fig_ex_STC_structs}(right) can be captured only using the ST-structures. Even more, this example can be obtained from the above when transforming an STC-structure into an ST-structure by removing (making empty) the $C$ component. From the above one can see that after such a translation (i.e., remove the third part) the sets $S_{k}^{1}$ and $S_{k}^{3}$ would become the same; and the same happens to $S_{k}^{2}$ and $S_{k}^{6}$. But this is natural when looking at the cyclic \HDA: there, when going from the \HDA\ of Figure~\ref{fig_ex_STC_structs}(middle-right) to the one in Figure~\ref{fig_ex_STC_structs}(right) two more identifications of cells are made, i.e., the two upper corners and the left with the right transitions.
\end{example}

\subsection{Correspondences with Chu spaces}

We first show the correspondence between ST-structures and the Chu spaces over 3 of Pratt \cite{Pratt00HDArev}. The latter can be represented in terms of the 3-2 logic over $E$, the set of events. Instead of two values for each event, $0$ not started and $1$ finished (or before and after), the 3-valued case introduces the value of \executing\ to stand for \textit{during}, or \textit{in transition}. These values are ordered as $0 < \executing\ < 1$, which extends to the whole $3^{E}$.

Note that configuration structures \cite{GlabbeekP95config,GlabbeekP09configStruct} correspond to Chu spaces over 2.

A \textit{Chu space over K} is a triple $\chuPrat=(A,r,X)$ with $A$ and $X$ sets and $r:A\times X\rightarrow K$ is an arbitrary function called the matrix of the Chu space, and K is in our examples a set with a partial order on it. Chu spaces can be viewed in various equivalent ways. For our setting we can take the view of $A$ as the set of events and the $X$ as the set of configurations. The set $K$ is representing the possible values the events may take: when $K=\{0,1\}$ is the classical case of an event being either not started or terminated, where an order of $0<1$ would be used to define the steps in the system. In general, the order on $K$ will be used to define \textit{the meaningful steps} in the Chu space.

The Chu space can be viewed as a matrix with entries from $K$ and rows representing the events of $A$ and columns representing the configurations of $X$. As an example, an entry $r((e,X))=0$ says that the event $e$ is not started yet in the configuration $X$.

In consequence, a Chu space can also be viewed as the structure $(A,X)$ where $X\subseteq K^{A}$. This very much resembles the configuration structures when $K$ is $2$.

\begin{proposition}[ST-structures and Chu spaces over 3]\label{prop_STstructChu3}
ST-structures \\and \emph{Chu spaces over 3} (cf.\ \cite[Sec.3]{Pratt00HDArev}) are isomorphic. 
\end{proposition}

\begin{proof}[sketch]
We provide an association between Chu spaces over 3 and ST-structures.
For an ST-structure $\mathsf{ST}$ construct the associated Chu space over 3 $\chu{E}{X}^{\mathsf{ST}}$ with $E$ the set of events from $\mathsf{ST}$ and $X\subseteq 3^{E}$ states of the system formed of valuations of the events into the set $3=\{0, \executing, 1\}$ as follows. For one ST-configuration $(S,T)$ build a state $x^{(S,T)}\in X$ by assigning to each $e\in E$: 
\begin{itemize}
\item $e \rightarrow 0$ iff $e\not\in S \wedge e\not\in T$;
\item $e \rightarrow \executing$ iff $e\in S \wedge e\not\in T$;
\item $e \rightarrow 1$ iff $e\in S \wedge e\in T$.
\end{itemize}
The possibility $e\!\not\in\!S \wedge e\!\in\!T$ is dismissed by the requirement $T\!\subseteq\!S$ of ST-structures.
\end{proof}

STC-structures can be put into one-to-one correspondence with the Chu spaces over 4; or in other words, are isomorphic to the 4-2 logic (or functions $f:4^{E}\rightarrow 2$), cf.\ \cite{Pratt03trans_cancel}.
The order on the 4 values $\{0,\executing,1,\cancellation\}$ extends that on three values by making $0 < \cancellation$ and incomparable with $\executing$ and $1$. As for ST-structures, this order gives the possible steps in an STC-structure. The same s-step and t-step are defined as for ST-structures. The cancellation steps are then added, where the two kinds defined for STC-structures always take into the cancellation set only events that have not been started yet. 
To also allow for steps that enable event we need to extend with $\cancellation < 0$; but not comparable with the other two values so that we cannot jump from canceled events directly to them being in execution or terminated.

\begin{proposition}[STC and Chu spaces over 4]\label{prop_STC_Chu4}
STC-structures are {iso}-\\{morphic} to \emph{Chu spaces over 4}.
\end{proposition}

\begin{proof}[sketch]
We provide a translation of an STC-structure into a Chu space over 4, denoted $(E,X)^{\STC}$. Take $E$ to be the events of \STC. For each STC-configuration $(S,T,C)\in \STC$ construct a state $x^{(S,T,C)}\in X$ by giving the following assignment to the events $e\in E$:
\begin{itemize}
\item $e \rightarrow 0$ iff $e\not\in S \wedge e\not\in T \wedge e\not\in C$;
\item $e \rightarrow \executing$ iff $e\in S \wedge e\not\in T \wedge e\not\in C$;
\item $e \rightarrow 1$ iff $e\in S \wedge e\in T \wedge e\not\in C$;
\item $e \rightarrow \cancellation$ iff $e\not\in S \wedge e\not\in T \wedge e\in C$.
\end{itemize}
One can check that these four choice of membership for $e$ are the only ones, as the rest are discarded by the first or the second restrictions of an STC-configuration.
\end{proof}

\section{Conclusion}\label{sec_conclusion}

The work reported here was started in \cite{P12turing} where the notion of ST-structures was first defined. Nevertheless, the work in \cite{P12turing} is mostly concerned with investigations into the higher dimensional modal logic with past modalities and its relations to \HDAs\ and their bisimulations; whereas ST-structures get only little attention. In contrast, the present paper concentrates solely on ST(C)-configuration structures, investigating their expressiveness and relationship with existing concurrency formalisms, including \HDAs\ \cite{pratt91hda,Glabbeek06HDA}, configurations structures \cite{GlabbeekP95config,GlabbeekG01refinement}, and general (or inpure as we call them) event structures \cite{GlabbeekP09configStruct}. We gave definitions of various notions for ST(C)-structures like steps and paths, bisimulations, or action refinement, and discussed their relationships with similar notions for existing models of concurrency.

Having a good understanding of ST-structures (and their extension STC-structures) would help tackle the problem posed by Pratt in \cite{Pratt00HDArev} of getting a better understanding of the cyclic structure of \HDAs\ wrt.\ event based models. This in turn would give a better understanding of the state-event duality in concurrency models described in \cite{Pratt02duality,Pratt03trans_cancel}.

Interesting further investigations will ask how the correlations of Sections~\ref{subsec_expressST} and \ref{sec_STCstruct} can be expressed with category theory, following the works of Winskel and Nielsen \cite{winskel95modelsCategory} and of Cattani and Sassone \cite{CattaniSassone96HDTS}; also trying to see the connections with the category of Chu spaces of Gupta \cite{gupta94phd_chu} and the categorical work on cubical sets of Goubault and Mimram \cite{Goubault12Category_Cubical}.

The results in Section~\ref{subsec_expressST} reveal connections and distinctions between the existing concurrency models of \HDAs, configuration structures, pure event structures, and inpure event structures (and eventually general Petri nets). Some of these results are useful because they show existing knowledge in the new light given by the ST-structures.
In particular, the Corollary~\ref{cor_cubicalProp_configST} shows the difference between configuration structures and inpure event structures to be the fact that configuration structures are filled-in in essence, whereas inpure event structures are not. This also explains the counter example of the filled in concurrency square and the empty version, where the latter cannot be captured by pure event structures, but only by the impure case.
But both pure and impure event structures are adjacent-closed. Because of this, when dropping the adjacent-closure constraint the ST-structures become more expressive, and the example from Fig.~\ref{fig_ex_winskel}(right) models a natural concurrent system that falls into this category.
Moreover, the correspondence between acyclic and non-degenerate \HDAs\ and ST-structures that are adjacent-closed comes to say that non-degeneracy and adjacent-closure are close connected. Moreover, coupled with the above it comes to say that these \HDAs\ can capture the impure event structures, as well as suggesting a tighter correlation between these two by allowing acyclic and non-degenerate \HDAs\ to be encoded into impure event structures.

There are thought various examples that break either the acyclic constraint (as the ones in the last part of the paper) or that break the non-degeneracy constraint (as the one from Fig.~\ref{fig_ex_hda}(right-most) or from Figure~\ref{fig_ex_STC_structs}(ii)). These examples find natural representations as ST-structures or as STC-structures. The geometric interpretation of these examples is still very natural, only that the geometric objects fall outside the definition of \HDAs\ as we gave here in either the sense that we do not work with cubes any more but with triangles, or the geometric objects are open as boundaries are missing. It would be useful to investigate more these degenerate or cyclic geometric structures in the same line as started here, by looking at the \HDA\ state-based model which is close to the standard finite state machines used in computer science, and looking at the event-based models of ST(C)-structures and the related configuration and event structures.

Another point that ST-structures and the results in Section~\ref{subsec_expressST} make (especially Corollary~\ref{cor_ev_intermediaryTrans} and the results relating to impure event structures) is that ST-structures make the transitions more fine-grained. This says that if for impure event structures a concurrent transition implies that all the possible interleavings exist, for ST-structures a concurrent transition just says that the respective events are running at the same time, i.e., they overlap at least on some part of their execution (this is represented by the fact that those events are in the started stage but not terminated yet). On top of this concurrency aspect more constraints can be put on which events start first and which end before which. Such fine-graining cannot be achieved with the other concurrency models that we compare with in this paper.

\subsection{Further remarks}

\begin{remark}\label{remark_weak_ST_constraint}
The constraint imposed on ST-structures in Definition~\ref{def_st_structs} can be seen as rather strong, and one may think of a weaker constraint with the similar intuitive purpose of ensuring that events that are started eventually are terminated. Such a weaker version of the constraint would be:
\[
\mbox{ if } (S,T)\in ST \mbox{ then }\exists S' \mbox{ s.t.\ }S\subseteq S'\wedge(S',S')\in ST.
\]
Many of the results in the paper use the constraint of Definition~\ref{def_st_structs}. It is not clear if (or which of) these results would still hold under the above weaker constraint.
\end{remark}

The following is a natural example, with a natural geometric interpretation, that breaks the constraint of Definition~\ref{def_st_structs} but respects the above weaker constraint.

\begin{example}\label{example_weakerConstraint_on_ST}
Consider two events $a$ and $b$ which may run concurrently but which are constrained such that the event $b$ should run ``inside'' event $a$. Intuitively, the event $a$ is like an environment providing resources (like an operating system), whereas event $b$ is a process that is to use resources from the environment provided by $a$. Action refinement would provide more structure to these two events, and thus give a more concrete description of this concurrent system. But for now, abstractly, the fact that $b$ is to run ``inside'' $a$ should be understood as: $b$ cannot start unless $a$ has started already, and $a$ cannot terminate unless $b$ has terminated already. But both events are allowed to run concurrently.

An ST-structure representing this would be: 
\[
(\{a,b\},\{(\emptyset,\emptyset),(a,\emptyset),(\{a,b\},\emptyset),(\{a,b\},b),(\{a,b\},\{a,b\})\}).
\]
This structure is rooted and connected though not adjacent-closed. Moreover, this structure breaks the constraint of Definition~\ref{def_st_structs}, but one can easily check that the weaker version of this constraint is respected.

There is also a natural geometric formalization of the above example. Take the filled in square of Figure~\ref{fig_ex_hda}(middle-left). Remove the left and right borders of the square, leaving the inside filled in, as well as all the corners and the two remaining upper and lower borders. This space is directed as before, and a path through this space can be taken as before, only that there are many paths missing now: all those paths that start with $b$ (i.e., on the left border first) and all those paths that end $a$ before $b$ (i.e., reaching the right border before reaching the final upper-right corner). This is not a closed geometric shape any more, as two borders are missing. This shape does not fit the definition of non-degenerate \HDA\ as we gave it here because we required that the source and target be \textit{maps}, whereas here we would require them to be only \textit{partial}. 
\end{example}

\begin{remark}\label{remark_redundancy}
There is an apparent redundancy in the definition of adjacent-closure because the forth constraint was never used in the proof of Proposition~\ref{prop_adj_equiv}. This too is due to the strong constraint on the definition of ST-structures, which is clearly visible in the second part of this proof.
Under the weaker version of this constraint on ST-structures (see Remark~\ref{remark_weak_ST_constraint}) the proof of this proposition still holds but the second part of the proof needs to be redone, using induction on the reachable path. All four constraints are then necessary in this proof.

We did not go for these complications; and moreover, the definition that we gave for adjacent-closure highlights more explicitly the properties of adjacent-closure. 
\end{remark}

\vspace{1ex}\noindent\textbf{Acknowledgements: }
The work on ST-structures started from a fruitful conversation with Luca Aceto.
Olaf Owe and reviewers from FoSSaCS'\!13 and CONCUR'\!13 helped improve this draft.


%
%
%
%
%

\bibliographystyle{eptcs}
\bibliography{bib}

\begin{thebibliography}{10}
\providecommand{\bibitemdeclare}[2]{}
\providecommand{\surnamestart}{}
\providecommand{\surnameend}{}
\providecommand{\urlprefix}{Available at }
\providecommand{\url}[1]{\texttt{#1}}
\providecommand{\href}[2]{\texttt{#2}}
\providecommand{\urlalt}[2]{\href{#1}{#2}}
\providecommand{\doi}[1]{doi:\urlalt{http://dx.doi.org/#1}{#1}}
\providecommand{\bibinfo}[2]{#2}

\bibitemdeclare{inproceedings}{BaldanC10concur}
\bibitem{BaldanC10concur}
\bibinfo{author}{Paolo \surnamestart Baldan\surnameend} \&
  \bibinfo{author}{Silvia \surnamestart Crafa\surnameend}
  (\bibinfo{year}{2010}): \emph{\bibinfo{title}{{A Logic for True
  Concurrency}}}.
\newblock In: {\sl \bibinfo{booktitle}{CONCUR'10}}, {\sl
  \bibinfo{series}{{LNCS}}} \bibinfo{volume}{6269},
  \bibinfo{publisher}{Springer}, pp. \bibinfo{pages}{147--161}.

\bibitemdeclare{inproceedings}{BoudolC88flowEventStruct}
\bibitem{BoudolC88flowEventStruct}
\bibinfo{author}{G{\'e}rard \surnamestart Boudol\surnameend} \&
  \bibinfo{author}{Ilaria \surnamestart Castellani\surnameend}
  (\bibinfo{year}{1989}): \emph{\bibinfo{title}{{Permutation of transitions: An
  event structure semantics for CCS and SCCS}}}.
\newblock In: {\sl \bibinfo{booktitle}{REX Workshop}}, {\sl
  \bibinfo{series}{{LNCS}}} \bibinfo{volume}{354},
  \bibinfo{publisher}{Springer}, pp. \bibinfo{pages}{411--427}.

\bibitemdeclare{inproceedings}{CattaniSassone96HDTS}
\bibitem{CattaniSassone96HDTS}
\bibinfo{author}{Gian~Luca \surnamestart Cattani\surnameend} \&
  \bibinfo{author}{Vladimiro \surnamestart Sassone\surnameend}
  (\bibinfo{year}{1996}): \emph{\bibinfo{title}{Higher Dimensional Transition
  Systems}}.
\newblock In: {\sl \bibinfo{booktitle}{11th Annual IEEE Symposium on Logic in
  Computer Science (LICS'96)}}, \bibinfo{publisher}{IEEE}, pp.
  \bibinfo{pages}{55--62}.
\newblock
  \urlprefix\url{http://doi.ieeecomputersociety.org/10.1109/LICS.1996.561303}.

\bibitemdeclare{misc}{glabbeek96histUnfold}
\bibitem{glabbeek96histUnfold}
\bibinfo{author}{Rob \surnamestart van Glabbeek\surnameend}
  (\bibinfo{year}{1996}): \emph{\bibinfo{title}{{History Preserving Process
  Graphs}}}.
\newblock \bibinfo{howpublished}{Stanford University}.

\bibitemdeclare{inproceedings}{Glabbeek99invitedCONCUR}
\bibitem{Glabbeek99invitedCONCUR}
\bibinfo{author}{Rob~J. \surnamestart van Glabbeek\surnameend}
  (\bibinfo{year}{1999}): \emph{\bibinfo{title}{Petri Nets, Configuration
  Structures and Higher Dimensional Automata}}.
\newblock In: {\sl \bibinfo{booktitle}{CONCUR'99}}, {\sl
  \bibinfo{series}{{LNCS}}} \bibinfo{volume}{1664},
  \bibinfo{publisher}{Springer}, pp. \bibinfo{pages}{21--27}.
\newblock \urlprefix\url{http://dx.doi.org/10.1007/3-540-48320-9_3}.
\newblock \bibinfo{note}{(invited talk)}.

\bibitemdeclare{article}{Glabbeek06HDA}
\bibitem{Glabbeek06HDA}
\bibinfo{author}{Rob~J. \surnamestart van Glabbeek\surnameend}
  (\bibinfo{year}{2006}): \emph{\bibinfo{title}{{On the Expressiveness of
  Higher Dimensional Automata}}}.
\newblock {\sl \bibinfo{journal}{Theor. Comput. Sci.}}
  \bibinfo{volume}{356}(\bibinfo{number}{3}), pp. \bibinfo{pages}{265--290}.

\bibitemdeclare{article}{GlabbeekG01refinement}
\bibitem{GlabbeekG01refinement}
\bibinfo{author}{Rob~J. \surnamestart van Glabbeek\surnameend} \&
  \bibinfo{author}{Ursula \surnamestart Goltz\surnameend}
  (\bibinfo{year}{2001}): \emph{\bibinfo{title}{Refinement of actions and
  equivalence notions for concurrent systems}}.
\newblock {\sl \bibinfo{journal}{Acta Informatica}}
  \bibinfo{volume}{37}(\bibinfo{number}{4/5}), pp. \bibinfo{pages}{229--327}.

\bibitemdeclare{inproceedings}{GlabbeekP95config}
\bibitem{GlabbeekP95config}
\bibinfo{author}{Rob~J. \surnamestart van Glabbeek\surnameend} \&
  \bibinfo{author}{Gordon \surnamestart Plotkin\surnameend}
  (\bibinfo{year}{1995}): \emph{\bibinfo{title}{Configuration Structures}}.
\newblock In: {\sl \bibinfo{booktitle}{LICS}}, pp. \bibinfo{pages}{199--209}.

\bibitemdeclare{article}{GlabbeekP09configStruct}
\bibitem{GlabbeekP09configStruct}
\bibinfo{author}{Rob~J. \surnamestart van Glabbeek\surnameend} \&
  \bibinfo{author}{Gordon \surnamestart Plotkin\surnameend}
  (\bibinfo{year}{2009}): \emph{\bibinfo{title}{{Configuration structures,
  event structures and Petri nets}}}.
\newblock {\sl \bibinfo{journal}{Theor. Comput. Sci.}}
  \bibinfo{volume}{410}(\bibinfo{number}{41}), pp. \bibinfo{pages}{4111--4159}.

\bibitemdeclare{article}{GlabbeekV97splitting}
\bibitem{GlabbeekV97splitting}
\bibinfo{author}{Rob~J. \surnamestart van Glabbeek\surnameend} \&
  \bibinfo{author}{Frits~W. \surnamestart Vaandrager\surnameend}
  (\bibinfo{year}{1997}): \emph{\bibinfo{title}{{The Difference between
  Splitting in n and n+1}}}.
\newblock {\sl \bibinfo{journal}{Information and Computation}}
  \bibinfo{volume}{136}(\bibinfo{number}{2}), pp. \bibinfo{pages}{109--142}.

\bibitemdeclare{article}{Goubault12Category_Cubical}
\bibitem{Goubault12Category_Cubical}
\bibinfo{author}{Eric \surnamestart Goubault\surnameend} \&
  \bibinfo{author}{Samuel \surnamestart Mimram\surnameend}
  (\bibinfo{year}{2012}): \emph{\bibinfo{title}{Formal Relationships Between
  Geometrical and Classical Models for Concurrency}}.
\newblock {\sl \bibinfo{journal}{Electronic Notes in Theoretical Computer
  Science}} \bibinfo{volume}{283}, pp. \bibinfo{pages}{77 -- 109}.

\bibitemdeclare{phdthesis}{gupta94phd_chu}
\bibitem{gupta94phd_chu}
\bibinfo{author}{Vincent \surnamestart Gupta\surnameend}
  (\bibinfo{year}{1994}): \emph{\bibinfo{title}{{Chu Spaces: A Model of
  Concurrency}}}.
\newblock Ph.D. thesis, \bibinfo{school}{Stanford University}.

\bibitemdeclare{article}{HoareMSW11CKA_foundationsJLAP}
\bibitem{HoareMSW11CKA_foundationsJLAP}
\bibinfo{author}{C.~A.~R. \surnamestart Hoare\surnameend},
  \bibinfo{author}{Bernhard \surnamestart M{\"o}ller\surnameend},
  \bibinfo{author}{Georg \surnamestart Struth\surnameend} \&
  \bibinfo{author}{Ian \surnamestart Wehrman\surnameend}
  (\bibinfo{year}{2011}): \emph{\bibinfo{title}{{Concurrent Kleene Algebra and
  its Foundations}}}.
\newblock {\sl \bibinfo{journal}{J. Log. Algebr. Program.}}
  \bibinfo{volume}{80}(\bibinfo{number}{6}), pp. \bibinfo{pages}{266--296}.
\newblock \urlprefix\url{http://dx.doi.org/10.1016/j.jlap.2011.04.005}.

\bibitemdeclare{inproceedings}{NielsenPW79eventstructures}
\bibitem{NielsenPW79eventstructures}
\bibinfo{author}{Mogens \surnamestart Nielsen\surnameend},
  \bibinfo{author}{Gordon \surnamestart Plotkin\surnameend} \&
  \bibinfo{author}{Glynn \surnamestart Winskel\surnameend}
  (\bibinfo{year}{1979}): \emph{\bibinfo{title}{Petri Nets, Event Structures
  and Domains.}}
\newblock In: {\sl \bibinfo{booktitle}{Semantics of Concurrent Computation}},
  {\sl \bibinfo{series}{{LNCS}}}~\bibinfo{volume}{70},
  \bibinfo{publisher}{Springer}, pp. \bibinfo{pages}{266--284}.

\bibitemdeclare{inproceedings}{phillips11express}
\bibitem{phillips11express}
\bibinfo{author}{Iain \surnamestart Phillips\surnameend} \&
  \bibinfo{author}{Irek \surnamestart Ulidowski\surnameend}
  (\bibinfo{year}{2011}): \emph{\bibinfo{title}{{A Logic with Reverse
  Modalities for History-preserving Bisimulations}}}.
\newblock In: {\sl \bibinfo{booktitle}{EXPRESS'11}}, {\sl
  \bibinfo{series}{EPTCS}}~\bibinfo{volume}{64}, pp. \bibinfo{pages}{104--118}.

\bibitemdeclare{misc}{Pratt96reconcilingevent}
\bibitem{Pratt96reconcilingevent}
\bibinfo{author}{Vaughan \surnamestart Pratt\surnameend}
  (\bibinfo{year}{1996}): \emph{\bibinfo{title}{Reconciling Event Structures
  and Higher Dimensional Automata}}.

\bibitemdeclare{inproceedings}{pratt91hda}
\bibitem{pratt91hda}
\bibinfo{author}{Vaughan~R. \surnamestart Pratt\surnameend}
  (\bibinfo{year}{1991}): \emph{\bibinfo{title}{Modeling Concurrency with
  Geometry}}.
\newblock In: {\sl \bibinfo{booktitle}{POPL'91}}, pp.
  \bibinfo{pages}{311--322}.

\bibitemdeclare{incollection}{pratt95chu}
\bibitem{pratt95chu}
\bibinfo{author}{Vaughan~R. \surnamestart Pratt\surnameend}
  (\bibinfo{year}{1995}): \emph{\bibinfo{title}{Chu Spaces and their
  Interpretation as Concurrent Objects}}.
\newblock In: {\sl \bibinfo{booktitle}{Computer Science Today: Recent Trends
  and Develop.}}, {\sl \bibinfo{series}{LNCS}} \bibinfo{volume}{1000},
  \bibinfo{publisher}{Springer}, pp. \bibinfo{pages}{392--405}.

\bibitemdeclare{article}{Pratt00HDArev}
\bibitem{Pratt00HDArev}
\bibinfo{author}{Vaughan~R. \surnamestart Pratt\surnameend}
  (\bibinfo{year}{2000}): \emph{\bibinfo{title}{Higher dimensional automata
  revisited}}.
\newblock {\sl \bibinfo{journal}{Math. Struct. Comput. Sci.}}
  \bibinfo{volume}{10}(\bibinfo{number}{4}), pp. \bibinfo{pages}{525--548}.

\bibitemdeclare{inproceedings}{Pratt02duality}
\bibitem{Pratt02duality}
\bibinfo{author}{Vaughan~R. \surnamestart Pratt\surnameend}
  (\bibinfo{year}{2002}): \emph{\bibinfo{title}{{Event-State Duality: The
  Enriched Case}}}.
\newblock In: {\sl \bibinfo{booktitle}{CONCUR'02}}, {\sl
  \bibinfo{series}{{LNCS}}} \bibinfo{volume}{2421},
  \bibinfo{publisher}{Springer}, pp. \bibinfo{pages}{41--56}.
\newblock \urlprefix\url{http://dx.doi.org/10.1007/3-540-45694-5_3}.

\bibitemdeclare{article}{Pratt03trans_cancel}
\bibitem{Pratt03trans_cancel}
\bibinfo{author}{Vaughan~R. \surnamestart Pratt\surnameend}
  (\bibinfo{year}{2003}): \emph{\bibinfo{title}{{Transition and Cancellation in
  Concurrency and Branching Time}}}.
\newblock {\sl \bibinfo{journal}{Math. Struct. Comput. Sci.}}
  \bibinfo{volume}{13}(\bibinfo{number}{4}), pp. \bibinfo{pages}{485--529}.

\bibitemdeclare{inproceedings}{P12turing}
\bibitem{P12turing}
\bibinfo{author}{Cristian \surnamestart Prisacariu\surnameend}
  (\bibinfo{year}{2012}): \emph{\bibinfo{title}{{The Glory of the Past and
  Geometrical Concurrency}}}.
\newblock In: {\sl \bibinfo{booktitle}{The Alan Turing Centenary Conference
  (Turing-100)}}, {\sl \bibinfo{series}{EPiC}}~\bibinfo{volume}{10}, pp.
  \bibinfo{pages}{252--267},
  \doi{http://www.easychair.org/publications/?page=1273355332}.

\bibitemdeclare{inproceedings}{Winskel86}
\bibitem{Winskel86}
\bibinfo{author}{Glynn \surnamestart Winskel\surnameend}
  (\bibinfo{year}{1986}): \emph{\bibinfo{title}{Event Structures}}.
\newblock In: {\sl \bibinfo{booktitle}{Advances in Petri Nets}}, {\sl
  \bibinfo{series}{{LNCS}}} \bibinfo{volume}{255},
  \bibinfo{publisher}{Springer}, pp. \bibinfo{pages}{325--392}.

\bibitemdeclare{incollection}{winskel95modelsCategory}
\bibitem{winskel95modelsCategory}
\bibinfo{author}{Glynn \surnamestart Winskel\surnameend} \&
  \bibinfo{author}{Mogens \surnamestart Nielsen\surnameend}
  (\bibinfo{year}{1995}): \emph{\bibinfo{title}{Models for Concurrency}}.
\newblock In \bibinfo{editor}{Samson \surnamestart Abramski\surnameend},
  \bibinfo{editor}{Dov~M. \surnamestart Gabbay\surnameend} \&
  \bibinfo{editor}{Tom~S.E. \surnamestart Maibaum\surnameend}, editors: {\sl
  \bibinfo{booktitle}{Handbook of Logic in Computer Science -- vol 4 --
  Semantic Modelling}}, \bibinfo{publisher}{Oxford University Press}, pp.
  \bibinfo{pages}{1--148}.

\end{thebibliography}

\end{document}